\documentclass[a4paper,twocolumn,11pt,unpublished]{quantumarticle}
\pdfoutput=1
\usepackage[utf8]{inputenc}
\usepackage[english]{babel}
\usepackage[T1]{fontenc}
\usepackage{amsmath}
\usepackage{amssymb}
\usepackage{amsthm}
\usepackage{hyperref}
\usepackage{physics}
\usepackage{algorithm}
\usepackage{algpseudocode}
\usepackage{paralist}

\usepackage[numbers,sort&compress]{natbib}
\usepackage{tensor}
\usepackage[normalem]{ulem}
\usepackage{tikz}
\usetikzlibrary{positioning, fit, calc, shapes, arrows, arrows.meta, backgrounds, patterns, matrix}
\usetikzlibrary{positioning}
\usepackage{quantikz}
\usepackage{lipsum}
\usepackage{xspace}

\usepackage[frozencache]{minted}
\usepackage[most,minted]{tcolorbox}

\usepackage{graphicx}
\usepackage[strings]{underscore}

\usepackage{bm}
 
\definecolor{myblue}{RGB}{68,114,196}
\definecolor{mygreen}{RGB}{112,173,71}
\definecolor{myyellow}{RGB}{255,192,0}
\definecolor{myred}{RGB}{255,0,0}
\definecolor{mypurple}{RGB}{112,48,160}
 
\definecolor{quantumblue}{HTML}{53257F}
\definecolor{quantumlightblue}{HTML}{DDC6F2}

\newcommand{\pythoninline}[1]{\mintinline{python}{#1}}

\setminted[python]{
    fontsize=\footnotesize,
    breaklines=true,
    breakafter={(=.;/\\-},
    breakbefore={\#\%)},
    style=friendly,
    autogobble=true
}

\setmintedinline[python]{
    fontsize=\normalsize
}

\tcbset{
    quantumstyle/.style={
        enhanced,
        breakable,
        colback=white,
        colframe=quantumblue,
        colbacktitle=quantumlightblue,
        coltitle=black,
        fonttitle=\bfseries,
        toptitle=1mm,
        bottomtitle=1mm,
        segmentation style={dashed, quantumblue},
        fontupper=\scriptsize,
        fontlower=\scriptsize\ttfamily,
        size=small
    }
}

\newcounter{quantumbox}[section]
\makeatletter
\renewcommand{\p@quantumbox}{\thesection.}
\makeatother

\newtcolorbox[use counter=quantumbox]{quantumstacked}[2][]{
    quantumstyle,
    label type=quantumbox,
    label={#1},
    title={Box~\thesection.\thequantumbox: #2}
}

\newtcolorbox[use counter from=quantumstacked]{quantumtwocol}[2][]{
    quantumstyle,
    label type=quantumbox,
    label={#1},
    title={Box~\thesection.\thequantumbox: #2},
    sidebyside,
    sidebyside align=top seam,
    lefthand width=0.43\linewidth
}

\usepackage{todonotes}

\usepackage{array}

\newcolumntype{C}[1]{>{\centering\arraybackslash}p{#1}}

\usepackage{dsfont}
\usepackage{mathtools}
\usepackage{tabularray}
\UseTblrLibrary{booktabs}

\newtheorem{definition}{Definition}

\newtheorem{remark}{Remark}

\newcommand{\ie}{i.e.\xspace}

\DeclareMathSymbol{\shortminus}{\mathbin}{AMSa}{"39}

\newcommand{\Id}{{I}}

\newcommand{\Ha}{{H}}

\newcommand{\Sa}{{S}} 

\newcommand{\CZ}{{CZ}}

\newcommand{\C}{{\sf C}}
\newcommand{\N}{{\sf N}}
\newcommand{\E}{{\sf E}}
\newcommand{\M}{{\sf M}}
\newcommand{\X}{{\sf X}} 
\newcommand{\Z}{{\sf Z}} 
 
\newcommand{\A}{{\sf A}}
\newcommand{\B}{{\sf B}}

\newcommand{\odd}{{\operatorname{Odd}}}

\newcommand{\planeXY}{{\rm XY}}
\newcommand{\planeXZ}{{\rm XZ}}
\newcommand{\planeYZ}{{\rm YZ}}
\newcommand{\axisX}{{\rm X}}
\newcommand{\axisY}{{\rm Y}}
\newcommand{\axisZ}{{\rm Z}}

\newcommand\Mdom[3]{\tensor[_{#3}]{\left[\M#1\right]}{^{#2}}}

\newcommand{\numpy}{NumPy\xspace}
\newcommand{\stim}{Stim\xspace}
\newcommand{\networkx}{NetworkX\xspace}
\newcommand{\perceval}{Perceval\xspace}
\newcommand{\numba}{Numba\xspace}
\newcommand{\pyzx}{PyZX\xspace}
\newcommand{\tket}{Tket\xspace}
\newcommand{\sympy}{SymPy\xspace}
\newcommand{\mypy}{Mypy\xspace}
\newcommand{\pyright}{Pyright\xspace}
\newcommand{\qasm}{OpenQASM\xspace}
\newcommand{\qiskit}{Qiskit\xspace}

\usepackage{tablefootnote}

 \tikzset{brickline/.style={dash pattern=on 3pt off 2pt, line width=0.8pt}}
 
\newcommand{\brickwork}[3]{%
  \begingroup
  \def\w{#1}%
  \def\h{#2}%
  \def\labels{#3}%

  \foreach \x in {0,...,\numexpr\w-1\relax} {%
    \foreach \y in {0,...,\h} {%
      \draw[brickline] (\x,\y) -- (\x+1,\y);
    }%
  }%
  \foreach \x in {1,...,\w} {%
    \foreach \y in {0,...,\numexpr\h-1\relax} {%
      \pgfmathtruncatemacro{\remx}{mod(\x,4)}%
      \pgfmathtruncatemacro{\cond}{mod(int((\x-1)/4),2)}%
      \pgfmathtruncatemacro{\remy}{mod(\y,2)}%
      \ifnum\remx=0
         \ifnum\remy=\cond
            \draw[brickline] (\x,\y) -- (\x,\y+1);
         \fi
      \else\ifnum\remx=2
         \ifnum\remy=\cond
            \draw[brickline] (\x,\y) -- (\x,\y+1);
         \fi
      \fi\fi
    }%
  }%
  \foreach [count=\row] \rowlabels in \labels {%
    \foreach [count=\col] \lab in \rowlabels {%
      \ifnum\col=1
        \node[draw, rectangle, fill=white, minimum size=16pt, inner sep=2pt]
          at (\col-1,\h-\row+1) {\(\lab\)};
      \else
        \node[draw, circle, fill=white, minimum size=16pt, inner sep=0pt]
          at (\col-1,\h-\row+1) {\(\lab\)};
      \fi
    }%
  }%
  \foreach \y in {0,...,\h} {%
    \node[draw, circle, fill=gray!30, minimum size=16pt, inner sep=0pt]
      at (\w, \y) {};
  }%
  \endgroup
}

\newcommand{\brickrow}[1]{%
  \begingroup
  \pgfmathtruncatemacro{\totalcols}{4*#1}%

  \foreach \x in {0,...,\numexpr\totalcols-1\relax} {%
    \foreach \y in {1,2} {%
      \draw[brickline] (\x,\y) -- (\x+1,\y);
    }%
  }%

  \foreach \x in {2,4,...,\totalcols} {%
    \draw[brickline] (\x,1) -- (\x,2);
  }%

  \foreach \x in {0,...,\totalcols} {%
    \foreach \y in {1,2} {%
      \ifnum\x=0
        \node[draw, rectangle, fill=white,
              minimum size=16pt, inner sep=0pt]
          at (\x,\y) {};
      \else
        \ifnum\x=\totalcols
          \node[draw, circle, fill=gray!30,
                minimum size=16pt, inner sep=0pt]
            at (\x,\y) {};
        \else
          \node[draw, circle, fill=white,
                minimum size=16pt, inner sep=0pt]
            at (\x,\y) {};
        \fi
      \fi
    }%
  }%
  \endgroup
}

\begin{document}

\title{Graphix: A software framework for Measurement-Based Quantum Computation}

\author{Mateo Uldemolins}
\affiliation{DIENS, \'Ecole Normale Sup\'erieure, PSL University, CNRS, INRIA, 45 rue d'Ulm, Paris 75005, France}

\author{Pranav Nair}
\affiliation{DIENS, \'Ecole Normale Sup\'erieure, PSL University, CNRS, INRIA, 45 rue d'Ulm, Paris 75005, France}

\author{Emlyn Graham}
\affiliation{DIENS, \'Ecole Normale Sup\'erieure, PSL University, CNRS, INRIA, 45 rue d'Ulm, Paris 75005, France}

\author{Shinichi Sunami}
\affiliation{Clarendon Laboratory, University of Oxford, Oxford OX1 3PU, United Kingdom}

\author{Thierry Martinez}
\email{thierry.martinez@inria.fr}
\affiliation{DIENS, \'Ecole Normale Sup\'erieure, PSL University, CNRS, INRIA, 45 rue d'Ulm, Paris 75005, France}

\author{Maxime Garnier}
\email{maxime.garnier@inria.fr}
\affiliation{DIENS, \'Ecole Normale Sup\'erieure, PSL University, CNRS, INRIA, 45 rue d'Ulm, Paris 75005, France}


\maketitle
\begin{abstract}
Measurement-based quantum computing (MBQC) is a powerful model for quantum computation, but dedicated software tools that bridge its theoretical foundations with practical research workflows remain limited. We present Graphix, a software framework for MBQC written in Python that provides a unified environment for developing, integrating, and exploring measurement-based protocols and algorithms. Graphix establishes a modular, extensible, and user-friendly architecture for the compilation and simulation of quantum computations in the MBQC model with abstractions closely aligned with the theoretical formulations of MBQC. Through examples drawn from recent research on MBQC, we demonstrate Graphix's ability to reproduce and extend previous results. Graphix thus provides a software infrastructure for MBQC, supporting education and research while enabling collaborative development and accelerating the transition toward practical implementations.
\end{abstract}



\section{Introduction}
\label{sec:intro}
Quantum computing primarily relies on the circuit model, where unitary quantum gates are applied sequentially to an input state. Its success stems from its intuitive graphical representation and its ability to provide a hierarchy of abstractions, from universal gate sets to hardware-aware compiled circuits. However, alternative computational models offer complementary perspectives and can motivate new theoretical and architectural developments. A notable example is the one-way model \cite{RB01:oneway,RBB03:computation,DKP07:calculus}, where computation proceeds through single-qubit measurements on a specific class of entangled resource states using 
classical feed-forward, i.e., the ability to make quantum operations dependent on classical signals obtained during the computation.

In this work, Measurement-Based Quantum Computing (MBQC) refers to the universal one-way model introduced by Raussendorf and Briegel \cite{RB01:oneway}. In this model, the resource state is a graph state, namely, a quantum state defined by a graph, with its vertices and edges representing, respectively, \(\ket+\) states and controlled Pauli-$Z$ ($CZ$) entangling  operations between the corresponding states. Other universal measurement-driven models have been proposed, based on different resource states and restricted single-qubit measurements \cite{Kissinger2019:parityphaseMBQC,RYA23:MBQCString} or on teleportation-based two-qubit measurements \cite{Aliferis04:equivalence}, but they share the same computational principles. 
We therefore adopt the one-way model as the reference framework throughout this work, as it is the most studied MBQC formulation and provides a simple yet expressive setting for theoretical developments. More generally, classical feed-forward defines the measurement-based paradigm beyond qubits, including qudits \cite{BKMM+23:quditMBQC,RD26} and continuous-variables models \cite{Menicucci06:CVMBQC,GWMRvL_09,Menicucci14:ftMBQC,BD2023:CVflow}.

Although universal, MBQC is not merely a reformulation of the unitary circuit model. Its native inclusion of ancilla qubits, mid-circuit measurements, and classical feed-forward enables computational processes beyond the circuit abstraction. This has motivated applications ranging from imaginary-time evolution and variational algorithms \cite{ARS24:nunit,F+21:mbqcvqa} to quantum machine learning \cite{SHG23,CFRB24:MBQML,M+24:vargenmod}, pseudo-randomness generation \cite{MGDM18:pseudo}, and Hamiltonian simulation \cite{KSNHB25}. Even in the unitary regime, MBQC provides advantages in depth complexity \cite{BKP10:MBQC_depth_complexity} and alternative approaches to circuit parallelization \cite{BK09:parallel,DKPP09:extended_calculus}.

Additionally, MBQC supports quantum error
correction and fault tolerance \cite{SDKO07,RHG07,BDCPS16,NB18,BR20} and has been at the forefront of
confidence-building for long-term applications such as secure delegated quantum computing \cite{BFK09:UBQC,FK17,LMKO21}. This has motivated its investigation for scalable architectures, particularly in photonic and distributed quantum computing settings \cite{B+23:FBQC,deFelice2026dataflowprogramming,S+25:mod,NST26}. As far as practical implementations are concerned, MBQC has been demonstrated through proof-of-concept experiments, notably in photonic and trapped-ion platforms \cite{Pan11,Barz12,Lanyon13,DNMN+24:verified,GLMG+25,HHEB26}. While recent experiments have demonstrated classical feed-forward capabilities \cite{S+26}, achieving fully integrated and scalable MBQC remains challenging because it requires measurement outcomes to be processed and used to condition subsequent quantum operations within the relevant time scales. Recent advances across photonic, superconducting, trapped-ion, and neutral-atom platforms have progressively improved the experimental capabilities required for such protocols, narrowing the gap between proof-of-principle demonstrations and scalable MBQC implementations \cite{P+07:opticalFF,B+24:dynamic,P+21:qccd,G+23:mcm_atom,R+26}. As these experimental platforms become more capable, corresponding software tools are needed to design, simulate, and validate increasingly complex MBQC protocols.

Software frameworks play a central role in the development of computational models. Beyond implementing theoretical concepts, they enable reproducible research, facilitate the comparison and validation of methods, and provide a common language through which ideas can be shared and extended. As a field matures, reliable software becomes a shared research infrastructure that allows independent contributions to accumulate into a coherent ecosystem.

Accordingly, mature quantum computing paradigms have developed dedicated software ecosystems adapted to their computational abstractions. The circuit model is supported by general-purpose frameworks such as Qiskit, PennyLane, Cirq, or Qibo \cite{Aleksandrowicz19:qiskit,qiskit24,PL22,Cirq25,Qibo21}, as well as specialized tools such as \stim for quantum error correction and fault-tolerant quantum computing \cite{Gidney21:stim}. Dedicated software has similarly emerged for analog quantum simulation and optical quantum computing, with examples including Bloqade and Perceval \cite{WWLL24:bloqade,H+23:perceval}. More generally, QuTiP provides a versatile environment for simulating quantum systems beyond a specific computational model \cite{L+26:qutip}.


Although MBQC can in principle be implemented within the circuit model (see, for example, \cite{XanaduMBQC}), such approaches do not naturally expose the abstractions that define the MBQC formalism necessary for research workflows centered around the model itself. While MBQC has benefited from a growing collection of dedicated software tools, which we discuss in detail below, the field still lacks a comprehensive software infrastructure that integrates its diverse theoretical developments into a coherent and extensible framework.

Graphix is designed to address this need by serving as a common software foundation for MBQC and an extensible integration target where new methods can be implemented, compared, combined, and built upon within a shared representation of the model. This work presents the current state of the library and details the contributions developed by the authors since June 2023, building upon the original framework introduced in \cite{SF22:graphix}. These contributions focus on three main objectives: (i) the conceptual unification and consolidation of the framework, (ii) the development and integration of state-of-the-art algorithms for compiling and analyzing MBQC computations, and (iii) the re-engineering of the codebase to improve software quality, reliability, usability, and maintainability. The current architecture results from practical experience using, extending, and refining the framework for MBQC research, with design choices informed by both successful approaches and lessons learned during its development. Together, these developments establish Graphix as a comprehensive framework for MBQC compilation, optimization, and simulation.

The core of Graphix is an MBQC-native representation based on the measurement calculus \cite{DKP07:calculus}, where measurement patterns are treated as first-class computational objects. Graphix extends this representation with local Clifford operations, open graphs, and flow-finding algorithms \cite{MB24:algebraic}, enabling compilation, optimization, simulation, and analysis directly within the MBQC formalism. Such MBQC-native representations are also relevant beyond direct MBQC execution, for instance through connections with the ZX calculus, where graphical formalisms provide intermediate representations for circuit optimization and transformation \cite{vandewetering20:zxcalculus,Duncan2012:graphical,BMBdF+21}. It also supports computations beyond the circuit model, including patterns whose deterministic implementation relies on mid-circuit measurements and classical feed-forward, as well as native state vector and noisy simulation capabilities, and an interface to \stim for stabilizer simulation \cite{Gidney21:stim}.

Beyond its built-in capabilities, Graphix is designed as a foundation for further developments. Its architecture enables specialized tools and research applications to build upon a shared MBQC infrastructure, including Veriphix \cite{veriphix2024,veriphix_paper2026}, which extends Graphix for implementing verification of secure quantum computation protocols in the measurement-based model \cite{BFK09:UBQC,FK17,LMKO21}, and QPatLib \cite{S26:qpatlib}, which reuses Graphix components for benchmarking MBQC optimization strategies. 

More broadly, Graphix illustrates the mutually reinforcing relationship between theory development and research software. By enabling new ideas to be implemented, tested, compared, and reused within a common framework, software becomes an integral part of the research process itself. Developed through continuous use in MBQC research, Graphix provides a robust and extensible foundation for future developments across theory, architecture, and applications, serving as a common integration target for the field.

\paragraph{Related works}
As a general-purpose MBQC library, Graphix overlaps with and extends existing MBQC software. We briefly review the most prominent projects and compare them to
Graphix.

Paddle Quantum incorporates an MBQC module within a larger quantum and
machine learning framework \cite{paddleQ}. It provides an ad-hoc circuit-to-pattern transpiler, a partial implementation of the
measurement calculus (it only supports the $\planeXY$ and $\planeYZ$ measurement planes, while the
extended measurement calculus also includes $\planeXZ$ \cite{DKPP09:extended_calculus}) enabling pattern
standardization, and a state vector simulator. In a similar spirit, the recent Python package DeepQuantum
\cite{HE25:deepQ} integrates machine learning frameworks with MBQC, partially building on existing functionality developed in
Graphix.

Authors in \cite{EOSR23:mcbeth} formalised the MBQC model as the quantum programming language MCBeth. This library
written in OCaml with a Python interface enables an MBQC-native implementation of quantum algorithms, supports pattern
standardization as well as weak and strong pattern simulation. While both tools allow to translate patterns to \qasm \cite{CJAdBBH22:qasm}
circuits with ancillas and feed-forward, the MCBeth library does not support circuit-to-pattern transpilation or unitary
extraction from patterns \cite{Simmons21}.

Packages such as the Pauli Tracking Library \cite{RD24} and t$\overline{\mbox{u}}$Q \cite{BD25:tuq} focus on the
compilation and optimization of cluster-state computations rather than providing comprehensive support for the
measurement calculus and simulation. Although some optimization techniques based on local-Clifford equivalence of graph
states are shared with Graphix \cite{KVP+24}, these tools primarily target compilation workflows and therefore have
limited functional overlap with Graphix.

Finally, the MentPy \cite{CFRB24:MBQML, MENTPY} library focuses on
MBQC-native variational algorithms and machine learning \cite{F+21:mbqcvqa,M+24:vargenmod,CFRB24:MBQML} which is
complementary to Graphix and paves the way for more developments in this direction.


\paragraph{Structure of the paper}

The remainder of this paper is structured as follows. In Section~\ref{sec:mbqc} we review the essential concepts of MBQC in a tutorial style with Graphix code examples. This serves two purposes: (i) it acts as a primer to
readers unfamiliar with MBQC, allowing them to skip the technical details and jump directly to the discussion of research workflows in Section~\ref{sec:applications} and/or directly start using Graphix, and (ii) it highlights the conceptual structure mirrored
in Graphix. Section~\ref{sec:graphix_lib} details the overall architecture of the library as well as several technical
and more advanced features. We conclude by summarizing the main
contributions of this work and mapping out several directions for future research, development and use.

\paragraph{Reproducibility} All the code and history of Graphix is available at \cite{graphix_repo}. All the code required to reproduce the present paper including the
data and plots contained in it are available at \cite{graphix_paper_repo}. Code examples use Graphix v0.4.

\section{MBQC fundamentals with Graphix}
\label{sec:mbqc}
In the MBQC model, a quantum computation is carried out in a fundamentally different way from the circuit model. Rather than applying a sequence of unitary gates on an input state, one first prepares an entangled multi-qubit state that serves as a computational resource --- with the input state embedded in it --- and then performs a sequence of single-qubit measurements. In the context of this work, the resource is always a graph state. To compensate for the inherent randomness of quantum measurements, Pauli corrections conditioned on previous measurement outcomes are applied. At the end, the initial resource state is ``consumed'' by the computation and only the output qubits encoding the computation result remain.

In the remainder of this section, we introduce the basic concepts and formal tools needed to describe MBQC, with a special emphasis on their representation in the Graphix software.


\subsection{Open graphs and correction functions}
\label{subsec:og}

The fundamental object underpinning the MBQC model is the \emph{labelled open graph} \(\Gamma = (G, I, O, \lambda)\) which consists of:
\begin{itemize}
    \item an undirected graph \(G = (V, E)\) with nodes $V$ and edges $E$ respectively representing the qubits and the entanglement structure of a graph state,
    \item two sequences \(I, O \in \bigcup_{n\geq 0}V^n\) without repetition denoting the input and output qubits of the computation,
    \item a map \(\lambda: O^c \to \Bqty{\planeXY, \planeXZ, \planeYZ} \) assigning a measurement plane to each non-output qubit, where $O^c := V\setminus O$.
\end{itemize}

A labelled open graph is a \emph{partial graph state}, that is, a graph state with full freedom over the state on the input nodes,
\begin{equation}
    \ket{\Gamma, \psi} := \prod_{(i,j)\in E} CZ_{ij}\pqty{\ket{+}^{\otimes|I^c|}\otimes\ket{\psi}_I},
    \label{eq:og_state}
\end{equation}
equipped with a measurement map $\lambda$. Here, \(CZ_{ij}\) denotes the controlled-$Z$ gate applied on qubits \(i\) and \(j\). Owing to the definition in Eq.~\eqref{eq:og_state}, we use the terms qubits and nodes interchangeably in the text. Furthermore, the ordering of nodes in $I$ and $O$ represents the order of the qubit registers in the input and output Hilbert spaces. Open graphs states are \emph{partial stabilizer states}, \(K_j \ket{\Gamma, \psi} = \ket{\Gamma, \psi}\), with $K_j := X_j \prod_{i \in N_G(j)} Z_i \; \forall j \in I^c$ the stabilizer generators and $\ket{\psi}_I$ an arbitrary state in the input Hilbert space, where $N_G(j)$ is the neighbourhood of node $j$ in the graph $G$ and $I^c := V\setminus I$ \cite{BKMP07:gflow}.

A labelled open graph is not sufficient to fully characterize measurement-based computations.
In full generality, any computation can be represented by the tuple \((\Gamma, \alpha, \bm{x}, \bm{z})\):

\begin{itemize}
    \item a labelled open graph \(\Gamma = (G, I, O, \lambda)\),
    \item a map \(\alpha: O^c \to \left[0,2\pi \right) \) assigning a measurement angle to each non-output qubit. When \(\alpha(i) \in \Bqty{0, \frac\pi2, \pi, \frac{3\pi}2 }\), the pair \(\lambda(i), \alpha(i)\) can be replaced by  an axis $\{\axisX, \axisY, \axisZ\}$ with a sign. Since axes belong to two planes, we chose the convention $(\planeXY, 0) := +\axisX$, $(\planeXY, \frac\pi2) := +\axisY$, $(\planeXZ, 0) := +\axisZ$, $(\planeXZ, \frac\pi2) := +\axisX$, $(\planeYZ, \frac{3\pi}2) := -\axisY$ and similarly for the other cases. These are denoted \emph{Pauli measurements} in contrast to the more general \emph{planar measurements}. 

    \item a correction strategy in the form of two functions \({\bm x}, \, {\bm z}: O^c \to \mathcal P(I^c) \) where \({\bm x}(i)\) is the set of nodes that receive a Pauli \(\X^{s_i}\) correction \emph{conditional} on a classical bit $s_i \in \{0, 1\}$. Conversely, \({\bm z}(i)\) is the set of nodes that receive a Pauli \(\Z^{s_i}\) correction. Here, $\mathcal P(I^c)$ denotes the power set of the non-input nodes. Crucially, for the associated computation to be \emph{runnable}, the transitive closure of $\Bqty{(i, j) \, \vert \,j \in \bm{x}(i) \cup \bm{z}(i)}$ must define a strict partial order $\prec$ on the set of measured nodes \cite{PS16, BMBdF+21}.
\end{itemize}

The plane-angle pair $(\lambda(i), \alpha(i))$ associated with qubit $i$ defines a single-qubit measurement as a projection onto the states \(\ket{\pm_{\lambda(i),\alpha(i)}}\). Specifically,
\begin{subequations}
\begin{align}
    \ket{\pm_{{\rm XY}, \alpha}} & = \frac{1}{\sqrt2}\left(\ket0 \pm e^{i\alpha}\ket1\right), \label{eq:meas_state_xy}\\
    \ket{\pm_{{\rm XZ}, \alpha}} & = t^{\alpha}_{\pm}\ket0 \pm t^{\alpha}_{\mp}\ket1,\\
    \ket{\pm_{{\rm YZ}, \alpha}} & = t^{\alpha}_{\pm}\ket0 \pm i t^{\alpha}_{\mp}\ket1,
\end{align}
\label{eq:meas_states}
\end{subequations}
with $t^{\alpha}_+ = \cos\left(\frac{\alpha}{2}\right)$ and $t^{\alpha}_- = \sin\left(\frac{\alpha}{2}\right)$. The classical bit \(s_i\) introduced in the definition of the correction strategy represents precisely the measurement outcome of qubit \(i\), with the convention that \(s_i = 0\) corresponds to projecting onto \(\ket{+_{\lambda(i),\alpha(i)}}\) and \(s_i = 1\) onto \(\ket{-_{\lambda(i),\alpha(i)}}\).

Due to the probabilistic nature of projective measurements, computations with $|O^c|$ measurements implement $2^{|O^c|}$ possible \emph{execution branches} defined by the collection of measurement outcomes throughout the computation, $s_i \in \{0, 1\}, \; \forall i \in O^c$. We define the desired quantum operation with respect to the $\bm{0}$-branch (i.e., $s_i = 0,\; \forall i \in O^c$), therefore, every occurrence of an outcome $1$ must be compensated by an appropriate correction. Since the measurement basis states in Eqs.~\eqref{eq:meas_states} are related by the single-qubit Pauli operator orthogonal to the measurement plane, $P_{\lambda^{\perp}} \ket{-_{\lambda,\alpha}} = \ket{+_{\lambda,\alpha}}, \; \forall \lambda, \alpha,$ these corrections can always be expressed in terms of Pauli $\X$ and $\Z$ operators, which therefore suffice as the elementary correction operations in MBQC.

\begin{figure}
    \centering
    \includegraphics[width=\columnwidth]{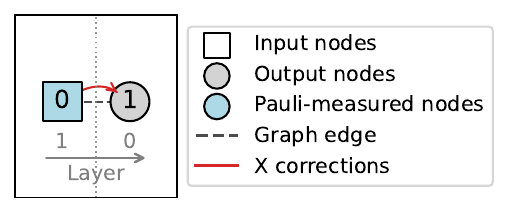}
    \caption{Measurement-based quantum computation implementing a Hadamard gate. A red arrow from node $i$ to node $j$ indicates a correction $X_j^{s_i}$. The partition of the nodes layers represents the partial order induced by the correction strategy.  Nodes are measured from left to right (in this basic example only node 0 is measured, and node 1 is an output node).}
    \label{fig:mbqc_naive_example}
\end{figure}

Let us consider the example shown in Fig.~\ref{fig:mbqc_naive_example} implementing the Hadamard gate. Here, the open graph consists of two entangled qubits 0 and 1 that are input and output nodes, respectively. The measurement maps are $\lambda(0) = \planeXY$ and $\alpha(0) = 0$, and the correction strategy is simply $\bm x(0) = \{1\}, \; \bm z(0) = \emptyset$. For an arbitrary input state on node 0, $\ket{\psi}_I = a \ket{0}_0 + b \ket{1}_0$, the computation proceeds as follows:
\begin{equation*}
\begin{split}
    \ket{\psi'}_O &\sim X_1^{s_0}\prescript{}{0}{\braket{\pm_{\planeXY, 0}}{\Gamma, \psi}}\\
    &\sim  X_1^{s_0}\prescript{}{0}{\bra{\pm_{\planeXY, 0}}} \left(a\ket{0}_0\ket{+}_1 + b \ket{1}_0\ket{-}_1\right)\\
    &\sim X_1^{s_0} \begin{cases}
        a \ket{+}_1 + b \ket{-}_1 \; \text{if} \; s_0 = 0,\\
        a \ket{+}_1 - b \ket{-}_1 \; \text{if} \; s_0 = 1,
        \end{cases}\\
    &\sim a \ket{+}_1 + b \ket{-}_1\\
    &\sim H \ket{\psi}_I,
\end{split}    
\end{equation*}
where equality is obtained upon normalizing the resulting state after the projective measurement.

In this simple example, we considered a computation implementing a single-qubit unitary transformation. In general, however, the input and output Hilbert spaces may have arbitrary and not necessarily equal dimensions. Moreover, as in this case, we often correct the $(s=1)$ measurement outcomes to implement a deterministic computation, i.e., a transformation between the input and output Hilbert spaces that is \emph{independent} of the intermediate measurement outcomes, but it is worth noting that the formalism perfectly accommodates non-deterministic computations too. We discuss determinism in MBQC in Subsection \ref{subsec:detflow}.

Graphix provides a native representation of arbitrary measurement-based quantum computations. A fundamental class is \pythoninline{OpenGraph} which bundles a \networkx undirected graph \cite{HSS18:netx}, two lists of integers -- one to characterize the input nodes $I$ and the other for the output nodes $O$ -- and a dictionary mapping non-output nodes to their measurement configuration. These measurement configurations could be given using solely \pythoninline{Plane} or \pythoninline{Axis} objects, in which case the \pythoninline{OpenGraph} instance simply represents a labelled open graph $\Gamma$. On the other hand, the measurements could be specified using \pythoninline{Measurement} objects, which either consist of a plane and an angle or an axis and a sign. In this case, the \pythoninline{OpenGraph} instance represents a labelled open graph accompanied by an angle mapping, $(\Gamma, \alpha)$. An MBQC computation \((\Gamma, \alpha, \bm{x}, \bm{z})\) is fully represented by the \pythoninline{XZCorrections} class, which we instantiate by specifying an \pythoninline{OpenGraph} object and the $\bm{x}$ and $\bm{z}$ correction maps. The class constructor ensures that the user-defined correction strategy is runnable and computes the induced partial order which can be visualized as a partition into layers of the open graph's nodes. This layering offers a representation of the measurement order: nodes in layer $l_i$ are measured after nodes in layer $l_j$ if $l_i < l_j$. Box \ref{box:xz_init} demonstrates how to manually construct an \pythoninline{XZCorrections} object representing the example in Fig.~\ref{fig:mbqc_naive_example}.

\begin{quantumstacked}[box:xz_init]{$\bm{xz}$-corrections instantiation}
\inputminted{python}
{scripts/2p1-xz_corrections_instantiation.py}
\end{quantumstacked}

\subsection{Patterns and the measurement calculus}
\label{par:comms}
Open graphs and correction strategies provide a simple representation of an MBQC computation. However, for more systematic reasoning within the formalism, it is convenient to formulate MBQC in terms of the measurement calculus \cite{DKP07:calculus}. The basic instructions -- called \emph{commands} -- are
\begin{itemize}
    \item \emph{preparation} \(\N_i\) of qubit \(i\),
    \item \emph{entanglement} \(\E_{ij}\) of qubits \(i\) and \(j\),
    \item \emph{destructive single-qubit measurement} \(\M_i^{\lambda,\alpha}\) of qubit \(i\) in the plane \(\lambda(i) \in \{\planeXY, \planeXZ, \planeYZ\}\) with angle \(\alpha(i) \in \left[0, 2\pi\right)\),
    \item \emph{Pauli corrections} \(\X_i^s\) and \(\Z_i^s\) of qubit \(i\) conditional on the classical bit \(s \in \{0,1\}\). The bit $s = \bigoplus_{j\in D} s_j$ combines the measurement outcomes of a set $D$ of previously measured nodes (the correction \emph{domain}).
\end{itemize}
We use sans serif font for MBQC commands and standard font for quantum gates.
In this setting, the combination of $\N$ and $\E$ commands offer a representation of the resource graph state, \(\lambda\) and \(\alpha\) are maps from the set of measured qubits $O^c$ to measurement planes and angles, analogous to the definition of open graphs, and the Pauli corrections implement the $\bm{xz}$-correction maps defined in the previous subsection. The semantics associated with this syntax are detailed in Table~\ref{tab:semantics}, where the states \(\ket{\pm_{\lambda,\alpha}}\) are defined in Eq.~\eqref{eq:meas_states}. Similarly to the graph-based representation of MBQC computations discussed earlier, when \(\alpha(i)\) is a Pauli angle, \(\Bqty{0, \frac\pi2, \pi, \frac{3\pi}2 }\), the pair \(\lambda(i), \alpha(i)\) can be replaced by an axis $\{\axisX, \axisY, \axisZ\}$ with a sign. It is worth emphasizing that this replacement promotes the measurement to a Pauli measurement allowing to take advantage of their specificities which is not the case for a \(\lambda(i), \alpha(i)\) with Pauli angle.

\begin{table}[ht]
\centering
\begin{tabular}{|c|c|c|c|c|}
\hline
\bf \footnotesize Syntax
& \footnotesize \(\N_i\)
& \footnotesize \(\E_{ij}\)
& \footnotesize \(\M^{\lambda,\alpha}_i\)
& \footnotesize \(\X_i^s,\ \Z_i^s\) \\ [0.5ex]
\hline
\bf \footnotesize Semantics
& \footnotesize \(\otimes \ket{+}_i\)
& \footnotesize \(CZ_{ij}\)
& \footnotesize \(\tensor[_i]{\bra{\pm_{\lambda(i),\alpha(i)}}}{}\)
& \footnotesize \(X_i^s,\ Z_i^s\) \\ [0.5ex]
\hline
\end{tabular}
\caption{Syntax and semantics of MBQC. We use $\lambda$ to denote the measurement planes, while $\alpha$ is used for measurement angles. These symbols may also denote the map from measured nodes to planes or angles. We use these two interchangeably, and the precise meaning will be clear from the context.}
\label{tab:semantics}
\end{table}

In the measurement calculus formalism, computations are expressed in the form of \emph{measurement patterns} (or simply \emph{patterns}): finite sequences of commands applied to a qubit set, paired with two specific qubit sets designating the input and output registers. The sequence of commands is constrained with additional \emph{runnability} conditions: commands cannot depend on outcomes not yet measured or act on measured or non-existent qubits, output qubits cannot be measured, a qubit cannot be prepared twice and a pair of qubits cannot be entangled twice, etc. This notion of runnability is equivalent to that of $\bm{xz}$-corrections discussed in Subsection \ref{subsec:og}. For formal definitions, we refer the reader to \cite{DKP07:calculus}.

One of the simplest examples is the pattern defined over the 2-node path graph with input node 0 and output node 1,
\begin{align}
    \mathcal{P}_H = \X_1^{s_0} \,\M_0^{\rm X}\,\E_{01} \,\N_1,
    \label{eq:pattern_example}
\end{align}
where the execution order of commands is read from right to left. This pattern represents the computation in Fig.~\ref{fig:mbqc_naive_example} and, as it was shown, corresponds to a Hadamard gate in the circuit model. Patterns are represented by the class \pythoninline{Pattern} in Graphix, which can be directly instantiated by providing a list of input nodes and a sequence of commands. \pythoninline{Pattern} objects can be simulated, providing a direct interface with the semantics of the measurement calculus (Table \ref{tab:semantics}). In Box \ref{box:pattern_init} we show how to instantiate $\mathcal{P}_H$. We take a more detailed look at pattern simulation in Section~\ref{subsec:pattern_simulation}.

\begin{quantumstacked}[box:pattern_init]{Pattern instantiation}
\begin{minted}{python}
from graphix import BasicStates, Pattern
from graphix.command import E, M, N, X

# Instantiate a pattern implementing
# a Hadamard gate.
p = Pattern(
    input_nodes=[0],
    cmds=[N(1), E((0, 1)), M(0), X(1, {0})])

# Commands are passed in execution order.

# By default, M commands represent a +X
# measurement and have empty signal domains.

# `X(1, {0})` acts on node 1 and its
# signal domain is the set of nodes `{0}`.

# Simulate the pattern with input state |0>.
s = p.simulate(input_state=BasicStates.ZERO)

print("Pattern: ", p)
print("Simulation output: ", s)
\end{minted}
\tcblower
Pattern: X(1,\{0\}) M(0) E(0,1) N(1)\\
Simulation output: sqrt(2)/2(|0> + |1>)
\end{quantumstacked}

Referring back to Section~\ref{subsec:og}, a pattern has a unique underlying open graph and defines a unique correction map. In other words, the graph representation of an MBQC computation \((\Gamma, \alpha, \bm{x}, \bm{z})\) captures the semantics of the corresponding pattern \cite{PS16}:
\begin{equation}
\mathcal{P} = \prod^{\prec}_{i\in O^c} \left(\X^{s_i}_{\bm x(i)} \Z_{\bm z(i)}^{s_i}\M^{\lambda,\alpha}_i \right)\prod_{(i,j)\in E} \E_{ij} \prod_{i\in I^c} \N_i.
\label{eq:pattern_xzcorr}
\end{equation}
The methods \pythoninline{Pattern.to_opengraph} and \pythoninline{Pattern.to_xzcorrections} in Graphix allow to transform patterns into their graph-based representations. 

The measurement calculus establishes a formal framework for pattern manipulation \cite{DKP07:calculus}. Specifically, it allows corrections on measured nodes to be absorbed into subsequent measurement commands, reflecting the measurement adaptivity inherent to MBQC. This is done as follows
\begin{align}
\tensor[_t]{\left[\M_i^{\lambda, \alpha}\right]}{^s} :=
    \M^{\lambda, \alpha}_i \; \X_i^s \, \Z_i^t = \M_i^{\lambda, \alpha'_{s,t}}
\label{eq:calc_domains_update}
\end{align}
with 
\begin{align}
    \alpha'_{s,t} = 
    \begin{cases}
        (-1)^s\,\alpha + t\pi , & \text{if} \;\lambda = \planeXY,\\
        (-1)^{s + t}\,\alpha + t\pi , & \text{if} \;\lambda = \planeXZ,\\
        (-1)^t\,\alpha  + s\pi, & \text{if} \;\lambda = \planeYZ.\\
    \end{cases}
\label{eq:alpha}
\end{align}
The bits $s := \bigoplus_{j\in D_X} s_j$ and $t := \bigoplus_{j\in D_Z} t_j$ are called \emph{signals} and encompass the previous measurement outcomes, and $D_X, D_Z \subseteq O^c$ are the measurement command \emph{domains}. In Graphix, measurement commands have the \({\tt s\_{domain}}\) and \({\tt t\_domain}\) attributes representing \(D_s\) and \(D_t\), respectively. When instantiating commands, we specify the correction signal with the set of nodes defining the domain.


The rules of the measurement calculus are built into Graphix so that \pythoninline{Pattern} objects can be easily brought into \emph{standard form}, starting with qubit preparations (\(\N \) commands), then entanglements (\(\E \) commands), then measurements (\(\M\) commands) and finally corrections on the output qubits (\(\X\) and \(\Z\) commands) \cite{DKP07:calculus}. See Appendix \ref{app:standardization} for additional details on the commutation rules and the algorithm implementation in Graphix. Alternatively, pattern commands can be rearranged in various ways -- for example, to minimize the size of the computational space (see Box \ref{box:pattern_std} and Section~\ref{subsec:patter_opt}).

\begin{quantumstacked}[box:pattern_std]{Standardization and optimization}
\begin{minted}{python}
from graphix import Measurement, Pattern
from graphix.command import E, M, N, X, Z

pattern = Pattern(cmds=
   [N(0), N(1), E((0, 1)), N(2), E((1, 2)),
    M(0, Measurement.XY(0.75)), Z(2, {0}),
    X(1, {0})])

# The M command represents a measurement
# on the XY plane with angle 3*PI/4.
# All angles are always in units of PI.

pattern.standardize()
print("Pattern in standard form:", pattern)
\end{minted}
\tcblower
Standard pattern: X(1,\{0\}) Z(2,\{0\}) M(0,3pi/4) E(1,2) E(0,1) N(2) N(1) N(0)\\
Space-optimal pattern: Z(2,\{0\}) X(1,\{0\}) E(1,2) N(2) M(0,3pi/4) E(0,1) N(1) N(0)
\end{quantumstacked}




\subsection{Determinism and flow}
\label{subsec:detflow}
As illustrated by the example in Fig.~\ref{fig:mbqc_naive_example}, applying specific outcome-dependent Pauli corrections ensures that the MBQC computation implements the same transformation, up to a global phase, for \emph{every} possible combination of measurement outcomes throughout the computation, i.e., for every execution branch. This property is denoted \emph{strong determinism}. The subset of strongly deterministic patterns is of particular importance, as they realize \emph{unitary} transformations and constitute the MBQC equivalent of quantum circuits.

A central question in the MBQC model is to determine which resource states support a unitary pattern (or, more generally, an isometry when $|O|>|I|$). Equivalently, given an open graph, one seeks to determine whether there exists a correction strategy that implements a strongly deterministic MBQC computation. Previous work has answered this question by identifying graph-theoretic conditions on labelled open graphs, known as \emph{flow conditions}. All types of flows are the same mathematical object -- a tuple of a correction function from measured to prepared qubits, and a partial order between nodes -- and can be found in polynomial time when they exist \cite{MP08:finding_flows,BKMP07:gflow,Simmons21,BMBdF+21,MB24:algebraic}. The simplest case is the \emph{causal flow}, defined on open graphs with measurements on the $\planeXY$ plane only:

\begin{definition}[Causal flow \cite{DK06:determinism}]
An open graph $(G, I, O, \lambda: O^c \rightarrow  {\planeXY})$ has causal flow if there exists a map $c: O^c \rightarrow I^c$ and a
strict partial order $\prec$ over $V$ such that for all $i \in O^c$:
\begin{itemize}
    \item (C1) Nodes $i$ and $c(i)$ are neighbors,
    \item (C2) $i \prec c(i)$,
    \item (C3) $i \prec j$ for all neighbors $j \neq i$ of node $c(i)$.
\end{itemize}
\label{def:causal_flow_maintext}
\end{definition}

Causal flow is a sufficient, but not necessary, condition for strong determinism. Indeed, there exist open graphs with measurements in planes other than $\planeXY$ (and therefore without causal flow) that nevertheless support patterns implementing unitary transformations. For open graphs with measurements in the $\planeXY$, $\planeXZ$, or $\planeYZ$ planes, strong determinism is characterized by the existence of a \emph{generalised flow} (or \emph{gflow}). Importantly, both causal flow and gflow guarantee strong determinism for \emph{any} choice of measurement angles, a property known as \emph{robust determinism}. This requirement can be relaxed by fixing some measurement angles to Pauli angles, i.e., integer multiples of $\frac{\pi}{2}$. At these angles, a measurement simultaneously belongs to two measurement planes, introducing additional structural freedom that can be exploited to establish strong determinism even when gflow is absent. This more general setting is captured by the notion of \emph{Pauli flow}, which strictly generalizes gflow by taking advantage of these fixed Pauli measurements (Table~\ref{tab:flows}). A recent extension, called \emph{Shadow Pauli flow} \cite{MPS22}, provides a complete characterization of robust determinism on patterns instead of open graphs: an MBQC pattern is robustly deterministic if and only if its correction strategy is consistent with a Shadow Pauli flow. Although Shadow Pauli flow can be computed in polynomial time, it has not yet been implemented in Graphix so we will not discuss it further. We refer the reader to Appendix~\ref{app:flow_definitions} for formal definitions of gflow and Pauli flow, and to Refs.~\cite{DK06:determinism,BKMP07:gflow, PS16} for a full
account of determinism in MBQC.

\begin{table*}
\centering
\begin{tblr}{
  colspec = {|c|c|c|c|},
  row{1} = {font=\bfseries},
  hlines, vlines,
}
Flow type & {Open graph\\measurements} & {Relationship to\\robust determinism} & {Finding-algorithm\\complexity} \\
Causal \cite{DK06:determinism} & \planeXY & Necessary & \(\order{\abs{V}^2}\) \cite{MP08:finding_flows} \\
Generalised \cite{BKMP07:gflow} & Planar & Necessary and sufficient & \(\order{\abs{V}^3}\) \cite{MB24:algebraic} \\
Pauli \cite{BKMP07:gflow} & Planar + Pauli & Necessary and sufficient & \(\order{\abs{V}^3}\) \cite{MB24:algebraic} \\
\end{tblr}
\caption{Main types of flows on open graphs and their properties. Each of them has a representation in Graphix.}
\label{tab:flows}
\end{table*}

Flow conditions on an open graph are not only a guarantee of the existence of a deterministic computation, but they also provide a recipe to obtain the deterministic pattern itself \cite{BKMP07:gflow}. In particular, the flow correction function and partial order prescribe a strongly deterministic pattern in the form of Eq.~\eqref{eq:pattern_xzcorr}, where the $\bm x$, $\bm z$ correction maps read as follows:
\begin{subequations}
    \begin{align}
    \text{Causal: }&\begin{cases}
        \bm x_{\text c}(i) = c(i),\\
        \bm z_{\text c}(i) = N_G(c(i)) \setminus \{i\},
    \end{cases} \label{eq:flow_to_corrections_causal}\\
    \text{Generalised: }&\begin{cases}
        \bm x_{\text g}(i) = g(i)\setminus \{i\},\\
        \bm z_{\text g}(i) = \operatorname{Odd}(g(i)) \setminus \{i\},
    \end{cases}\label{eq:flow_to_corrections_g}\\
    \text{Pauli: }&\begin{cases}
        \bm x_{\text p}(i) = p(i)\cap \{j | i \prec j\},\\
        \bm z_{\text p}(i) = \operatorname{Odd}(p(i)) \cap \{j | i \prec j\}, \label{eq:flow_to_corrections_pauli}
    \end{cases}
    \end{align}
\label{eq:flow_to_corrections}
\end{subequations}
where $g$ and $p$ are the gflow and Pauli flow correction functions (analogous to $c$ for the causal flow) and $\operatorname{Odd}(S) = \{v | N_G(v) \cap S = 1 \operatorname{mod} 2\}$ denotes the odd neighbourhood of a set $S$ in the open graph.

Graphix enables users to analyze the determinism of computations on open graphs using flows. The main types of flows discussed in the literature are represented by the classes \pythoninline{CausalFlow}, \pythoninline{GFlow} and \pythoninline{PauliFlow}. We implement the best-known complexity flow-finding algorithms so that all three types of flow objects can be extracted from a given \pythoninline{OpenGraph} instance, provided they exist (see Table \ref{tab:flows} and Section \ref{subsec:flows} for additional discussion).

\begin{quantumstacked}[box:causal_flow_corrections]{Extracting a flow from an open graph}
\begin{minted}{python}
import networkx as nx
from graphix import Measurement, OpenGraph

og = OpenGraph(
    graph=nx.Graph(
        [(0, 2), (1, 3), (2, 3),
         (2, 4), (3, 5)]),
    input_nodes=[0, 1],
    output_nodes=[4, 5],
    measurements={
        0: Measurement.XY(0.1),
        1: Measurement.XY(0.2),
        2: Measurement.XY(0.3),
        3: Measurement.XY(0.4)})

cf = og.to_causalflow() # Flow finding.
xz = cf.to_xzcorrections() # Eq. (7a)
p = xz.to_pattern() # Eq. (4)

# This workflow is wrapped in
# `og.to_pattern` for convenience.

cf.draw()
xz.draw()
print(p)
\end{minted}
\tcblower
\vspace{-0.5cm}
\begin{figure}[H]
\centering
\includegraphics[width=1.0\textwidth]{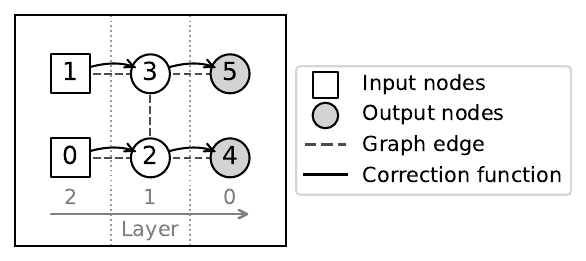}
\end{figure}
\vspace{-1cm}
\begin{figure}[H]
\centering
\includegraphics[width=1.0\textwidth]{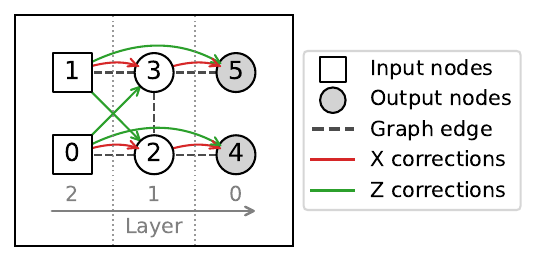}
\end{figure}
\vspace{-0.5cm}
Pattern:

X(5,\{3\}) M(3,2pi/5) X(4,\{2\}) M(2,3pi/10)

X(3,\{1\}) Z(5,\{1\}) Z(2,\{1\}) M(1,pi/5)

X(2,\{0\}) Z(4,\{0\}) Z(3,\{0\}) M(0,pi/10)

E(3,5) E(1,3) E(2,4) E(2,3) E(0,2)

N(5) N(4) N(3) N(2)
\end{quantumstacked}

Further, Graphix leverages the relationships in Eq.~\eqref{eq:flow_to_corrections} to convert any flow instance into an \pythoninline{XZCorrections} object representing the $\X$ and $\Z$ corrections induced by the flow. From an \pythoninline{XZCorrections} instance we then extract a \pythoninline{Pattern} (see Eq.~\eqref{eq:pattern_xzcorr}). This chain of conversions from flow to pattern complements the flow extraction functionality and provides a straightforward way to obtain a robustly deterministic pattern, when it exists, from an open graph (see Box \ref{box:causal_flow_corrections}).

Finally, we mention that flow objects are equipped with the method \pythoninline{check_well_formed} which, if called on an incorrect flow, raises an exception indicating the proposition in the flow definition that is violated (see Box \ref{box:flow_exception} and Appendix \ref{app:flow_definitions}). This functionality allows to verify the compatibility of a given correction function and partial order with an open graph, which serves both as a pedagogical utility and a debugging tool for developing new flow extraction algorithms.

\begin{quantumstacked}[box:flow_exception]{Testing flow correctness}
\begin{minted}{python}
import networkx as nx
from graphix import GFlow, OpenGraph, Plane

og = OpenGraph(
    graph=nx.Graph(
        [(0, 3), (0, 4), (1, 4), (2, 4)]),
    input_nodes=[0],
    output_nodes=[3, 4],
    measurements={0: Plane.XY, 1: Plane.YZ,
                  2: Plane.XZ})

# Measurements can be specified with `Plane`
# objects too, but then conversion
# to `Pattern` is not possible.

# Initialize an incorrect gflow.
gf = GFlow(
    og=og,
    correction_function=
    {0: {3, 4}, 1: {1}, 2: {2, 3, 4}},
    partial_order_layers=
    [{3, 4}, {1}, {2}, {0}])

gf.check_well_formed()
\end{minted}
\tcblower
...\\
graphix.flow.exceptions.FlowPropositionError: G3: nodes measured on plane XY cannot be in their own correcting set and must belong to the odd neighbourhood of their own correcting set.
Error found at c(0) = \{3, 4\}.
\end{quantumstacked}

\subsection{Universality and relationship to the circuit model}\label{subsec:universality}

The universality of the MBQC model can be seen through its equivalence to the circuit model. A universal set of gates, for example $J(\alpha) := H R_Z(\alpha)$, where $R_Z(\alpha) := \exp(-i\frac{\alpha}{2} Z)$ and $H$ is the Hadamard gate, and a controlled-$Z$ gate, can be represented in pattern form (see Fig.~\ref{fig:mbqc_universality}). Since patterns can be composed, any unitary transformation can be constructed by composing the patterns that correspond to a universal gate set. This also means that circuits can easily be transpiled into patterns by converting them to this universal gate set~\cite{DKP07:calculus}.

\begin{figure}[H]
    \centering
    \resizebox{\linewidth}{!}{\usetikzlibrary{positioning,calc}

\begin{tikzpicture}[
  every node/.style={inner sep=1pt, outer sep=0pt}
]

\node (hadamard) {
  \begin{quantikz}[column sep=0.25cm, row sep=0.2cm]
    & \gate{R_Z(\alpha)} & \gate{H} & \qw
  \end{quantikz}
};

\node[right=0.25cm of hadamard] (arrow_hadamard) {$\Leftrightarrow$};

\node[right=0.25cm of arrow_hadamard] (pattern_hadamard) {
$\X_1^{s_0}\M_0^{\planeXY,-\alpha}\E_{01}\N_1$
};

\node[below=0.25cm of hadamard] (cz) {
\begin{quantikz}[column sep=0.25cm, row sep=0.2cm]
 & \ctrl{1} & \qw \\
 & \control{} & \qw
\end{quantikz}
};

\node (arrow_cz) at ($(arrow_hadamard |- cz)$) {$\Leftrightarrow$};

\node[right=0.25cm of arrow_cz] (pattern_cz) {
$\E_{01}$
};

\end{tikzpicture}}
    \caption{Equivalence between a universal gate set and MBQC patterns. In the first pattern we have $I = \{0\}, O = \{1\}$, and in the second $I = O = \{0, 1\}$.}
    \label{fig:mbqc_universality}
\end{figure}





At the same time, every pattern can be represented as a circuit augmented with ancillary qubits, mid-circuit measurements, and feed-forward Pauli corrections. The reverse also holds, implying that the MBQC model inherently includes this extension of unitary quantum circuits (see Fig.~\ref{fig:mbqc_ff}).

\begin{figure}[H]
    \centering
    \resizebox{\linewidth}{!}{\usetikzlibrary{positioning,calc}

\begin{tikzpicture}[
  every node/.style={inner sep=1pt, outer sep=0pt}
]

      \node (pattern) {
        {$\X_1^{s_0}\, \M_0^{\planeXY,-\alpha}\, \E_{01}\,\N_1$}
      };

      \node[right=0.25cm of pattern] (arrow) {$\Leftrightarrow$};

      \node[right=0.25cm of arrow] (circuit) {
        \begin{quantikz}[column sep=0.25cm, row sep=0.2cm]
          & \ctrl{1} & \meter{\ket{\pm_{\planeXY, -\alpha}}}\vcw{1} \\
          \lstick{\(\ket{+}\)} & \control{} & \gate{X} & \qw
        \end{quantikz}
      };
    \end{tikzpicture}}
    \caption{MBQC patterns in the circuit model with ancillas, mid-circuit measurements and feed-forward. This pattern implements an $H$ gate if $\alpha = 0$.}
    \label{fig:mbqc_ff}
\end{figure}

Graphix provides an interface with the circuit model by implementing a circuit-to-pattern transpiler and an \qasm 3.0 exporter to convert a pattern to a circuit with ancillas and classical control. Further, it also implements the circuit extraction algorithm presented in Ref.~\cite{Simmons21} which extracts a pattern's underlying unitary transformation if it exists and compiles it into a circuit without ancillas or feed-forward (see Box \ref{box:pattern_transpilation} and Section~\ref{subsec:from_to_circuits} for an in-depth discussion).\\

\begin{quantumstacked}[box:pattern_transpilation]{Pattern-Circuit conversions}
\begin{minted}{python}
from graphix import BasicStates, Circuit

qc = Circuit(2)
qc.cz(0, 1)
qc.rz(0, -0.25) # Angle = -PI/4
qc.h(0)

p = qc.transpile().pattern
print(p)

# Export as a circuit with ancillas
# and feed-forward to a QASM file.
p.to_qasm3("pattern.qasm",
           input_state=BasicStates.ZERO)

# Extract the pattern's unitary.
qc_ext = p.to_opengraph().to_circuit()
print(qc_ext)

# `qc` and `qc_ext` realize the same
# unitary but with different gates.
\end{minted}
\tcblower
X(4,{3}) Z(4,{2}) {0}[M(3,0)]{2} E(3,4) N(4) [M(2,0)]{0} E(2,3) N(3) M(0,pi/4) E(0,2) N(2) E(0,1)\\

Circuit(width=2, instr=[H(1), CNOT(1, 0), H(1), H(0), RX(0, -pi/4)])
\end{quantumstacked}

\subsection{MBQC + local Clifford operations}
\label{subsec:lc_frag}
All stabilizer states are local Clifford (LC)-equivalent to graph states, where LC operations are tensor products of single-qubit unitaries in the Clifford group \cite{vdNDdM04}. This equivalence permits graph-theoretic transformations that can be exploited for pattern optimization \cite{Elliot09:graphicalmeas,BMBdF+21}. To represent such transformations, it is useful to extend the measurement calculus with \emph{unconditional} Clifford operations.
Graphix implements this MBQC+LC extension through the command \({\C_i^c}\), where \(i \in V\) denotes a node and \(c\) specifies one of the 24 elements of the single-qubit Clifford group (see Appendix~\ref{app:clifford_gates} for a list). Internally, Clifford operations are represented as products of the gates \(H\), \(S\), and \(Z\) for convenience.
Appendix \ref{app:standardization} presents the commutation rules for MBQC+LC commands and describes how Clifford operations acting on measured nodes can be absorbed into measurement commands. These rewrite rules are used during pattern standardization and are required to establish the correspondence between deterministic MBQC+LC patterns and open graphs. Accordingly, the open-graph representation in Graphix is extended with an additional attribute, \pythoninline{output_cliffords}, which maps output nodes to Clifford operations. In Graphix \(\C_i^c\) commands are at the very end of standardised patterns. This convention enables the flow-analysis algorithms described in Section~\ref{subsec:flows} to work equally well on patterns with Clifford commands: indeed, since Clifford gates on output nodes do not change whether a pattern is deterministic or not, we can simply drop the final Clifford commands in the flow analysis.

\section{An in-depth look into Graphix: architecture and algorithms} 
\label{sec:graphix_lib}
\subsection{Architecture and class structure}
\label{subsec:architecture}
\begin{figure*}
    \centering
    \resizebox{0.8\textwidth}{!}{\input{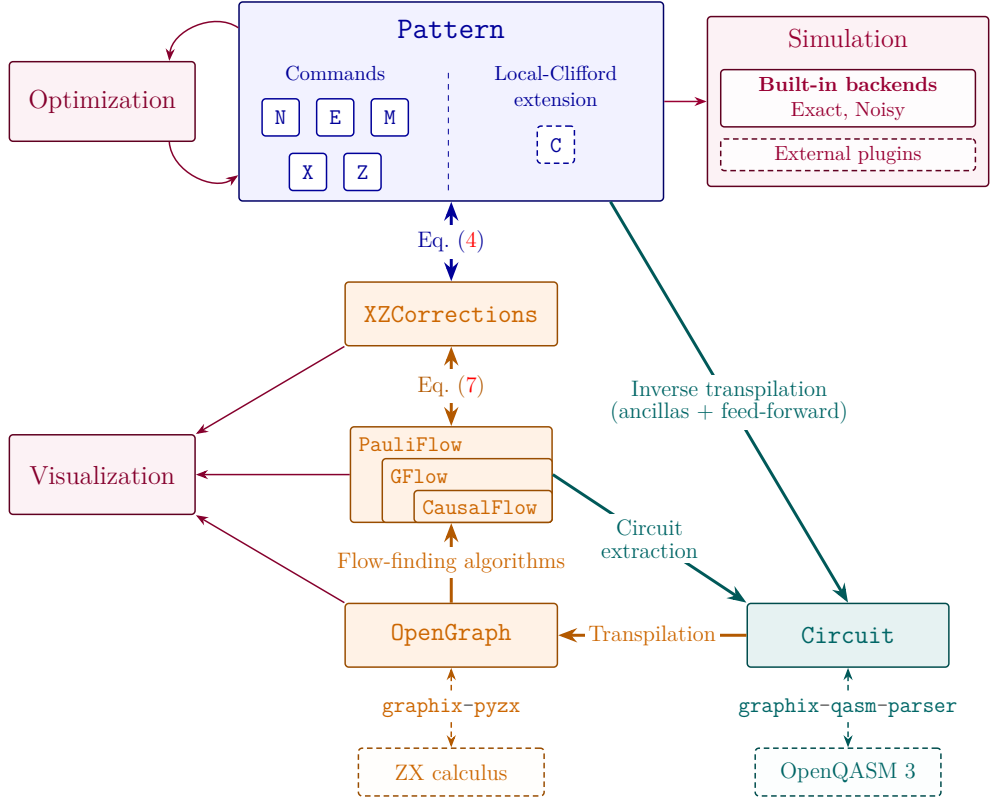}}
    \caption{Illustration of Graphix' software architecture. Core MBQC objects are centered in blue and yellow, interfaces in red, circuits in green, and external packages with dashed borders.}
    \label{fig:architecture}
\end{figure*}

The code examples in the previous section demonstrate how Graphix was designed to mirror the theory of MBQC. In this way, central concepts in MBQC (e.g., measurements, commands, patterns, open graphs, flows, etc.) become full-fledged class abstractions in the software library. Furthermore, the fundamental results in the field are translated into operations on these building blocks --- the proof of universality becomes a transpiler from circuits to patterns, the results on determinism provide an array of methods to convert between various representations of MBQC computations, local-Clifford complementation serves as the basis for pattern optimization methods, etc. This design choice ultimately results in a modular architecture that allows users to input a quantum computation in alternative languages (the measurement calculus, open graph-based MBQC or the circuit model), interchange between them, and operate on the most suitable representation (Fig.~\ref{fig:architecture}).  This flexibility has also led to an adaptable structure: Graphix is equipped with a collection of plugins which target additional representations and tooling. For example, \pythoninline{OpenGraph} objects can be translated to and from ZX-calculus diagrams via \pyzx \cite{KvdW20}, a \pythoninline{Pattern} can be written to and from an \qasm file, and the built-in simulators can be readily substituted for third-party backends (see Appendix~\ref{app:plugins} for the full list of existing plugins). All in all, this portability yields an extensible package that is well connected with the quantum software ecosystem.

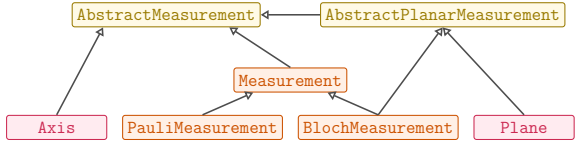
\begin{figure}
    \centering
    \resizebox{\linewidth}{!}{\newcommand{\pythoninlinefs}[2]{%
  \mintinline[fontsize=\fontsize{#2}{#2}]{python}{#1}%
}

\definecolor{amber}{rgb}{1.0, 0.8, 0.0}
\definecolor{amber2}{rgb}{1.0, 0.4, 0.0}
\definecolor{amber3}{rgb}{1.0, 0.2, 0.4}

\definecolor{abscolor}{named}{amber}
\definecolor{meascolor}{named}{amber2}
\definecolor{measlabelcolor}{named}{amber3}

\newcommand{\toolfontsize}{\fontsize{14pt}{14pt}\selectfont}
\newcommand{\arrowfontsize}{\fontsize{9pt}{9pt}\selectfont}

\begin{tikzpicture}[
    font=\rmfamily,
    >=Stealth,
    box/.style={draw, rounded corners=2pt, align=center, thick},
    smallbox/.style={draw, rounded corners=2pt, align=center, inner sep=2pt, fill=white, thick},
    absBox/.style={box, fill=abscolor!10, draw=abscolor!60!black, minimum width=2.7cm, minimum height=0.65cm},
    measBox/.style={box, fill=meascolor!10, draw=meascolor!80!black, minimum width=2.7cm, minimum height=0.65cm},
    measlabelBox/.style={box, fill=measlabelcolor!10, draw=measlabelcolor!80!black, minimum width=2.7cm, minimum height=0.65cm},
    arr/.style={-Triangle[open], very thick},
]


    \newcommand{\xshift}{0.2cm}
    \newcommand{\yshift}{0.6cm}

    \node[measBox, text=meascolor!80!black](meas) at (0,0) {\pythoninlinefs{Measurement}{14pt}};

    \node[absBox, text=abscolor!60!black, anchor=south east](abs) at ([xshift=-4*\xshift, yshift=1.8*\yshift]meas.north) {\pythoninlinefs{AbstractMeasurement}{14pt}};
    \node[absBox, text=abscolor!60!black, anchor=south west](absplanar) at ([xshift=4*\xshift, yshift=1.8*\yshift]meas.north){\pythoninlinefs{AbstractPlanarMeasurement}{14pt}};
    
    \node[measBox, text=meascolor!80!black, anchor=north east](measpauli) at ([xshift=-\xshift, yshift=-\yshift]meas.south) {
      \pythoninlinefs{PauliMeasurement}{14pt}
    };

    \node[measBox, text=meascolor!80!black, anchor=north west](measbloch) at ([xshift=\xshift, yshift=-\yshift]meas.south) {
      \pythoninlinefs{BlochMeasurement}{14pt}
    };

    \node[measlabelBox, text=measlabelcolor!80!black, anchor=east](axis) at ([xshift=-2*\xshift]measpauli.west) {\pythoninlinefs{Axis}{14pt}};
    \node[measlabelBox, text=measlabelcolor!80!black, anchor=west](plane) at ([xshift=2*\xshift]measbloch.east) {\pythoninlinefs{Plane}{14pt}};

    \draw[arr, draw=black!70] (meas.north) -- ([xshift=1.7cm]abs.south);
    \draw[arr, draw=black!70] (absplanar.west) -- (abs.east);
    \draw[arr, draw=black!70] (axis.north) -- ([xshift=-1.7cm]abs.south);
    \draw[arr, draw=black!70] (plane.north) -- (absplanar.south);
    \draw[arr, draw=black!70] (measpauli.north) -- ([xshift=-1cm]meas.south);
    \draw[arr, draw=black!70] (measbloch.north) -- ([xshift=1cm]meas.south);
    \draw[arr, draw=black!70] (measbloch.north) -- (absplanar.south);


    
\end{tikzpicture}}
    \caption{A class diagram of measurements in Graphix. An arrow from \pythoninline{B} to \pythoninline{A} indicates that \pythoninline{B} is a subtype of \pythoninline{A}. MBQC objects \pythoninline{O} (e.g., \pythoninline{OpenGraph}, \pythoninline{PauliFlow}) are covariant in the measurement labels, meaning that \pythoninline{O[B]} is also a subtype of \pythoninline{O[A]}. Orange and red boxes correspond to classes representing measurement configurations with and without information about the angle, respectively. Yellow boxes correspond to abstract classes which cannot be directly instantiated. \pythoninline{AbstractPlanarMeasurement} narrows the type \pythoninline{AbstractMeasurement} allowing to distinguish between planar (\pythoninline{Plane}, \pythoninline{BlochMeasurement}) and Pauli (\pythoninline{Axis}, \pythoninline{PauliMeasurement}) measurement configurations.}
    \label{fig:meas_types}
\end{figure}

The theory-inspired design of Graphix is also rooted in a robust, custom class hierarchy. A prime illustration of this is the representation of measurements. The class relationships shown in Fig.~\ref{fig:meas_types} reflect the intrinsic differences between various measurement configurations --- specifically, between planar and Pauli measurements, or between measurements with and without specified angles. Classes representing MBQC concepts in the library (\pythoninline{OpenGraph}, \pythoninline{PauliFlow}, etc.) are parametrized by specific measurement classes, thereby enforcing their properties at the type level.
Let us consider an example. \pythoninline{OpenGraph[T]} is the type of all open graphs whose measurements are of type \pythoninline{T}. For instance, \pythoninline{OpenGraph[Plane|Axis]} is the type of open-graph objects whose measurements are specified only by a plane or an axis on the Bloch sphere, without an angle.
Given an instance of \pythoninline{OpenGraph[Plane|Axis]}, we can attempt to extract a \pythoninline{PauliFlow} but not a \pythoninline{GFlow} because gflows are not defined on open graphs with Pauli measurements. Similarly, we can construct a \pythoninline{Pattern} from \pythoninline{XZCorrections[BlochMeasurement]} but not from \pythoninline{XZCorrections[Plane]}, as the latter lacks information about the measurement angles. 

Type constraints allow users to restrict their operational space to a certain subset of measurement types, with guarantees that it will be preserved by any allowed transformation. We mention in passing that Graphix features the manipulation and simulation of symbolic patterns (i.e., patterns whose measurement angles are not floating-point numbers but affine algebraic expressions, see Subsection \ref{subsec:symbolic} for an example), and there is ongoing work to incorporate this distinction in the measurement types. Graphix' underlying class hierarchy is not only a reflection of MBQC principles, but also provides the basis for robust extensibility. For instance, backends are parametric in the type of the internal quantum state (e.g., \pythoninline{Statevector}, \pythoninline{DensityMatrix}, etc.) which allows users to confidently extend simulation capabilities with specialized toolkits.

Crucially, the codebase is fully type-checked which enables type-checkers such as \mypy and \pyright to exploit the custom type structure described in this subsection. In addition to type-checking, development of Graphix adheres to modern best-practice: the codebase is extensively tested and all random processes can be seeded, permitting reproducible numerical experiments (particularly important for simulation).


\subsection{Flow and flow-finding algorithms}
\label{subsec:flows}
Flows in MBQC not only are crucial to characterize determinism, but, as it will be clear in subsequent sections, they are the basis for other MBQC utilities, such as circuit transpilation and extraction, and space minimization. We recall that in Graphix we represent the main types of flows by the classes \pythoninline{CausalFlow}, \pythoninline{GFlow} (parent class of \pythoninline{CausalFlow}) and \pythoninline{PauliFlow} (parent class of \pythoninline{GFlow}). The inheritance relation reflects that gflows are a special case of Pauli flows (with planar measurements only), and causal flows are a special case of gflows (with $\text{XY}$-measurements only). Flow objects encompass three data structures:
\begin{itemize}
    \item The \pythoninline{OpenGraph} object on which the flow is defined.
    \item A dictionary mapping between measured qubits and sets of prepared qubits representing the flow correction function  (see Definition \ref{def:causal_flow_maintext} and Appendix \ref{app:flow_definitions}).
    \item A partition of the open graph's nodes in a sequence of sets (denoted \emph{layers}) representing the flow partial order. By convention, layer 0 contains the output nodes if there are any. For all other layers, if $l_i < l_j$, then nodes in layer $l_j$ are measured before those in layer $l_i$. Equivalently, nodes in layer $l_j$ precede nodes in layer $l_i$ in the flow partial order. This partial order is analogous to that discussed in the context of a graph-representation of an MBQC computation and included in an \pythoninline{XZCorrections} object. Furthermore, we note that the flow's correction function is enough to characterize a flow as it defines a partial order \cite{MB24:algebraic}, but storing the partial order in a layer form as an attribute is practical to convert the flow to an \pythoninline{XZCorrections} instance or visualize it.
\end{itemize}

The \pythoninline{OpenGraph} class features specific instance methods for granular control over the target flow class (\pythoninline{to_causalflow}, \pythoninline{to_gflow}, \pythoninline{to_pauliflow}) since different flow types may have different correction functions or partial order on the same open graph. The associated routines do not require information on the measurement angles, reflecting the \emph{uniformity} of robust determinism. At the same time, Graphix can seamlessly cast planar measurements with a Pauli angle into Pauli measurements, which can allow to extract a Pauli flow where a gflow does not exist or a flow with lower depth (see Box~\ref{box:flow_extraction}). This functionality illustrates the power of the class hierarchy describing measurement configurations presented in Fig.~\ref{fig:meas_types}.  

\begin{quantumstacked}[box:flow_extraction]{Pauli and gflow extraction}
\begin{minted}{python}
import networkx as nx
from graphix import Measurement, OpenGraph

og = OpenGraph(
    graph=nx.Graph(
        [(0, 1), (1, 2), (3, 4), (4, 5),
         (6, 7), (7, 8), (1, 3), (4, 6)]),
    input_nodes=[0, 3, 6],
    output_nodes=[2, 5, 8],
    measurements=dict.fromkeys(
        (0, 1, 3, 4, 6, 7),
        Measurement.XY(0)))

gf = og.to_gflow()
# Cast `Measurement.XY(0)` (planar meas.)
# into `+Measurement.X` (Pauli meas.)
# and extract Pauli flow.
pf = og.infer_pauli_measurements().\
     to_pauliflow()

gf.draw()
pf.draw()
\end{minted}
\tcblower
\begin{figure}[H]
\centering
\includegraphics[width=0.8\textwidth]{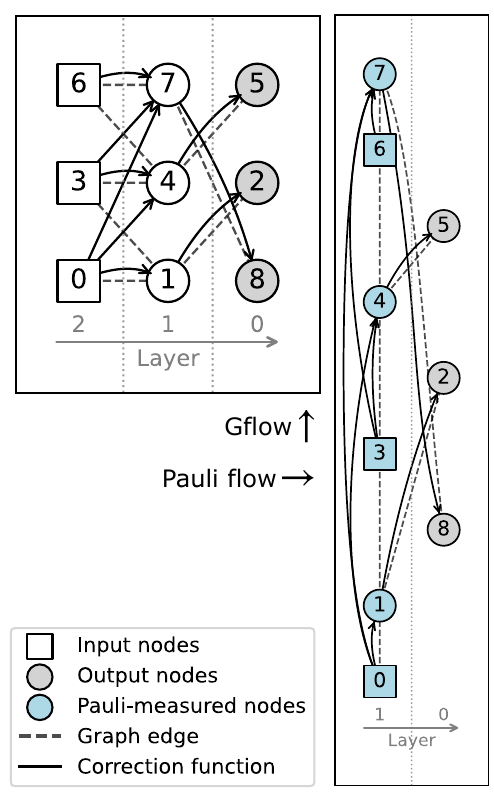}
\end{figure}
\end{quantumstacked}

\begin{figure}[H]
    \centering
    \includegraphics[width=1.0\linewidth]{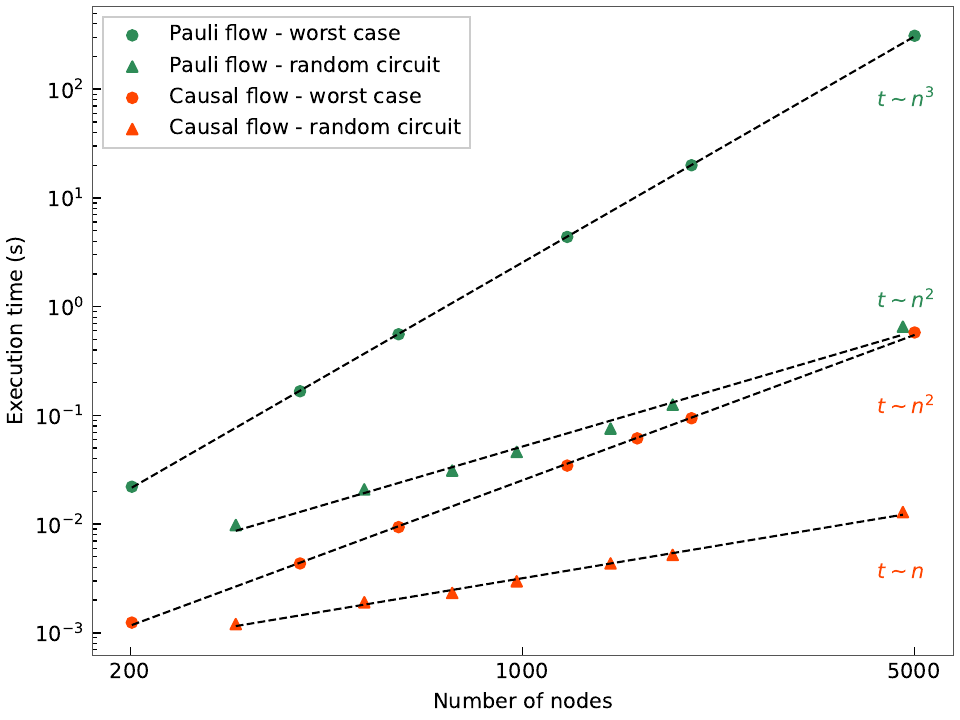}
    \caption{Scaling of flow-finding algorithms. Dashed lines represent a fit to $t = a \cdot n^c$, with $a, c$ the free parameters. Labels in the plot show $c$ rounded to the closest integer. Worst-case open graphs for the Pauli algorithm are obtained from random non-singular $\mathbb{F}_2$ matrices $M$ with size $(|I|, |O|)$  \cite{R93:rnd} such that there is an edge between the input node $i$ and output node $j$ iff $M_{ij} =1$, and input nodes are measured in $\planeXY$ plane \cite{M25:pc}. The worst case for the causal flow algorithm are linear open graphs. Random circuits are generated using Graphix' \pythoninline{rand_circuit} function with 20 qubits and increasing depth, and are subsequently transpiled into open graphs.}
    \label{fig:flow-scaling}
\end{figure}

The causal flow extraction routine implements the layer-by-layer algorithm introduced in Ref.~\cite{MP08:finding_flows} which runs in $O(|V|^2)$ time on the number of nodes $|V|$ and outputs a maximally delayed causal flow when successful. On the other hand, the gflow and Pauli flow extraction routines implement the $O(|V|^3)$ algorithm presented in Ref.~\cite{MB24:algebraic} improving on past $O(|V|^4)$ and $O(|V|^5)$ bounds respectively \cite{BMBdF+21, Simmons21}. In addition to its improved efficiency, this algorithm yields \emph{focused} flows (i.e., flows whose correction function has additional structure, see \cite{MB24:algebraic} for a formal definition), which is crucial for the circuit extraction algorithm (Subsection \ref{subsec:extraction}) or to compute the MBQC fidelity of a resource graph state (Subsection \ref{subsec:fidelity}). The procedure follows an algebraic approach whereby the open graph is expressed in terms of two matrices $M, N$ derived from its adjacency matrix, the correction function is obtained from the right inverse of $M$, $C := M^{-1}$, and the partial order is encoded in $NC$, the adjacency matrix of a directed acyclic graph (DAG). The extraction of a gflow or Pauli flow proceeds the same way since Pauli and planar measurements are treated on the same footing when constructing the $M$ and $N$ matrices (although, naturally, different measurement configurations result in different matrices). Within this algebraic interpretation, the flow-extraction algorithm amounts to performing matrix inversion, multiplication and Gaussian elimination in $\mathbb{F}_2$. To that purpose, Graphix implements a lightweight, performant, just-in-time compiled $\mathbb{F}_2$ linear algebra module based on the \numba library \cite{numba2015} which allows computing the flow on open graphs with thousands nodes in a few seconds on a 2026 laptop. In Fig.~\ref{fig:flow-scaling} we show that for typical open graphs extracted from quantum circuits, the flow-finding algorithms perform significantly better than their upper complexity bound. We also compute the execution time scaling for worst-case examples, thereby providing a numerical validation of the main results in \cite{MP08:finding_flows, MB24:algebraic}.

Graphix also allows to extract the flow implemented by a specific (deterministic) correction strategy on a given open graph via the \pythoninline{XZCorrections} instance methods  \pythoninline{to_causalflow}, \pythoninline{to_gflow} and  \pythoninline{to_pauliflow}. These routines do not rely on the open-graph based flow-extraction algorithms. Instead, they exploit Eq.~\eqref{eq:flow_to_corrections} to extract the flow's correction function from the $\mathbf x$ and $\mathbf z$ correction maps. This distinction is crucial: in general, it is not guaranteed that a gflow or Pauli flow of the underlying open graph of the $\mathbf{xz}$-correction strategy given by the flow-finding algorithms (Table \ref{tab:flows}) generates the $\mathbf{xz}$-correction strategy itself because the gflow or Pauli flow of an open graph is not unique. Further, in the case of causal and gflow, the $\mathbf{xz}$-corrections-to-flow algorithms are linear in the number of elements in the correction maps. For Pauli flows, since the anachronistic corrections do not appear in the $\mathbf{x}_p$ and $\mathbf{z}_p$ maps [recall Eq.~\eqref{eq:flow_to_corrections_pauli}], the procedure requires solving a linear equation over $\mathbb{F}_2$ with cubic worst-case complexity. We provide additional details on these procedures in Appendix \ref{app:flow_from_xz}. Finally, we mention that Graphix offers equivalent flow-extraction methods at the level of the \pythoninline{Pattern} class which proceed by first standardizing the pattern to univocally determine an $\mathbf{xz}$-correction strategy. Pattern standardization is crucial to reconstruct a graph representation of an MBQC computation from a pattern following the prescription in Eq.~\eqref{eq:pattern_xzcorr}, since commuting entanglement commands with $\X$- and $\Z$-corrections may generate additional corrections and, consequently, introduce further constraints on the underlying partial order. This observation shows that implementing the standardization procedure in the library is not only a requisite to provide full support of the measurement calculus, but it is also crucial to transform between different MBQC representations.

\subsection{Pattern optimizations}
\label{subsec:patter_opt}
The measurement calculus provides a formal framework for reasoning about quantum computation beyond the circuit model. Therefore, any MBQC software must support manipulation and optimization at the level of patterns. In this subsection, we present three fundamental such tools implemented in Graphix. Pauli pushing that enables the systematic propagation of Pauli operators through a pattern, Pauli measurement removal that reduces the command count using graph-complementation techniques, and space minimization that reduces the number of qubits required during execution, which is essential for achieving optimal resource usage in simulation and on hardware.

\subsubsection{Pauli pushing}
\label{subsec:paulipush}

Early work mentioned that, due to their specific commutation relations, the Pauli measurements in any MBQC pattern could always be moved before the planar measurements \cite{RB00:measurements, RBB03:computation} without ever formalizing and implementing this procedure. It was not until recently that \cite{MPS22} formally defined this ``Pauli pushing'' transformation and proved that it preserves robust determinism. Concurrently, Graphix implemented such a procedure which is crucial for optimizing the pattern by removing the Pauli measurements.

Pauli pushing relies on commuting Pauli corrections~($\X^s$, $\Z^s$) through the operations of an MBQC pattern while preserving determinism. The commutation relations required to do this are defined in the measurement calculus~\cite{DKP07:calculus} and outlined in Section \ref{par:comms} and Appendix \ref{app:standardization}. Using these relations, ~\cite{MPS22} defined a general rule for shifting a Pauli measurement across a planar measurement. Iteratively applying this operation until it halts returns a pattern in which all Pauli measurements precede all planar measurements. 





Formally, measurement commands with empty correction domains commute with one another; however, \(\X_i^s\)
and \(\Z_i^t\) correction commands introduce dependencies, requiring that the
measurement of node \(i\) follows the measurements of
nodes in the domains \(D_X\) and \(D_Z\), which produce the signals $s := \bigoplus_{j\in D_X} s_j$ and $t := \bigoplus_{j\in D_Z} t_j$, respectively. What distinguishes Pauli
measurements from planar measurements with an arbitrary angle is that they are either
invariant under these corrections or undergo a simple bit flip of their output,  depending on their axis.
More precisely, for \(\lambda \in \{\axisX, \axisY, \axisZ\}\), the corrected Pauli measurement \(\Mdom{_i^{\pm\lambda}}{s}{t}\) is equivalent
to the bare measurement \(\M_i^{\pm\lambda}\), where the outcome is flipped if there is an odd number of 1s among the outcomes of the nodes in \(\sigma(\lambda, D_X, D_Z)\), with
\begin{equation}
\sigma(\lambda, D_X, D_Z) =
\begin{cases}
D_Z & \text{if \(\lambda = \axisX\)},\\
D_X \bigtriangleup D_Z & \text{if \(\lambda = \axisY\)},\\
D_X & \text{if \(\lambda = \axisZ\)},
\end{cases}
\end{equation}
where $\bigtriangleup$ denotes the symmetric difference between two sets. Whenever \(i\) belongs to the domain of a subsequent measurement,
\(\sigma(\lambda, D_X, D_Z)\) is incorporated into that domain via symmetric
difference to ensure that the outcome of node \(i\) is inverted if necessary.

The Pauli pushing algorithm is implemented in the \pythoninline{perform_pauli_pushing} method of \pythoninline{StandardizedPattern}. This class represents a pattern in standardized form and provides direct access to partitioned lists of commands categorized by type. For convenience, the method is also exposed via the \pythoninline{Pattern} class, which handles the round-trip conversion between \pythoninline{Pattern} and \pythoninline{StandardizedPattern}. The algorithm operates through a single pass over the measurements of the standardized pattern.

The algorithm initializes the following data structures:
\newcommand\MB{\mathcal M_{\mathrm B}}
\newcommand\MP{\mathcal M_{\mathrm P}}
\begin{itemize}
\item A mapping \(\mathcal S\) from nodes to set of nodes, initially empty;
\item A list \(\MP\) of Pauli measurements, initially empty;
\item A list \(\MB\) of non-Pauli measurements, initially empty.
\end{itemize}
The pattern measurements are partitioned between \(\MP\) and \(\MB\).
\(\mathcal S\) keeps track of the signals carried by the Pauli measurements that
have been moved.
\newcommand\Symdiff{\raisebox{-0.3ex}{\mbox{\Large \ensuremath{\bigtriangleup}}}}
More precisely, we define \(\epsilon(D) = D \bigtriangleup \Symdiff_{n \in D} \mathcal S(n)\), $\tilde{s} := \bigoplus_{j\in \epsilon(D_X)} s_j$, and $\tilde{t} := \bigoplus_{j\in \epsilon(D_Z)} t_j$. For each measurement \(\Mdom{_i^{\lambda,\alpha}}{s}{t}\) in the pattern taken in execution order (right to left):
\begin{itemize}
\item If \(\lambda \in \{\axisX, \axisY, \axisZ\}\) (Pauli):
\begin{itemize}
\item \(\MP \leftarrow \M_i^{\pm\lambda} \cdot \MP\);
\item \(\mathcal S(i) \leftarrow \sigma(\lambda, \mathcal \epsilon(D_X), \mathcal \epsilon(D_Z))\).
\end{itemize}
\item If \(\lambda \in \{\planeXY, \planeYZ, \planeXZ\}\) (non-Pauli):
\[\MB \leftarrow \Mdom{_i^{\lambda,\alpha}}{\tilde{s}}{\tilde{t}} \cdot \MB.\]
\end{itemize}

After processing all measurement commands the resulting pattern is \(
  \mathcal C \cdot \left(\prod_{\X_i^s \in \mathcal X}\X_i^{\tilde s}\right)\cdot\left(\prod_{\Z_i^t \in \mathcal Z}\Z_i^{\tilde t}\right) \cdot
  \MB \cdot \MP
  \cdot
  \mathcal E
  \cdot
  \mathcal N
  \), where  $\mathcal N, \mathcal E, \mathcal X, \mathcal Z, \mathcal C $ are, respectively, the sets of preparation, entanglement, correction and Clifford commands of the input pattern in standard form. This algorithm has a linear time complexity relative to the length of the pattern. The resulting pattern is equivalent to the input pattern, up to Pauli measurement flips, which can be corrected using \(\mathcal S\) to recover the original outcomes. Next, we explain how Graphix enables us take advantage of this transformation to remove most of the Pauli measurements in the pattern.



\subsubsection{Pauli measurement removal}
\label{subsec:pauli_removal}

The method \pythoninline{remove_pauli_measurements}, implemented at the level of \pythoninline{StandardizedPattern} and also exposed in \pythoninline{Pattern}, removes most Pauli measurements in a given pattern (and, therefore, the corresponding nodes in the associated open graph), at the expense of adding, at most, a Clifford command on each output node (see Box \ref{box:pauli_removal}).

\begin{quantumstacked}[box:pauli_removal]{Pauli removal}
\begin{minted}{python}
from graphix import Circuit

circuit = Circuit(2)
circuit.rzz(0, 1, 0.6)
circuit.s(0)

# Transpile the circuit into a pattern and
# cast all planar (Bloch) measurements
# into Pauli measurements.
pattern = circuit.transpile().pattern.\
          infer_pauli_measurements()

# Draw the pattern before and after removing
# Pauli measurement commands showing the LC
# commands on the output nodes. 
pattern.draw(local_clifford=True)
pattern.remove_pauli_measurements()
pattern.draw(local_clifford=True)

# By default, `pattern.draw` shows the
# pattern's open graph and flow.
# To show the xz-correction maps,
# pass the parameter
# `annotations=DrawPatternAnnotations.XZCorrections`
\end{minted}
\tcblower
\begin{figure}[H]
\centering
\includegraphics[width=\textwidth]{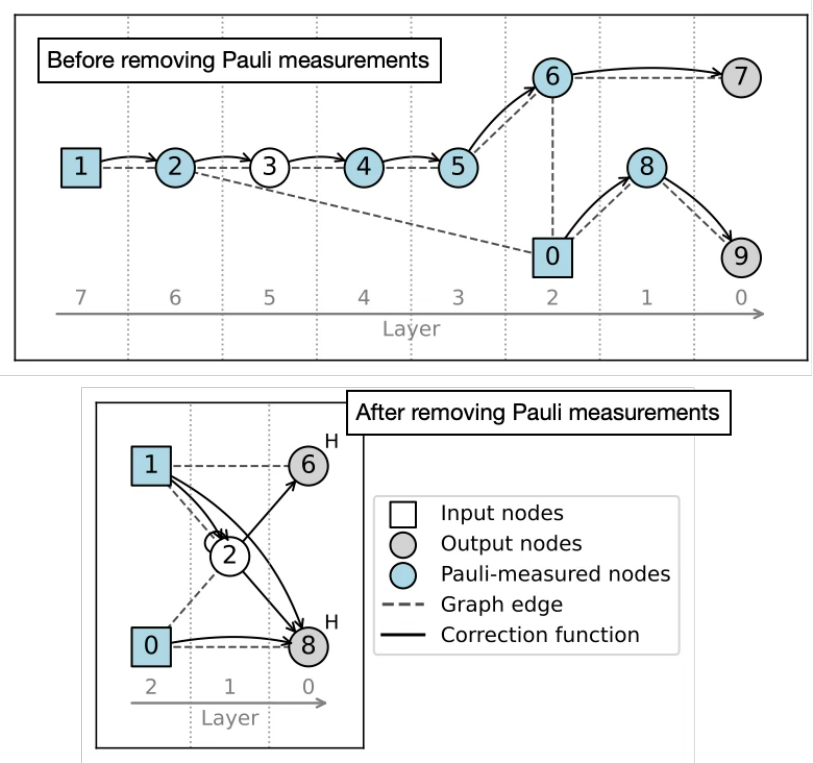}
\end{figure}
\end{quantumstacked}

This method preserves the semantics of the \textbf{0}-branch and preserves generalized flows and Pauli flows, if present. However, some measurements may leave the XY-plane in the procedure, which implies that the causal flow of the initial pattern is not conserved in general. Losing the causal flow property hinders the ability to optimally reduce the pattern's space (see Subsection~\ref{subsec:space_minimization}). Determining whether a subset of the Pauli measurements can be removed while preserving the $\planeXY$-plane measurements and the causal flow is an open problem. In general all Pauli measurements on non-input nodes are removed if the initial pattern has any kind of flow. This method implements the algorithm described in \cite{BMBdF+21}, Section~4.1, adapting the operations described on ZX diagrams to the measurement calculus.

The algorithm maintains a graph state together with the measurement calculus information attached to each remaining vertex (\emph{measurement calculus annotations}): correction domains, local Clifford operators and, when applicable, the associated Pauli measurement. Graph operations such as local complementation and pivoting are lifted to this enriched representation by updating both the graph structure and the attached measurement calculus annotations. In particular, local Clifford operations are propagated through measurement commands and correction domains according to the measurement calculus commutation rules, ensuring that the transformed pattern remains semantically equivalent to the original one.

Vertex removals, pivot-induced exchanges of vertices, and graph relabellings are tracked through an explicit mapping from the original pattern to the transformed graph. This allows the optimization result to be reconstructed directly as a valid standardized pattern.

The removal procedure follows Theorem 4.12 of \cite{BMBdF+21}. Vertices measured in the $Z$ basis are removed immediately, applying the Clifford $Z$ to the neighbour nodes if the measurement is $-Z$. Vertices measured in the $Y$ basis are turned into the $Z$ basis using local complementation, and then $Z$-removal is applied. Vertices measured in the $X$ basis are removed whenever they have an internal neighbour, using pivoting followed by $Z$-removal. If an $X$-measured vertex is connected only to an output vertex that is not also an input, a pivot converts the measurement into a removable $Z$ measurement, after which the algorithm restarts. The process terminates when no further Pauli measurements can be eliminated. The proof of Theorem 4.12 leaves some freedom in the interleaving of the different removal steps. The implementation adopts one particular strategy, but each type of removal is implemented in a separate method, making it straightforward to implement alternative strategies or perform only partial removal if desired.

An important property of this lifting is that it preserves the correction strategy. Classical dependencies are updated explicitly rather than recomputed, the resulting pattern remains standardized, and the graph transformations preserve the existence of generalized flow. Consequently, for patterns admitting flow, all non-input Pauli measurements are eliminated while preserving deterministic execution.

From a software engineering perspective, the implementation separates the algorithm into two components. The first is a class that maintains the graph structure and the annotations representing the correction strategies. This class exposes methods implementing the graph transformations for pivoting and removing of individual nodes. This class also provides a method for reconstructing an optimized pattern at any point during the transformation. The second component is a function that uses this class to apply all possible graph transformations before reconstructing the final pattern.

\subsubsection{Space minimization}
\label{subsec:space_minimization}

The \emph{max-space} of a pattern is the maximum number of simultaneously allocated qubits during execution. At the beginning of the execution, the input nodes are prepared simultaneously. Then each preparation command increases the number of prepared qubits by one, while each measurement command decreases it by one.

Once the order of measurements is fixed, we can compute a pattern with optimal max-space by preparing each qubit as late as possible, that is, immediately before one of its neighbors is measured. This strategy is implemented by the method \pythoninline{StandardizedPattern.to_pattern}, which returns a pattern that realizes the optimal max-space for the given measurement order by rearranging the \(\N\) and \(\E\) commands. This algorithm has linear time complexity in the length of the pattern. The difficulty lies in finding the total measurement order compatible with a given pattern such that the corresponding optimal max-space is the smallest possible max-space for that pattern.

If the pattern has causal flow, any total measurement order extending the flow partial order is optimal, provided that we compute the corresponding pattern as described in the preceding paragraph. As stated by Theorem 4.1 in \cite{HHF18}, if the graph has \(m\) qubits and the same number \(n\) of input and output nodes, then the max-space is \(\min(n + 1, m)\). The claim can be stated more generally: any pattern with causal flow has a max-space equal to the number of \emph{output} nodes plus one (regardless of the number of input nodes), unless there are isolated output nodes, in which case the max-space is simply the number of output nodes (see Appendix \ref{app:max_space_cf} for a proof). The case where the numbers of input and output nodes are equal is indeed a special case: the causal flow induces an injection from the input nodes to the output nodes. Since the two sets have the same finite cardinality, this injection is a bijection, so there are no isolated output nodes. By default, the method \pythoninline{Pattern.minimize_space} detects whether a pattern has causal flow and, if so, returns an optimal pattern (recall Box \ref{box:pattern_std}).

Finding the optimal measurement order for an arbitrary pattern is NP-hard. Indeed, given an arbitrary graph, a space-optimal pattern that prepares and measures all the nodes of the graph has a max-space equal to one plus the interval thickness of the graph \cite{EGM25}. Also known as the pathwidth of the graph, this graph invariant is NP-hard to compute~\cite{KF79}. Consequently, the method \pythoninline{Pattern.minimize_space} uses a heuristic approach in the general case.

The heuristic currently implemented in Graphix greedily selects the node of minimum degree in the graph obtained after removing previously measured nodes, while preserving runnability. If the heuristic fails to find a measurement order better than the original one, then the measurement order is left untouched. There is ongoing work to develop improved heuristics.

The close interplay between space-minimization algorithms and flows illustrates that native support for MBQC concepts is not only valuable for representing the formalism, but also for developing efficient simulation pipelines.


\subsection{Patterns from and to circuits}
\label{subsec:from_to_circuits}
The MBQC and the circuit models offer two alternatives to perform quantum computations. As such, it is of fundamental interest to translate from one representation to another. Graphix allows us to transpile quantum circuits into MBQC patterns and, conversely, extract a unitary evolution from a pattern into ancilla-free quantum circuits. This enables users to cast problems posed in the circuit picture into MBQC, manipulate them in the MBQC context, and export them as transformed circuits. At the same time, the method \pythoninline{Pattern.to_qasm3} shown in Box \ref{box:pattern_transpilation} allows to express a pattern as a quantum circuit with ancilla qubits, mid-circuit measurements and feed-forward. This is crucial for connecting both representations, but here we focus on unitary extraction, as the underlying logic is richer.

\subsubsection{Circuit transpilation}
\label{subsubsec:transpilation}
\begin{figure*}
\begin{center}
\resizebox{\textwidth}{!}{
\begin{tikzpicture}[
    nodebase/.style={font=\sffamily, inner sep=1.5pt, minimum size=6mm},
    inputnode/.style={nodebase, rectangle, draw=black, fill=white, text=black},
    measurednode/.style={nodebase, circle, draw=black, fill=white, text=black},
    outputnode/.style={nodebase, circle, draw=black, fill=lightgray, text=black},
    inputoutputnode/.style={nodebase, rectangle, draw=black, fill=white, text=black},
    >=stealth'
]
 
    
    \node[anchor=east, font=\Large] (psi1) at (0.4, 1.5) {$|i_0\rangle$};
    \node[anchor=east, font=\Large] (psi2) at (0.4, 0) {$|i_1\rangle$};
    
    \node[anchor=west, font=\Large] (o1) at (3.8, 1.5) {$|o_0\rangle$};
    \node[anchor=west, font=\Large] (o2) at (3.8, 0) {$|o_1\rangle$};
 
    \draw[line width=2.5pt] (psi1.east) -- (o1.west);
    \draw[line width=2.5pt] (psi2.east) -- (o2.west);
 
    \node[draw=mygreen, fill=white, line width=1.2pt, minimum height=0.8cm, minimum width=1.1cm] at (1.5, 1.5) {\Large $R_z(\alpha)$};
    \node[draw=myyellow, fill=white, line width=1.2pt, minimum height=0.8cm, minimum width=0.8cm] at (1.5, 0) {\Large $H$};
 
    \filldraw[mypurple] (2.9, 1.5) circle (4pt);
    \draw[mypurple, line width=2.5pt] (2.9, 1.5) -- (2.9, 0.3);
    \draw[mypurple, line width=2.5pt] (2.9, 0) circle (0.3cm);
    \draw[mypurple, line width=2.5pt] (2.9, 0.3) -- (2.9, -0.3);
 
    \node at (5.55, 0.75) {\Huge $\mathbf{\rightarrow}$};

    
    \draw[mygreen, line width=1.5pt, rounded corners=2pt, opacity=0.6] 
        (7.2, 1.1) rectangle (10.35, 1.9);
        
    \draw[myyellow, line width=1.5pt, rounded corners=2pt, opacity=0.6] 
        (7.2, -0.4) rectangle (9.2, 0.4);
        
    \draw[mypurple, line width=1.5pt, rounded corners=4pt, opacity=0.6]
        (9.55, 0.45) -- (9.55, 1.95) -- (10.45, 1.95) -- (10.45, 0.45) --
        (11.65, 0.45) -- (11.65, -0.45) --
        (8.35, -0.45) -- (8.35, 0.45) -- cycle;
 
    \draw[black, dashed, line width=1.5pt] (7.6, 1.5) -- (8.8, 1.5) -- (10.0, 1.5);
    \draw[black, dashed, line width=1.5pt] (7.6, 0) -- (8.8, 0) -- (10.0, 0) -- (11.2, 0);
    \draw[black, dashed, line width=1.5pt] (10.0, 1.5) -- (10.0, 0);  
    
    \node[inputnode] (n0) at (7.6, 1.5) {0};
    \node[inputnode, fill=cyan!15] (n1) at (7.6, 0) {1};
    \node[measurednode, fill=cyan!15] (n2) at (8.8, 1.5) {2};
    \node[measurednode, fill=cyan!15] (n4) at (8.8, 0) {4};
    \node[measurednode, fill=cyan!15] (n5) at (10.0, 0) {5};
    \node[outputnode] (n3) at (10.0, 1.5) {3}; 
    \node[outputnode] (n6) at (11.2, 0) {6}; 

    \draw [->, line width=1pt] (n0) to [out=-30,in=-150] (n2);
    \draw [->, line width=1pt] (n1) to [out=30,in=150] (n4);
    \draw [->, line width=1pt] (n4) to [out=-30,in=-150] (n5);
    \draw [->, line width=1pt] (n2) to [out=30,in=150] (n3);
    \draw [->, line width=1pt] (n5) to [out=30,in=150] (n6);
 
    \node[left=0.15cm of n0, font=\Large] {$|i_0\rangle$};
    \node[left=0.15cm of n1, font=\Large] {$|i_1\rangle$};
    \node[right=0.15cm of n3, font=\Large] {$|o_0\rangle$};
    \node[right=0.15cm of n6, font=\Large] {$|o_1\rangle$};

    \node (e2) at (13.05, 0.75) {\Huge $\mathbf{\rightarrow}$};

    
   \node[anchor=north west, inner sep=0pt, yshift=-0.2em] (pat1) at (14, 2) {$
  {\color{mypurple}\X_6^{s_5}\,\Z_6^{s_4}}\,
  {\color{mygreen}\X_3^{s_2}}\,
  \Z_3^{{\color{mygreen}s_0}\oplus {\color{mypurple}s_4}}\,
  \tensor[_{{\color{myyellow}s_1}\oplus\color{mygreen}s_2}]{\left[\M_5^{\rm X}\right]}{^{\color{mypurple}s_4}}
$};

    \node[anchor=north west, inner sep=0pt, yshift=-0.2em] (pat2) at (pat1.south west) {$
      {\color{mypurple}\E_{35}\,\E_{56}\,\N_6}\,
      \color{mypurple}\tensor[_{}]{\left[\M_4^{\rm X}\right]}{^{\color{myyellow}s_1}}
      {\color{mypurple}\E_{45}\,\N_5}
    $};
    
    \node[anchor=north west, inner sep=0pt, yshift=-0.2em] (pat3) at (pat2.south west) {$
      {\color{mygreen}\tensor[_{}]{\left[\M_2^{\rm X}\right]}{^{s_0}}}\,
      {\color{mygreen}\E_{23}\,\N_3}\,
      {\color{myyellow}\M_1^{\rm X}}\,
      {\color{myyellow}\E_{14}\,\N_4}
    $};
    
    \node[anchor=north west, inner sep=0pt, yshift=-0.3em] (pat4) at (pat3.south west) {$
      {\color{mygreen}\M_0^{\planeXY,\,-\alpha}}\,
      {\color{mygreen}\E_{02}\,\N_2}
    $};
\end{tikzpicture}
}
\end{center}
\caption{Transpilation of circuit gates into an open graph and pattern using the $\left\{J(\alpha),\,CZ\right\}$ gate set. \textit{Left:} initial circuit. \textit{Middle:} resulting open graph after the transpilation. Arrows represent the causal flow of the open graph and numbers represent node indices. Measurements are in the $\planeXY$ plane with angle $\theta = 0$ except node 0 with angle $\theta=-\alpha$. \textit{Right:} The corresponding transpiled pattern, with commands and signals highlighted by the gate they were transpiled from. The seemingly counter-intuitive $\Z$ commands stem from applying Eqs.~\eqref{eq:pattern_xzcorr} and \eqref{eq:flow_to_corrections_causal} on the causal flow in the middle panel. Here and in the rest of the figures we follow Graphix convention by which input nodes are represented with squares, output nodes are shaded in gray, and Pauli-measured nodes are filled in light blue.}
\label{fig:jcz}
\end{figure*}
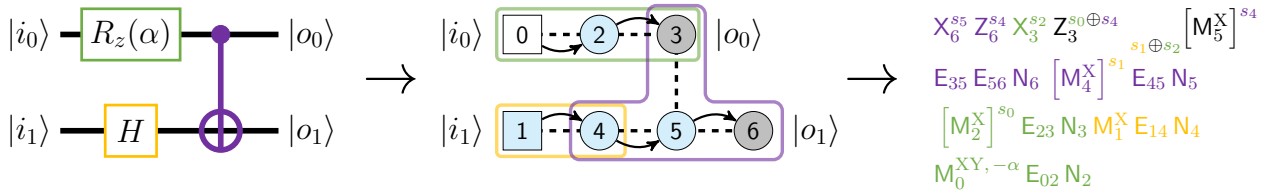

As described in Section \ref{subsec:universality}, a universal set of gates in the circuit model can be chosen and converted to their pattern form, giving a way to transpile circuits to patterns. The default method of transpilation in Graphix proceeds by rewriting the circuit instructions in the $\left\{J(\alpha),\,CZ\right\}$ universal gate set \cite{DKP07:calculus,DKPP09:extended_calculus}. Then, the prescription in Fig.~\ref{fig:mbqc_universality} is used to construct an open graph with causal flow where the flow correction function is given by the $\X$ corrections. The latter is written into a pattern following Eqs.~\eqref{eq:pattern_xzcorr} and \eqref{eq:flow_to_corrections_causal}. Passing through the universal $\left\{J(\alpha),\,CZ\right\}$ gate set instead of hard-coding the transpilation of specific selected gates \cite{paddleQ} often results in longer patterns. However, it has two crucial advantages: first, transpiled patterns are guaranteed to have causal flow and hence are apt for performing subsequent optimization manipulations; second, extending the transpiler only requires rewriting the new gates in the universal gate set and it is therefore a concern independent of the circuit-to-MBQC transpilation. An example of transpilation is demonstrated in Fig.~\ref{fig:jcz}.

The default $JCZ$-transpiler generates computation-dependent graphs which might not be a desired feature in some applications such as blind quantum computation~\cite{BFK09:UBQC} where universal graphs \ie, that can encode any computation, are needed. Hence, Graphix has been written to allow for alternative transpilation methods. 

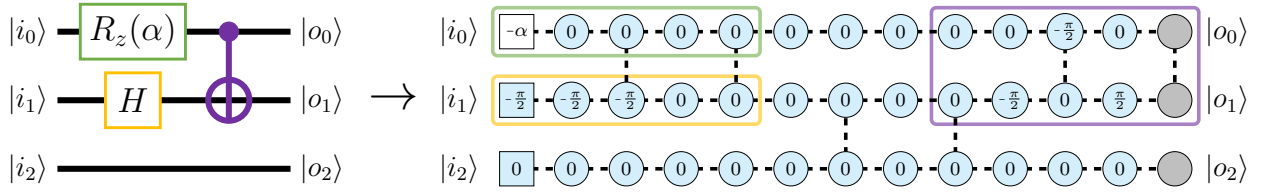
\begin{figure*}
\begin{center}
\resizebox{\textwidth}{!}{
\begin{tikzpicture}[
    nodebase/.style={font=\sffamily\scriptsize, inner sep=1pt, minimum size=5mm},
    inputnode/.style={nodebase, rectangle, draw=black, fill=white, text=black},
    measurednode/.style={nodebase, circle, draw=black, fill=white, text=black},
    outputnode/.style={nodebase, circle, draw=black, fill=lightgray, text=black}
]

    \def\dx{0.8}   
    \def\dy{1.0}   
    \def\bp{0.35}  

    \foreach \n in {0,...,38}{\expandafter\gdef\csname ang\n\endcsname{$0$}}
    \expandafter\gdef\csname ang0\endcsname{$-\alpha$}
    \foreach \n in {1,4,7,28,30}{%
        \expandafter\gdef\csname ang\n\endcsname{\sffamily\tiny$\shortminus \tfrac{\pi}{2}$}}
    \foreach \n in {34}{%
        \expandafter\gdef\csname ang\n\endcsname{\sffamily\tiny$\tfrac{\pi}{2}$}}
    \node[anchor=east, font=\large] (psi1) at (-6.7, 1)  {$|i_0\rangle$};
    \node[anchor=east, font=\large] (psi2) at (-6.7, 0)  {$|i_1\rangle$};
    \node[anchor=east, font=\large] (psi3) at (-6.7, -1) {$|i_2\rangle$};

    \node[anchor=west, font=\large] (o1) at (-3.3, 1)  {$|o_0\rangle$};
    \node[anchor=west, font=\large] (o2) at (-3.3, 0)  {$|o_1\rangle$};
    \node[anchor=west, font=\large] (o3) at (-3.3, -1) {$|o_2\rangle$};

    \draw[line width=2.5pt] (psi1.east) -- (o1.west);
    \draw[line width=2.5pt] (psi2.east) -- (o2.west);
    \draw[line width=2.5pt] (psi3.east) -- (o3.west);

    \node[draw=mygreen,  fill=white, line width=1.2pt,
          minimum height=0.8cm, minimum width=1.1cm] at (-5.6, 1) {\Large $R_z(\alpha)$};
    \node[draw=myyellow, fill=white, line width=1.2pt,
          minimum height=0.8cm, minimum width=0.8cm] at (-5.6, 0) {\Large $H$};

    \filldraw[mypurple] (-4.2, 1) circle (4pt);
    \draw[mypurple, line width=2.5pt] (-4.2, 1) -- (-4.2, 0.3);
    \draw[mypurple, line width=2.5pt] (-4.2, 0) circle (0.3cm);
    \draw[mypurple, line width=2.5pt] (-4.2, 0.3) -- (-4.2, -0.3);

    \node at (-1.8, 0) {\huge $\rightarrow$};

    \draw[mygreen, line width=1.5pt, rounded corners=2pt, opacity=0.6]
        ({-\bp}, {\dy-\bp}) rectangle ({4*\dx+\bp}, {\dy+\bp});
    \draw[myyellow, line width=1.5pt, rounded corners=2pt, opacity=0.6]
        ({-\bp}, {-\bp}) rectangle ({4*\dx+\bp}, {\bp});
    \draw[mypurple, line width=1.5pt, rounded corners=2pt, opacity=0.6]
        ({8*\dx-\bp}, {-\bp}) rectangle ({12*\dx+\bp}, {\dy+\bp});

    \foreach \i in {0,...,11}{
        \draw[black, dashed, line width=1.5pt] ({\i*\dx}, \dy)  -- ({(\i+1)*\dx}, \dy);
        \draw[black, dashed, line width=1.5pt] ({\i*\dx}, 0)    -- ({(\i+1)*\dx}, 0);
        \draw[black, dashed, line width=1.5pt] ({\i*\dx}, -\dy) -- ({(\i+1)*\dx}, -\dy);
    }
    \foreach \c in {2,4,10,12}{
        \draw[black, dashed, line width=1.5pt] ({\c*\dx}, \dy) -- ({\c*\dx}, 0);
    }
    \foreach \c in {6,8}{
        \draw[black, dashed, line width=1.5pt] ({\c*\dx}, -\dy) -- ({\c*\dx}, 0);
    }

    \node[inputnode] (n0) at (0, \dy)  {$\shortminus \alpha$};
    \node[inputnode, fill=cyan!15] (n1) at (0, 0)    {\sffamily\tiny$\shortminus \tfrac{\pi}{2}$};
    \node[inputnode, fill=cyan!15] (n2) at (0, -\dy) {$0$};

    \node[outputnode] (n36) at ({12*\dx}, \dy)  {};
    \node[outputnode] (n37) at ({12*\dx}, 0)    {};
    \node[outputnode] (n38) at ({12*\dx}, -\dy) {};

    \foreach \c in {1,...,11}{
        \pgfmathtruncatemacro{\nl}{3*\c}
        \node[measurednode, fill=cyan!15] (n\nl) at ({\c*\dx}, \dy) {\csname ang\nl\endcsname};
    }
    \foreach \c in {1,...,11}{
        \pgfmathtruncatemacro{\nl}{3*\c+1}
        \node[measurednode, fill=cyan!15] (n\nl) at ({\c*\dx}, 0) {\csname ang\nl\endcsname};
    }
    \foreach \c in {1,...,11}{
        \pgfmathtruncatemacro{\nl}{3*\c+2}
        \node[measurednode, fill=cyan!15] (n\nl) at ({\c*\dx}, -\dy) {\csname ang\nl\endcsname};
    }

    \node[left=0.15cm of n0, font=\large] {$|i_0\rangle$};
    \node[left=0.15cm of n1, font=\large] {$|i_1\rangle$};
    \node[left=0.15cm of n2, font=\large] {$|i_2\rangle$};

    \node[right=0.05cm of n36, font=\large] {$|o_0\rangle$};
    \node[right=0.05cm of n37, font=\large] {$|o_1\rangle$};
    \node[right=0.05cm of n38, font=\large] {$|o_2\rangle$};
\end{tikzpicture}
}
\end{center}
\caption{Representation of the transpilation of circuit gates into a brickwork open graph using the $\left\{J(\alpha),\,CNOT\right\}$ gate set. An extra qubit has been added to showcase the brickwork structure. Contrary to convention, angles have been labelled (all measured in the $\planeXY$ plane) to highlight the placement of transpiled gates.}
\label{fig:brickwork}
\end{figure*}

One example of universal graph is the brickwork state \cite{BFK09:UBQC}. The \pythoninline{graphix-brickwork-transpiler} plug-in provides an alternative transpiler which also decomposes circuits using the $\left\{J(\alpha),\,CNOT\right\}$ universal gate set but maps them to a brick (a \(2 \times 5\) subgraph) of the brickwork state as shown in Fig.~\ref{fig:brickwork}.

\subsubsection{Circuit extraction}
\label{subsec:extraction}

W.~Simmons introduced in \cite{Simmons21} a circuit extraction algorithm allowing to re-express a deterministic pattern as a sequence of gates without mid-circuit measurements or feed-forward. In essence, this method amounts to writing planar measurements as rotated Pauli measurements, then propagating these rotations to determine their effect on the output nodes, and finally characterizing the remaining stabilizer process. 

\begin{figure}[H]
    \centering
    \resizebox{\linewidth}{!}{\newcommand{\pythoninlinefs}[2]{%
  \mintinline[fontsize=\fontsize{#2}{#2}]{python}{#1}%
}

\definecolor{graphcolor}{named}{orange}
\definecolor{circuitcolor}{named}{teal}
\definecolor{toolcolor}{named}{purple}

\newcommand{\toolfontsize}{\fontsize{14pt}{14pt}\selectfont}
\newcommand{\arrowfontsize}{\fontsize{9pt}{9pt}\selectfont}

\newcommand{\og}{
    \node[inputnode] (n0) at (0,0) {$\alpha_0$};
    \node[inputnode] (n1) at ([yshift=-1cm]n0) {$\alpha_1$};
    \node[measurednode] (n2) at ([xshift=1cm]n0) {$\alpha_2$};
    \node[measurednode] (n3) at ([xshift=1cm]n1)  {$\alpha_3$};
    \node[outputnode] (n4) at ([xshift=1cm]n2) {}; 
    \node[outputnode] (n5) at ([xshift=1cm]n3) {}; 

    \draw[black, densely dashed, line width=1pt] (n0) -- (n2) -- (n4);
    \draw[black, densely dashed, line width=1pt] (n1) -- (n3) -- (n5);
    \draw[black, densely dashed, line width=1pt] (n2) -- (n3);
 
    \node[left=0.15cm of n0, font=\Large] {$|i_0\rangle$};
    \node[left=0.15cm of n1, font=\Large] {$|i_1\rangle$};
    \node[right=0.15cm of n4, font=\Large] {$|o_0\rangle$};
    \node[right=0.15cm of n5, font=\Large] {$|o_1\rangle$};
}

\newcommand{\circuit}{
\begin{quantikz}[column sep=0.3cm, row sep=0.001cm]
\lstick{$q_0$} & \gate[2]{\mathcal{C}} & \gate[2]{P_{\text{exp}}(\boldsymbol{\alpha})} & \qw \\
\lstick{$q_1$} &                        & \qw      & \qw
\end{quantikz}
}

\begin{tikzpicture}[
    font=\rmfamily,
    >=Stealth,
    box/.style={draw, rounded corners=2pt, align=center, thick},
    smallbox/.style={draw, rounded corners=2pt, align=center, inner sep=2pt, fill=white, thick},
    patternBox/.style={box, fill=blue!5, draw=blue!40!black, minimum width=2.7cm, minimum height=0.65cm},
    graphBox/.style={box, fill=graphcolor!10, draw=graphcolor!60!black, minimum width=3cm, minimum height=2cm},
    flowBox/.style={box, fill=graphcolor!10, draw=graphcolor!60!black, minimum width=3cm, minimum height=0.65cm},
    circuitBox/.style={box, fill=circuitcolor!10, draw=circuitcolor!60!black, minimum width=3cm, minimum height=1.9cm},
    extBox/.style={box, fill=graphcolor!1, draw=graphcolor!60!black,minimum width=4.5cm, minimum height=3cm},
    extInBox/.style={box, fill=graphcolor!1, draw=graphcolor!60!black, densely dashed, minimum width=1cm, minimum height=0.5cm},
    arr/.style={->, ultra thick},
    nodebase/.style={font=\sffamily, inner sep=1.5pt, minimum size=6mm},
    inputnode/.style={nodebase, rectangle, draw=black, fill=white, text=black},
    measurednode/.style={nodebase, circle, draw=black, fill=white, text=black},
    outputnode/.style={nodebase, circle, draw=black, fill=lightgray, text=black},
    inputoutputnode/.style={nodebase, rectangle, draw=black, fill=white, text=black},
]


    \node[graphBox, text=graphcolor!80!black, anchor=north](og) at (0,0) {};
    \node[text=graphcolor!80!black, anchor=north](oglabel) at ([yshift=-0.15cm]og.north) {\pythoninlinefs{OpenGraph}{14pt}};

    \node (og_cartoon) at ([yshift=-0.9cm]oglabel) {
    \begin{tikzpicture}[scale=0.6, transform shape]
        \og
    \end{tikzpicture}
    };
    
    \node[flowBox, text=graphcolor!80!black,](pauliflow) at ([yshift=-1.7cm]og.south) {\pythoninlinefs{PauliFlow}{14pt}};

    \node[extBox, text=graphcolor!80!black, anchor=north east](ext) at ([xshift=-2.8cm]pauliflow.north west) {};
    \node[text=graphcolor!80!black, anchor=north](extlabel) at ([yshift=-0.25cm]ext.north) {\pythoninlinefs{ExtractionResult}{14pt}};
    
    \node[extInBox, text=graphcolor!80!black, anchor=north](pexp) at ([yshift=-0.4cm]extlabel.south) {\pythoninlinefs{PauliExponentialDag}{11pt}};
    \node[extInBox, text=graphcolor!80!black, anchor=north](cm) at ([yshift=-0.3cm]pexp.south) {\pythoninlinefs{CliffordMap}{11pt}};

    \node[circuitBox, text=circuitcolor!80!black, anchor=south] (circuit) at (pauliflow |- ext.south) {};
    \node[text=circuitcolor!80!black, anchor=north](circuitlabel) at ([yshift=-0.1cm]circuit.north) {\pythoninlinefs{Circuit}{14pt}};

    \node (circuit_cartoon) at ([yshift=-0.9cm]circuitlabel) {
        \scalebox{0.8}{$\circuit$}
    };

    \node[patternBox, text=blue!60!black](pattern) at (ext |- og){\pythoninlinefs{Pattern}{14pt}};
    

    \draw[arr, draw=graphcolor!70!black] (pattern.east) -- node[fill=white, font=\arrowfontsize, inner sep=2pt, text=graphcolor!80!black, align=center] {$\N$, $\E$ and $\M$\\commands\\from std.~form} (og.west);

    \draw[arr, draw=graphcolor!70!black] (og.south) -- node[fill=white, font=\arrowfontsize, inner sep=2pt, text=graphcolor!80!black, align=center, yshift=0.15cm] {Flow-finding\\algorithm} (pauliflow.north);

    \draw[arr, draw=graphcolor!70!black] (pauliflow.west) -- node[fill=white, font=\arrowfontsize, inner sep=2pt, text=graphcolor!80!black, align=center] {Extraction} (pauliflow.west -| ext.east);

    \draw[arr, draw=circuitcolor!70!black] (circuit.west -| ext.east) -- node[fill=white, font=\arrowfontsize, inner sep=2pt, text=circuitcolor!80!black, align=center]  {Compilation} (circuit.west);
    
\end{tikzpicture}}
    \caption{Diagram illustrating the circuit extraction pipeline in Graphix. An \pythoninline{ExtractionResult} instance is obtained with the method chain  \pythoninline{Pattern.to_opengraph().to_pauliflow().extract_circuit()}, and can be converted to a \pythoninline{Circuit} by calling the method \pythoninline{to_circuit}. The full extraction routine is wrapped in the instance method \pythoninline{OpenGraph.to_circuit}.}
    \label{fig:circuit_ext}
\end{figure}
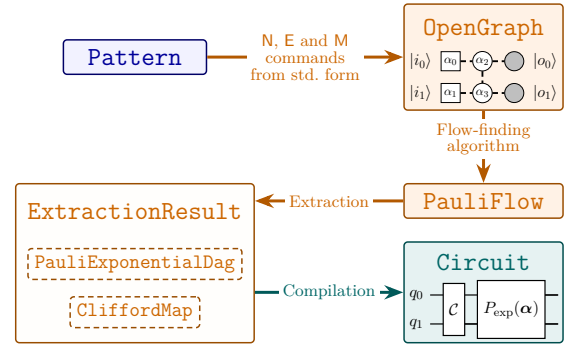

Graphix implements this procedure in three modular steps (Fig.~\ref{fig:circuit_ext}): first, we find a focused Pauli flow on the underlying open graph of the pattern of interest. From this flow we construct an ordered product of Pauli exponentials $P_{\text{exp}}(\bm{\alpha})$  and a Clifford map $\mathcal C$ such that the product
\begin{equation}
L_{\mathcal P} (\bm{\alpha}) = P_{\text{exp}}(\bm{\alpha}) \, {\mathcal C}    
\end{equation}
represents the pattern's linear transformation. In Graphix, this intermediate representation (\pythoninline{ExtractionResult}) bundles two stand-alone classes. Namely,  $P_{\text{exp}}(\bm{\alpha}) = \prod_{j}^{\prec} e^{i \alpha_j P_j/2}$ is encoded as a \pythoninline{PauliExponentialDAG} object which consists of a mapping between planar-measured nodes $j$ to the pair measurement angle $\alpha_j$ and Pauli string $P_j$ on the output nodes, and a partial order $\prec$ over the open graph's nodes representing the order of application of the Pauli exponentials. The Pauli string is the stabilizer represented by the Pauli flow's correction function $p(j)$, and the partial order is precisely the flow's partial order which in this context implies that exponentials corresponding to measured nodes in the same layer commute with each other. Considering a focused flow correction function is critical to guarantee that the associated Pauli strings act on output nodes \cite{BMBdF+21, Simmons21}, but crucially, the algebraic flow-finding algorithm always returns a focused Pauli (or g-) flow \cite{MB24:algebraic}. On the other hand, the object \pythoninline{CliffordMap} encodes the Clifford transformation $\mathcal{C}$ in the stabilizer formalism as two maps ($X$ and $Z$) from input nodes to Pauli strings on the output nodes \cite{AG04:imp}.

On the last step of the extraction process, the $P_{\text{exp}}(\bm{\alpha})$ and $\mathcal C$ are compiled separately and merged into a single circuit. Currently, Graphix implements two compilation passes \cite{NC10:quantum, vanB21:simple} which allow to natively extract a circuit from unitary patterns, thereby achieving full interoperability between the MBQC and the circuit models. In addition, the modularity of the circuit extraction process allows users to design their own custom passes targeting \pythoninline{PauliExponentialDAG} and \pythoninline{CliffordMap}, possibly interfacing with specialized third-party libraries (e.g., \tket or \stim), which opens a route to exploit the MBQC formalism for circuit optimization. Finally, we mention that the extraction procedure (\pythoninline{Pattern} $\rightarrow$ \pythoninline{ExtractionResult}) also supports isometries (i.e., deterministic patterns with $|O| > |I|$). In the short term, we plan to introduce compilation passes to synthesise the corresponding circuits.

Circuit extraction has been discussed by the seminal work \cite{BMBdF+21} in the context of ZX calculus for open graphs with gflow. However, the algorithm introduced in \cite{Simmons21} and incorporated in Graphix also targets open graphs with Pauli flow and it is amenable to a native MBQC implementation.

\subsection{Symbolic manipulation}
\label{subsec:symbolic}

Modern quantum software often allows to represent continuous variables (e.g., angles in rotation gates) as symbolic expressions. This parametric view of quantum computations eases the implementation of variational algorithms, which have also been discussed in the context of MBQC~\cite{F+21:mbqcvqa}. In Graphix, all angles (measurement angles in patterns and open graphs, and angles in rotation gates) can be represented as affine algebraic expressions on \pythoninline{Placeholder} objects. Crucially, pattern manipulations such as standardization or space minimization can be applied to patterns with placeholders. Further, all transformations between the core objects in the library (transpilation, circuit extraction, generation of patterns from flows, etc.) support symbolic measurement angles which can be substituted by numerical values in any representation (see Box \ref{box:symb_manip}). 

\begin{quantumstacked}[box:symb_manip]{Symbolic manipulation}
\begin{minted}{python}
import networkx as nx
from graphix import Measurement, OpenGraph, Placeholder

# Initialize placeholders
alphas = [Placeholder(f"alpha({i})")
          for i in range(3)]

og = OpenGraph(
    graph=nx.Graph(
        [(0, 3), (0, 4), (1, 4), (2, 4)]),
    input_nodes=[0],
    output_nodes=[3, 4],
    measurements={
        0: Measurement.XY(alphas[0]),
        1: Measurement.YZ(alphas[1]),
        2: Measurement.XZ(alphas[2])})

# Convert open graph to pattern
# (recall Box 2.4)
p_symb = og.to_pattern()

# Substitution relies on object identity.
# Provide the same `Placeholder` instance 
# stored in the open graph.
p = p_symb.with_parameters(
    dict(zip(alphas, [0.25, 0.5, 0.75])))

print("Symbolic pattern: ", p_symb)
print("Pattern: ", p)
\end{minted}
\tcblower
Symbolic pattern:

Z(4,\{1\}) M(1,YZ,pi*alpha(1))

X(4,\{2\}) X(3,\{2\}) Z(4,\{2\}) Z(1,\{2\}) M(2,XZ,pi*alpha(2))

X(3,\{0\}) M(0,pi*alpha(0))

E(4,2) E(4,1) E(0,4) E(0,3) N(4) N(3) N(2) N(1)

Pattern:

Z(4,\{1\}) M(1,YZ,pi/2)

X(4,\{2\}) X(3,\{2\}) Z(4,\{2\}) Z(1,\{2\}) M(2,XZ,3pi/4)

X(3,\{0\}) M(0,pi/4)

E(4,2) E(4,1) E(0,4) E(0,3) N(4) N(3) N(2) N(1)
\end{quantumstacked}

\subsection{Pattern simulation}
\label{subsec:pattern_simulation}
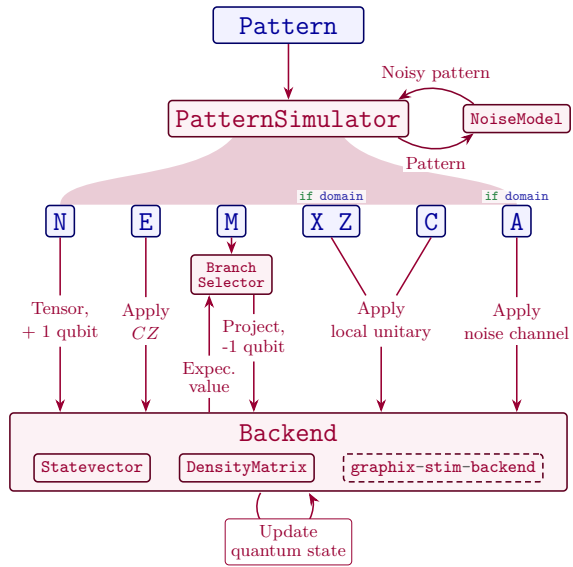
\begin{figure}
    \centering
    \resizebox{\linewidth}{!}{\usetikzlibrary{decorations}
\usetikzlibrary{positioning, fit, calc, shapes, arrows.meta, backgrounds}

\newcommand{\pythoninlinefs}[2]{%
  \mintinline[fontsize=\fontsize{#2}{#2}]{python}{#1}%
}

\definecolor{toolcolor}{named}{purple}

\newcommand{\toolfontsize}{\fontsize{14pt}{14pt}\selectfont}
\newcommand{\arrowfontsize}{\fontsize{9pt}{9pt}\selectfont}

\begin{tikzpicture}[
    font=\rmfamily,
    >=Stealth,
    box/.style={draw, rounded corners=2pt, align=center, thick},
    patternBox/.style={box, fill=blue!5, draw=blue!40!black, minimum width=2.7cm, minimum height=0.65cm},
    cmdBox/.style={box, fill=blue!5, draw=blue!40!black},
    simBox/.style={box, fill=toolcolor!5, draw=toolcolor!50!black, minimum width=1.5cm, minimum height=0.65cm},
    arr/.style={->, thick},
]


    \node[patternBox, text=blue!60!black](pattern) at (0, 0){\pythoninlinefs{Pattern}{14pt}};

    \node[simBox, text=toolcolor!80!black, anchor=north](psim) at ([yshift=-1cm]pattern.south) {\pythoninlinefs{PatternSimulator}{14pt}};

    \matrix (allcmds) [
    matrix of nodes,
    nodes={cmdBox,text=blue!60!black, align=center},
    column sep=1cm,
    anchor=center,
    ] at ([yshift=-1.5cm]psim.south)
    {
        \pythoninlinefs{N}{14pt}&
        \pythoninlinefs{E}{14pt}&
        \pythoninlinefs{M}{14pt}&
        \pythoninlinefs{X Z}{14pt}&
        \pythoninlinefs{C}{14pt}&
        \pythoninlinefs{A}{14pt}\\
    };

    \node[simBox, text=toolcolor!80!black, anchor=north, minimum width=1cm](bs) at ([yshift=-0.3cm]allcmds-1-3.south) {\pythoninlinefs{Branch}{8pt}\\[-0.2cm]\pythoninlinefs{Selector}{8pt}};

    \node[simBox, minimum width=10cm, anchor = north, minimum height=1.4cm](backend) at ([yshift=-3cm]allcmds.south) {};
    \node[text=toolcolor!80!black, anchor=north, yshift=-2pt] at (backend.north) {\pythoninlinefs{Backend}{14pt}};

    \matrix (backends) [
    matrix of nodes,
    nodes={simBox, text=toolcolor!80!black, align=center, minimum height = 0.5cm},
    column sep=0.5cm,
    anchor=center,
    text depth=0.2ex,
    ] at ([yshift=-1cm]backend.north)
    {
        \pythoninlinefs{Statevector}{9pt}&
        \pythoninlinefs{DensityMatrix}{9pt}&
        \node[densely dashed]{\pythoninlinefs{graphix-stim-backend}{9pt}};\\
    };

    \draw[arr, toolcolor!80!black](backend.south) ++(-0.5cm,-0.02cm) 
  arc[start angle=150, end angle=390, radius=0.5cm];

    \node[fill=white, draw=toolcolor!80!black, inner sep=3pt, font=\arrowfontsize, align=center, text=toolcolor!80!black, rounded corners=2pt] at ([yshift=-.9cm]backend.south) {Update\\[0.02cm] quantum state};

    \fill[toolcolor!80!black, opacity=0.2]
      ($(psim.south) + (-1,0)$)
      .. controls ($(psim.south) + (-1.7,-0.8)$) and ($(allcmds-1-1.north) + (0, 0.6)$) ..
      (allcmds-1-1.north)
      .. controls (allcmds-1-1.north)  ..
      (allcmds-1-2.north)
      .. controls (allcmds-1-2.north) ..
      (allcmds-1-3.north)
      .. controls (allcmds-1-3.north) ..
      (allcmds-1-4.north)
      .. controls (allcmds-1-4.north) ..
      (allcmds-1-5.north)
      .. controls (allcmds-1-5.north) ..
      (allcmds-1-6.north)
      .. controls ($(allcmds-1-6.north) + (0, 0.6)$) and ($(psim.south) + (1.7,-0.8)$) ..
      ($(psim.south) + (1,0)$)
      -- cycle;

    \draw[arr, draw=toolcolor!80!black](pattern.south) -- (psim.north);
    
    \node[fill=white, inner sep=1pt, font=\arrowfontsize, align=center, text=blue!60!black, fill opacity=0.8](ifXZ) at ([yshift=0.15cm]allcmds-1-4.north) {\pythoninlinefs{if}{7pt} \pythoninlinefs{domain}{7pt}};
    
    \node[fill=white, inner sep=1pt, font=\arrowfontsize, align=center, text=blue!60!black, fill opacity=0.8](ifA) at ([yshift=0.15cm]allcmds-1-6.north) {\pythoninlinefs{if}{7pt} \pythoninlinefs{domain}{7pt}};
    

    \node[simBox, text=toolcolor!80!black, minimum height=0.5cm](ns) at (allcmds-1-6.north |- psim.east) {\pythoninlinefs{NoiseModel}{9pt}};

    \coordinate (I) at ($(psim.east)!0.5!(ns.west)$);

    \node[fill=white, inner sep=1pt, font=\arrowfontsize, align=center, text=toolcolor!80!black](nslabel1) at ([yshift=0.8cm]I) {Noisy pattern};
    \node[fill=white, inner sep=1pt, font=\arrowfontsize, align=center, text=toolcolor!80!black](nslabel1) at ([yshift=-0.8cm]I) {Pattern};

    \draw[arr, draw=toolcolor!70!black]
            ([xshift=-0.2cm]psim.south east)
            to[bend right=35]
            ([xshift=0.2cm]ns.south west);

    \draw[arr, draw=toolcolor!70!black]
            ([xshift=0.2cm]ns.north west)
            to[bend right=35]
            ([xshift=-0.2cm]psim.north east);

    
    \node[fill=white, inner sep=3pt, font=\arrowfontsize, align=center, text=toolcolor!80!black](nlabel) at ([yshift=-1.5cm]allcmds-1-1.south) {Tensor,\\[0.1cm]+ 1 qubit};
    
    \node[fill=white, inner sep=3pt, font=\arrowfontsize, align=center, text=toolcolor!80!black](elabel) at ([yshift=-1.5cm]allcmds-1-2.south) {Apply\\[0.1cm]$CZ$};

    \node[fill=white, inner sep=3pt, font=\arrowfontsize, align=center, text=toolcolor!80!black](mlabel1) at ([yshift=-2.5cm, xshift=-0.4cm]allcmds-1-3.south) {Expec.\\value};
    \node[fill=white, inner sep=3pt, font=\arrowfontsize, align=center, text=toolcolor!80!black](mlabel2) at ([yshift=-1.8cm, xshift=0.4cm]allcmds-1-3.south) {Project,\\[0.03cm]-1 qubit};

    \coordinate (J) at ($(allcmds-1-4.south)!0.5!(allcmds-1-5.south)$);
    \node[fill=white, inner sep=3pt, font=\arrowfontsize, align=center, text=toolcolor!80!black](xzclabel) at ([yshift=-1.5cm]J.south) {Apply\\[0.1cm]local unitary};
    
    \node[fill=white, inner sep=3pt, font=\arrowfontsize, align=center, text=toolcolor!80!black](alabel) at ([yshift=-1.5cm]allcmds-1-6.south) {Apply\\[0.1cm]noise channel};
    
    \draw[-, thick, draw=toolcolor!80!black](allcmds-1-1.south) -- (nlabel);
    \draw[-, thick, draw=toolcolor!80!black](allcmds-1-2.south) -- (elabel);
    \draw[-, thick, draw=toolcolor!80!black](allcmds-1-4.south) -- (xzclabel);
    \draw[-, thick, draw=toolcolor!80!black](allcmds-1-5.south) -- (xzclabel);
    \draw[-, thick, draw=toolcolor!80!black](allcmds-1-6.south) -- (alabel);
    \draw[-, thick, draw=toolcolor!80!black]($(mlabel2.north |- bs.south)$) -- (mlabel2);
    \draw[-, thick, draw=toolcolor!80!black](mlabel1) -- ($(mlabel1 |- backend.north)$);

    \draw[arr, draw=toolcolor!80!black](allcmds-1-3.south) -- (bs);
    \draw[arr, draw=toolcolor!80!black](nlabel) -- ($(nlabel |- backend.north)$);
    \draw[arr, draw=toolcolor!80!black](elabel) -- ($(elabel |- backend.north)$);
    \draw[arr, draw=toolcolor!80!black](xzclabel) -- ($(xzclabel |- backend.north)$);
    \draw[arr, draw=toolcolor!80!black](alabel) -- ($(alabel |- backend.north)$);
    \draw[arr, draw=toolcolor!80!black](mlabel1) -- ($(mlabel1.north |- bs.south)$);
    \draw[arr, draw=toolcolor!80!black](mlabel2) -- ($(mlabel2 |- backend.north)$);
    
\end{tikzpicture}}
    \caption{Schematic for Graphix' simulation stack. The \pythoninline{PatternSimulator} is instantiated with a \pythoninline{Backend}, a \pythoninline{BranchSelector} and, optionally, a \pythoninline{NoiseModel}. It handles the MBQC logic while the \pythoninline{Backend} updates a representation of the quantum state according to the commands' semantics.}
    \label{fig:simulationflow}
\end{figure}

Graphix provides a modular framework for pattern simulation (Fig.~\ref{fig:simulationflow}), supporting multiple built-in simulation backends, configurable measurement branch selection, custom qubit preparation and measurement methods\footnote{Preparation and measurement methods allow to customize the semantics of $\N$ and $\M$ commands. We will not discuss this in detail here, but this is crucial for simulating secure delegated quantum computation. See the Veriphix package \cite{veriphix2024,veriphix_paper2026}.}, and integration with third-party quantum software development kits. Graphix simulator follows a two-tier design: the top layer processes the pattern commands sequentially, and is in charge of computing outcome probabilities and handling the feed-forward logic. Simultaneously, the simulator's bottom layer (\emph{backend}) manages the commands semantics, i.e., it applies quantum operations (entanglement, Pauli operations, measurement, etc.) on a certain representation of the quantum state. This separation between the MBQC control logic and the quantum-state simulation enables alternative backends to be used interchangeably with minimal changes to the simulation workflow. Different quantum-state representations, such as state vectors, density matrices, or tensor networks, can therefore be employed transparently, and the same interface can target external simulators or quantum hardware. Existing plugins (listed in Appendix \ref{app:plugins}) include interfaces to IBM's \qiskit, and Quandela's \perceval frameworks \cite{Aleksandrowicz19:qiskit, qiskit24, H+23:perceval}.

The core simulation functionality of Graphix comprises its built-in state vector and density matrix backends, both written in \numpy \cite{numpy2020}. A tensor-network backend with limited functionality is currently available, and further development is ongoing. Their implementation is similar to standard circuit-based quantum simulators \cite{McGRI25}, except for one crucial difference which motivates the need for dedicated MBQC simulation backends: MBQC simulations require a dynamically-sized register that increases under the application of $\N$ commands and decreases under the application of $\M$ commands. The basic usage for pattern simulation is \pythoninline{Pattern.simulate(backend=backend)} with \pythoninline{backend="statevector"} or \pythoninline{backend="densitymatrix"}. For convenience, Graphix also includes an equivalent \pythoninline{Circuit.simulate} to do an exact simulation of a circuit that has not yet been transpiled to a pattern.

\subsubsection{Noisy simulation} \label{sec:backends}




Graphix supports noisy simulation through the \pythoninline{NoiseModel} interface. At the top level of the simulation stack, a noise model acts as a transpilation step from patterns to noisy patterns (formally, a transduction) by introducing a new command \(\A_{\vb i}^s\pqty{\cal E}\), whose semantics are: apply the completely positive trace-preserving (CPTP) map ${\cal E}$ to the set of nodes ${\vb i}$, conditioned on the signal $s$. This mechanism allows to replace any command \(\B_{\vb i}^s\) by a noisy counterpart of the form \(\A_{\vb j}^s\pqty{{\mathcal E}_2}\, \B_{\vb i}^s \,\A_{\vb k}^s\pqty{{\mathcal E}_1}\), subject only to dimensional consistency and runnability constraints. Conditioning the noise operations on the same signal as the original command is essential to preserve the execution semantics of the pattern. In addition, a \pythoninline{NoiseModel} can specify noise acting on the input state and stochastic flips of measurement outcomes. The \pythoninline{NoiseModel} interface is designed to support flexible and composable noise descriptions, enabling the modelling of device-specific effects such as correlated errors or cross-talk. 

An example of a pattern simulation in the density matrix backend using the built-in depolarizing noise model is shown in Box~\ref{box:noisy_sim}.


\begin{quantumstacked}[box:noisy_sim]{Simulation of noisy patterns}
\begin{minted}{python}
from graphix import BasicStates, Circuit, DepolarisingNoiseModel

circuit = Circuit(1)
circuit.h(0)
pattern = circuit.transpile().pattern

state = pattern.simulate(
    input_state=BasicStates.ZERO)

noisy_state = pattern.simulate(
    backend="densitymatrix",
    input_state=BasicStates.ZERO,
    noise_model=DepolarisingNoiseModel(
        x_error_prob=0.1,
        entanglement_error_prob=0.1))

# Computes 〈state|noisy_state|state〉
f = noisy_state.fidelity(state)
assert 0 < f < 1
\end{minted}
\end{quantumstacked}

\subsubsection{\stim backend}

Measurement patterns including solely Pauli measurements implement Clifford operations. By the Gottesman-Knill theorem \cite{G97:stabilizer}, these patterns admit efficient classical simulation when the input state is a stabilizer state. To this purpose, Graphix provides an interface with the \stim library \cite{Gidney21:stim} via the \pythoninline{graphix-stim-backend} plug-in, which leverages \stim's efficient stabilizer simulation and sampling strategies. Crucially, since noise models are defined as a compilation step on measurement patterns, they are independent of the choice of backend, provided the backend supports the required operations. Consequently, any stochastic Pauli noise specified in Graphix can also be simulated using the \pythoninline{graphix-stim-backend} plug-in.

\subsubsection{Branch selection}
As discussed in Section \ref{subsec:detflow}, each possible measurement outcome generates a new execution branch in the pattern. Graphix provides several branch-selection strategies that can be used with any simulation backend, thanks to the separation between the MBQC control logic and the quantum-state evolution in the simulator architecture. By default, the built-in simulator uses the \pythoninline{RandomBranchSelector}, which computes the probability of each measurement outcome and samples branches accordingly. For strongly deterministic patterns, however, all branches are equivalent up to a global phase and occur with equal probability~\cite{MMPST14}. In this case, the branch selector can instead sample outcomes uniformly at random, thereby reducing the simulation cost, since it avoids explicit probability calculations and the associated exponentially-expensive computation of expectation values. This illustrates that branch selectors are not only a mechanism for handling MBQC post-selection, but also enable specialized simulation strategies.

Graphix additionally provides two alternative branch-selection strategies. The \pythoninline{FixedBranchSelector} allows to supply predefined measurement outcomes, thereby selecting a specific execution branch in advance. As shown in Box~\ref{box:bs}, this is particularly useful for analyzing non-deterministic patterns. The \pythoninline{ConstBranchSelector} fixes all measurement outcomes to the same value (0 or 1), enabling exploration of the reference branch, or its complement, without prior knowledge of the pattern structure.

\begin{quantumstacked}[box:bs]{Branch selector}
\begin{minted}{python}
from graphix import FixedBranchSelector, Pattern
from graphix.command import E, M, N, Z

# Non-deterministic pattern:
# Prepare |+⟩|0⟩ + |-⟩|1⟩, 
# measure qubit 0 along X,
# apply Z on qubit 1 conditionally
# on the measurement outcome
p = Pattern(cmds=
    [N(0), N(1), E((0, 1)), M(0), Z(1, {0})]) 

bs_0 = FixedBranchSelector(results={0: 0})
bs_1 = FixedBranchSelector(results={0: 1})

s0 = p.simulate(branch_selector=bs_0)
s1 = p.simulate(branch_selector=bs_1)

print("Statevector for branch 0:", s0)
print("Statevector for branch 1:", s1)
\end{minted}
\tcblower
Statevector for branch 0: |0>

Statevector for branch 1: -|1>
\end{quantumstacked}

\section{Reproducible research workflows with Graphix}
\label{sec:applications}
In this section, we exemplify the use of the Graphix library by re-implementing and complementing published work. The topics considered are efficient measurement-based pseudorandomness \cite{MGDM18:pseudo}, relating resource-state quality to MBQC computation fidelity \cite{SFF26} and implementing non-unitary gates with a focus on imaginary-time evolution \cite{ARS24:nunit}.
This shows the quality of the design of the library as well as the use of a generic MBQC framework to encompass realistic research-level examples. All the code used to generate these examples is available on the GitHub repository \cite{graphix_paper_repo}.

\subsection{Efficient pseudorandomness}
\label{subsec:pseudorand}
Efficient pseudorandomness is an important topic of research in the field of quantum information processing. Objects of central interest are unitary and state \(t\)-designs, namely, ensembles of unitaries or quantum states that reproduce the first \(t\) moments of the corresponding Haar-random ensembles. These ensembles provide practical approximations to Haar randomness while avoiding the complexity of exact Haar sampling. Due to its intrinsic randomness, MBQC is a natural framework to construct such designs. Ref.~\cite{MGDM18:pseudo} introduces a measurement-based unitary \(t\)-design. We focus here on the state \(t\)-design resulting from applying the unitary \(t\)-design on a fixed state. The closeness of an ensemble of states \({\sf S}\) to a state \(t\)-design can be characterized by the state frame potential \cite{NTKD24}
\begin{align}
    F_t(\mu) \coloneqq \mathbb{E}_{\ket \psi, \ket \phi \sim \mu}\left[\abs{\braket{\psi}{\phi}}^{2t}\right],
    \label{eq:state_frame_pot_def}
\end{align}
where \(\mu\) is the probability distribution associated to \({\sf S}\) over quantum states living in a Hilbert space of dimension \(d\). Importantly, $F_t(\mu) \in \left[\binom{d + t - 1}{t}^{-1} , 1\right]$ where equality with the lower bound signals an exact state $t$-design \cite{NTKD24}. Following \cite{NTKD24}, we denote \(d_t := \binom{d + t - 1}{t}
\).

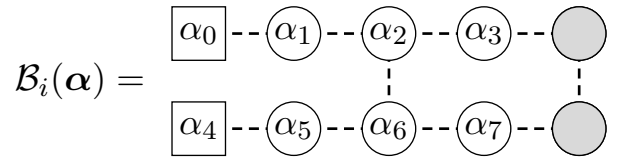
\begin{figure}[h!]
    \centering
    \resizebox{\linewidth}{!}{\begin{tikzpicture}
    \brickwork{4}{1}{
        {\alpha_0,\alpha_1,\alpha_2,\alpha_3,},
          {\alpha_4,\alpha_5,\alpha_6,\alpha_7,},}
    \node[anchor=east, xshift=-4pt] at (current bounding box.west) {$ \mathcal{B}_i(\vb* \alpha) = $};
    \end{tikzpicture}}
    \caption{Brickwork labelled open graph: node labels are measurement angles, all measurements are in the \(\planeXY\) plane. \(\mathcal{B}_1\) is obtained by setting \({\vb* \alpha} = (\alpha_0, \dots, \alpha_7) = (\frac{\pi}{4},0,\frac{\pi}{4},0, 0,\frac{\pi}{4},0,\frac{\pi}{2}) \) and \(\mathcal{B}_2\) has angles \({\vb* \alpha} = (0, \frac{\pi}{4}, 0, \frac{\pi}{2}, \frac{\pi}{4}, 0, \frac{\pi}{4}, 0) \).}
    \label{fig:first_pseudorand}
\end{figure}

The building block of the construction is \(\mathcal{B}_i (\vb* \alpha)\), the elementary ``brick'' of the brickwork graph (recall Subsection~\ref{subsubsec:transpilation}) defined via the labelled open graph displayed in Fig.~\ref{fig:first_pseudorand} and parametrized by the vector of angles \({\vb* \alpha}\).

Since this open graph has causal flow, the patterns with correction functions given by Eq.~\eqref{eq:flow_to_corrections_causal}
implement unitaries \(U\pqty{\vb* \alpha}\) for all measurement angles values \({\vb* \alpha} \in \left[0, 2\pi \right)^8\) due to robust determinism. Thus, in order to generate randomness, we instead extract patterns with trivial correction functions \({\vb x}\left(i\right) = {\vb z}\left(i\right) = \varnothing\), \(\forall\, i \) in the measured qubits set (i.e., we remove all corrections from the pattern). This generates an ensembles of unitaries since obtaining outcome $s_i = 1$ for measurement $i$ with angle $\alpha_i$ corresponds to obtaining the $s_i=0$ outcome for the pattern with angle $\alpha_i + \pi$, and hence implementing the unitary \(U\pqty{\alpha_0, \dots, \alpha_{i-1}, \alpha_i + \pi, \alpha_{i+1}, \dots }\). Then each correctionless \(\mathcal{B}_i\) pattern implements $2^8$ unitaries for a choice of \(\vb* \alpha\). Additionally, since strong determinism implies the equiprobability of all measurement branches \cite{MMPST14}, the pattern generates the uniform distribution over its set of unitaries. Note however that this does not necessarily imply that the generated distribution of states obtained by applying the different unitaries to a fixed state is uniform.

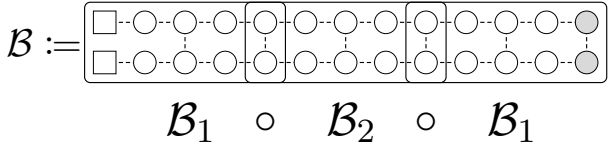
\begin{figure}[h!]
    \centering
    \resizebox{\columnwidth}{!}{
    \begin{tikzpicture}
        \brickrow{3}
        \node[anchor=east, xshift=-4pt] at (current bounding box.west){\scalebox{2.5}{$\mathcal{B} \coloneqq $}};
        \draw[rounded corners, thick] (-0.5, 0.5) rectangle (4.5, 2.5);
        \node at (2.15, -.5) {\scalebox{3}{\({\mathcal B}_1\)}};
        \draw[rounded corners, thick] (3.5, 0.5) rectangle (8.5, 2.5);
        \node at (6.15, -.5) {\scalebox{3}{\({\mathcal B}_2\)}};
        \draw[rounded corners, thick] (7.5, 0.5) rectangle (12.5, 2.5);
        \node at (10.15, -.5) {\scalebox{3}{\({\mathcal B}_1\)}};
        \node at (4, -.5) {\scalebox{3}{\(\circ\)}};
        \node at (8, -.5) {\scalebox{3}{\(\circ\)}};
    \end{tikzpicture}
}
    \caption{Measurement pattern \({\cal B}\). Measurement angles are omitted for clarity.}
    \label{fig:Bgraph}
\end{figure}

Following the prescription in Ref.~\cite{NTKD24}, we choose two bricks \({\cal B}_1\) and \(\mathcal{B}_2\) with vector angles \( {\vb* \alpha_1} = (\frac{\pi}{4},0,\frac{\pi}{4},0, 0,\frac{\pi}{4},0,\frac{\pi}{2}) \) and \({\vb* \alpha_2} = (0, \frac{\pi}{4}, 0, \frac{\pi}{2}, \frac{\pi}{4}, 0, \frac{\pi}{4}, 0) \). These bricks are composed into a new pattern \({\cal B} \coloneqq {\cal B}_1 \circ {\cal B}_2 \circ {\cal B}_1\) by ``plugging'' the output nodes of \({\cal B}_1\) into the input nodes of \({\cal B}_2\) while keeping the ordering as shown in Fig.~\ref{fig:Bgraph}.

Pattern \({\cal B}\) is then composed in a brickwork fashion as shown in Fig.~\ref{fig:full_pseudorand_pattern}. The total number of input qubits is denoted as $n$ and a layer consists of two horizontal \({\cal B}\) patterns.

\newlength{\BpWidth}
\newlength{\BpHeight}
\newlength{\BpRadius}
\newlength{\BpFontSize}
\newlength{\BpColSep}
\newlength{\BpRowSep}
\newlength{\BpGap}

\setlength{\BpWidth}{1.4cm}   

\newcommand{\BpSetDims}{%
  \BpHeight=0.39\BpWidth
  \BpRadius=0.08\BpWidth
  \BpFontSize=0.45\BpWidth
  \BpColSep=0.75\BpWidth      
  \BpGap=0.12\BpWidth
  \BpRowSep=\BpHeight
  \advance\BpRowSep by \BpGap
}
\BpSetDims

\newcommand{\BpAt}[2]{%
  \draw[rounded corners=\BpRadius, thick]
    (#1, #2) rectangle +(\BpWidth, \BpHeight);
  \node at ({#1 + 0.5\BpWidth}, {#2 + 0.5\BpHeight})
    {\fontsize{\BpFontSize}{\BpFontSize}\selectfont$\mathcal{B}$};
}

\newcommand{\BrickBlock}[1]{%
  \foreach \col in {0,...,\numexpr#1-1\relax} {
    \pgfmathsetlengthmacro{\xpos}{\col * \BpColSep}
    \pgfmathsetlengthmacro{\yoff}{mod(\col,2) * (-0.5 * \BpRowSep)}
    \BpAt{\xpos}{\yoff}
    \BpAt{\xpos}{\yoff + \BpRowSep}
    \pgfmathsetlengthmacro{\dotsx}{\xpos + 0.5*\BpWidth}
    \pgfmathsetlengthmacro{\dotsy}{\yoff - 0.55*\BpRowSep}
    \node at (\dotsx, \dotsy) {$\vdots$};
  }%
}

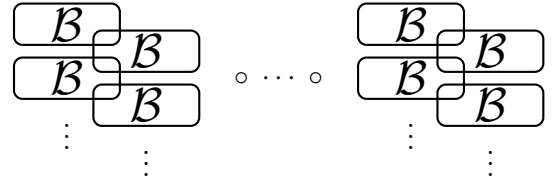
\begin{figure}[h!]
  \centering
  \begin{tikzpicture}[x=1pt, y=1pt, baseline=(current bounding box.center)]
    \BrickBlock{2}
    \pgfmathsetlengthmacro{\blockRightEdge}{1*\BpColSep + \BpWidth}
    \pgfmathsetlengthmacro{\dotsGap}{1.5*\BpWidth}
    \pgfmathsetlengthmacro{\rightX}{\blockRightEdge + \dotsGap}
    \pgfmathsetlengthmacro{\dotsX}{\blockRightEdge + 0.5*\dotsGap}
    \pgfmathsetlengthmacro{\dotsY}{0.5*\BpHeight}
    \node[anchor=center] at (\dotsX, \dotsY) {$\circ\;\cdots\;\circ$};
    \begin{scope}[xshift=\rightX]
      \BrickBlock{2}
    \end{scope}
  \end{tikzpicture}
  \caption{Final structure. \({\cal B}\) graphs are composed in a brickwork fashion. We call a layer a column made of two \({\cal B}\) bricks in width.}
  \label{fig:full_pseudorand_pattern}
\end{figure}

\begin{figure}[h]
    \centering
    \includegraphics[width=\columnwidth]{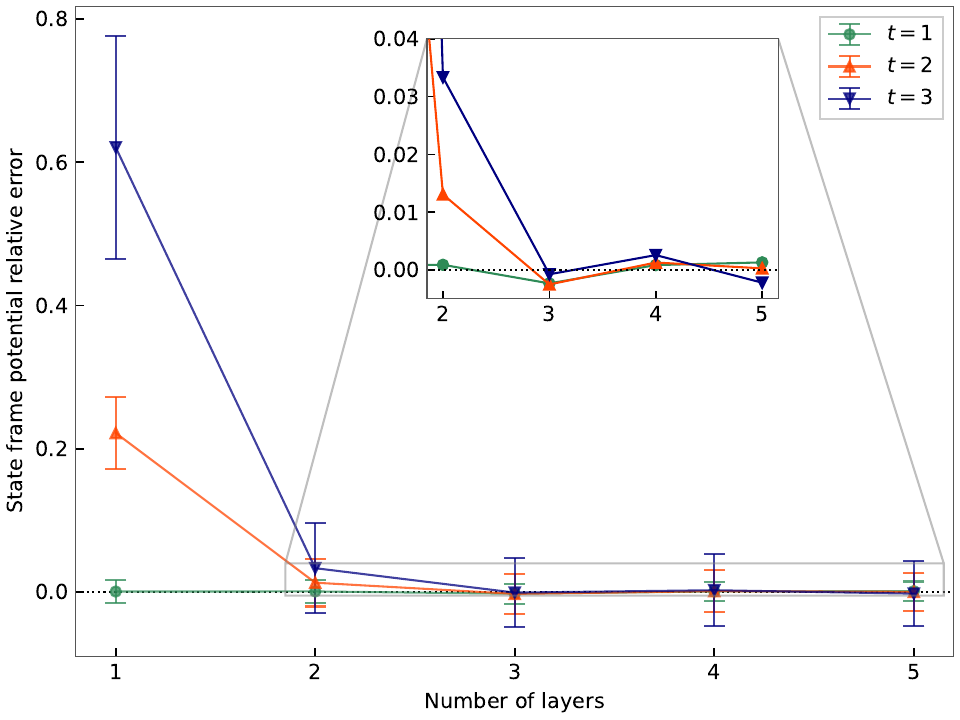}
    \caption{Mean state frame potential relative error \(\overline{F_t}/d_t^{-1} - 1\) for 4 qubits, a number of layers between 1 and 5, \(t \in \{1,2,3\}\). Individual results for the relative error are obtained by sampling 100 times from the distribution. This is repeated 100 times to gather statistics and compute the mean and error bars. Small negative values are due to numerical errors.}
    \label{fig:pseudorand_results}
\end{figure}

To implement the pattern in Fig.~\ref{fig:full_pseudorand_pattern} in Graphix, we make use of the \pythoninline{pattern_A.compose(pattern_B,{B_input:A_output})} method which allows to compose two patterns by mapping a subset of the outputs of one pattern to a subset of the inputs of the other. Note that Graphix also allows for the composition of open graphs which allows more flexibility than the composition of patterns. Once the pattern is defined, it can easily be simulated yielding an output state. Since states are the natural output of the simulation, we focus on characterizing how close the generated ensembles of states are to a $t$-design using the state frame potential defined in Eq.~(\ref{eq:state_frame_pot_def}). We numerically estimate the state frame potential for a particular $t$ by generating a given number of states from the pattern with fixed input state (here a product of $\ket+$ states) and compute all pairwise terms $\abs{\braket{\psi}{\phi}}^{2t}$. We then compute the relative difference of the estimated state frame potential \(\overline{F_t}\) with the analytical lower bound \(d_t^{-1}\) as \(\overline{F_t}/d_t^{-1} - 1\). Repeating this procedure yields an estimate of the mean relative error. The results for 4 qubits, varying number of layers between 1 and 5, 100 samples drawn from the distribution and 100 repetitions are displayed in Fig.~\ref{fig:pseudorand_results}.



This figure shows the convergence of the state frame potential to the analytical lower bound as a function of the number of layers for different values of \(t\): the higher the value of \(t\), the slower the convergence.
This shows that Graphix provides the necessary tools for creating and simulating MBQC patterns of relevance for numerically investigating pseudorandom properties beyond exact analytical results often based on the Haar measure. Being out of the scope of the present paper, we leave it to interested user to refine this investigation.

\subsection{MBQC fidelity}
\label{subsec:fidelity}
Very recently, authors in \cite{SFF26} introduced the notion of \emph{average MBQC fidelity} to quantify how noise in the resource graph state affects the output of measurement-based quantum computations. Specifically,
\begin{equation}
    \overline{F}_{\text{MBQC}}(\rho) \equiv \int \frac{d\boldsymbol{\theta}}{(2\pi)^M}\sum_{\boldsymbol{s}} p_{\boldsymbol{s}} \bra{\psi(\boldsymbol{\theta})}\sigma_{\boldsymbol{s}}(\boldsymbol{\theta})\ket{\psi(\boldsymbol{\theta})},
    \label{eq:fidelity_sim}
\end{equation}
where $\ket{\psi(\boldsymbol{\theta})}$ and $\sigma_{\boldsymbol{s}}(\boldsymbol{\theta})$ denote, respectively, the output state obtained on the ideal and noisy resource state for an MBQC computation defined by the measurement angles $\boldsymbol{\theta}$. The sum runs over all execution branches $\boldsymbol{s}$ weighted by their probability $p_{\boldsymbol{s}}$. We then average over all choices of measurement angles, which amounts to averaging over all unitary transformations implementable on the resource state under consideration. Following \cite{SFF26}, here and in the rest of the subsection we restrict ourselves to open graphs with $\planeXY$ measurements only. The authors show that the average MBQC fidelity can be expressed as a correlation function on the noisy resource state $\rho$,
\begin{equation}
    \overline{F}_{\text{MBQC}}(\rho) = \tr \rho \, \Omega,
    \label{eq:fidelity_samp}
\end{equation}
where $\Omega = \sum_{g\in \mathcal{G}^{XY}} p_g \, g$ with $\mathcal{G}^{XY}$ the subset of all stabilizers of the resource state that act as only $I$, $X$, $Y$ on measured qubits, and $p_g$ a weight depending on the structure of the resource state such that $\sum_{g\in \mathcal{G}^{XY}} p_g = 1$.

\begin{figure}
    \centering
    \includegraphics[width=1.0\linewidth]{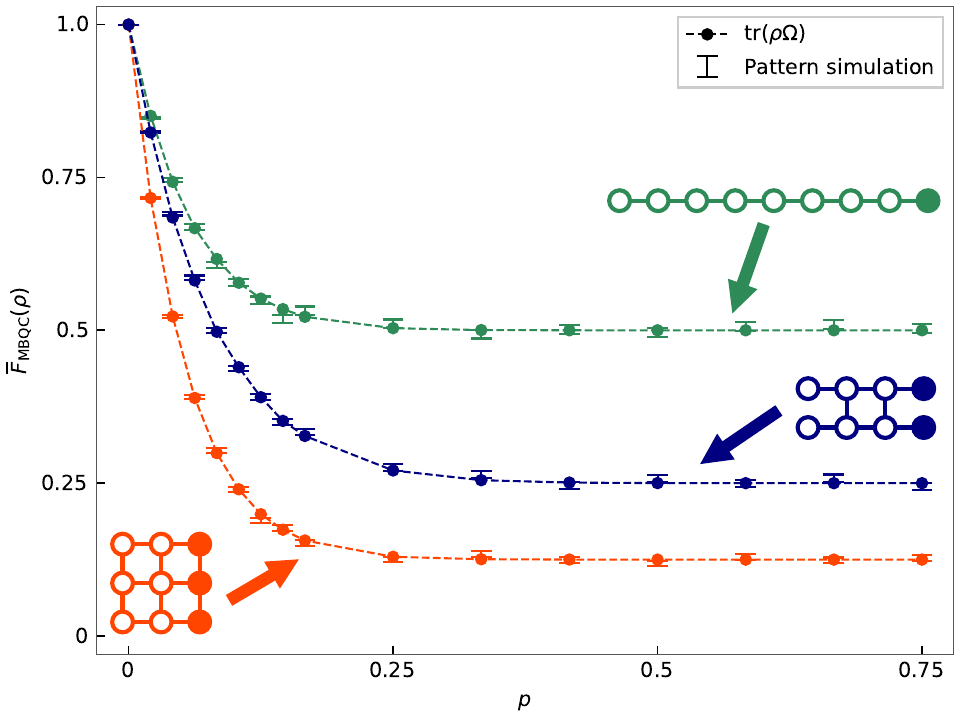}
    \caption{Average MBQC fidelity as a function of the depolarizing parameter $p$. Round markers were obtained by averaging the trace of the matrix product $\rho g$ over 5000 stabilizers $g\in \mathcal{G}^{XY}$ drawn according to $p_g$, with $\rho$ the density matrix of the noisy resource state. Error-bar markers indicate the mean estimator of Eq.~\eqref{eq:fidelity_sim} with $n_{\boldsymbol{\theta}} = 50$ sample realizations. Each curve corresponds to a different resource graph state as represented by the schematics in the plot. Hollow and solid nodes respectively represent measured and output qubits. Dashed lines are drawn to guide the eye.}
    \label{fig:mbqc-fidelity}
\end{figure}

Graphix's tools allow to readily compute both expressions of $\overline{F}_{\text{MBQC}}(\rho)$ for arbitrary resource graph states and noise models. Starting from a resource state in the form of an \pythoninline{OpenGraph[Measurement]}, the flow machinery yields the corresponding deterministic pattern. By simulating this pattern for multiple random $\boldsymbol{\theta}$ with a noise model in the density matrix backend, we obtain an estimate of the integral in Eq.~\eqref{eq:fidelity_sim}. We chose a depolarising channel for which all branches are equally probable, and branch sampling is unnecessary. However, branch sampling is straightforward in Graphix since by default each measurement outcome probability is computed and branches are sampled accordingly. On the other hand, since the algebraic flow-finding algorithm outputs a focused flow \cite{MB24:algebraic}, we have direct access to the $R$-stabilizers defined in Ref.~\cite{SFF26} which are the fundamental ingredient to sample $g\in \mathcal{G}^{XY}$ according to the probability distribution $p_g$ \cite{SFF26}, and we can efficiently estimate Eq.~\eqref{eq:fidelity_samp}. 

In Fig.~\ref{fig:mbqc-fidelity} we compare Eq.~\eqref{eq:fidelity_sim} and Eq.~\eqref{eq:fidelity_samp} for 1D and 2D cluster states with a depolarising noise model on both qubit preparation and entanglement, showing consistency with the results of \cite{SFF26}. This preliminary analysis indicates that some resource-state topologies exhibit greater resilience to noise than others, even with comparable number of qubits and entanglement operations, demonstrating the usefulness of Graphix for numerical investigations in MBQC.

\subsection{Non-unitary gates and imaginary-time evolution}
\label{subsec:nonunitary}
While quantum computation primarily relies on unitary operations, it has been shown that non-unitary gates can also be an important tool in studying topics such as entanglement phase transitions, imaginary time evolution, and solving NP-complete problems \cite{ARS24:nunit, Mao_23:ite, Motta_19:qite}. The authors in \cite{ARS24:nunit} introduce a way to realize non-unitary gates using MBQC patterns and demonstrate how they can be used in particular to perform imaginary time evolution. In this section, we reproduce and extend their analysis using Graphix.

We know from Section \ref{subsec:detflow} that strongly deterministic patterns realize unitary transformations. It is shown in \cite{ARS24:nunit} that a deterministic pattern containing measurements only along the X axis and in the XY plane can be transformed into a non-deterministic pattern by flipping the planar measurements onto the XZ plane. In this section, we will examine one such non-deterministic pattern with one node measured along X and one node measured in the XZ plane:
\begin{equation} \label{eq:nunit_pattern}
    \mathcal{P}_\text{NU} = \M_{1}^{\X}\,\M_{0}^{\planeXZ,\theta}\,\E_{12}\,\E_{01}\,\N_{2}\,\N_{1}.
\end{equation}
Note that we do not include any corrections in the above pattern --- the transformation applied by $\mathcal{P}_\text{NU}$ depends on the measurement outcomes $\{s_1, s_2\}$ on nodes $0$ and $1$ respectively\footnote{For convenience, we follow the notation in the original paper \cite{ARS24:nunit}. With the notation employed in the rest of this manuscript, we would denote the measurement outcomes $\{s_0, s_1\}$.}, and is equivalent to applying the non-unitary gate
\begin{equation} \label{eq:nunit_gate}
    M_{s_1, s_2} = \frac{1}{\sqrt{2}} X^{s_2} M_{s_1},
\end{equation}
where 
\begin{equation} \label{eq:nunit_subgates}
\begin{split}
    M_0 &= \frac{1}{\sqrt{1 + a^2}} \begin{pmatrix}
        a & 0 \\
        0 & 1
    \end{pmatrix},\\
    M_1 &= \frac{1}{\sqrt{1 + a^2}} \begin{pmatrix}
        1 & 0 \\
        0 & -a
    \end{pmatrix},
\end{split}
\end{equation}
and $a = \frac{\sin{\theta}}{1 - \cos{\theta}}$, with $\theta \in (0, \frac{\pi}{2})$ the measurement angle in the XZ plane with respect to the Z axis \cite{ARS24:nunit}.

Starting from a graphical representation, Graphix allows us to first extract a deterministic pattern and then transform it to a non-deterministic one by modifying the appropriate measurement and removing the corrections. Alternatively, we can directly instantiate the desired non-deterministic pattern by specifying the precise sequence of commands. Once we have constructed $\mathcal{P}_\text{NU}$, simulating it for any arbitrary input state and each of the possible outcomes $\{s_1, s_2\}$ shows its equivalence to the non-unitary gate $M_{s_1, s_2}$ given by Eq.~\eqref{eq:nunit_gate}.

\begin{figure}[H]
    \centering
    \includegraphics[width=1.0\linewidth]{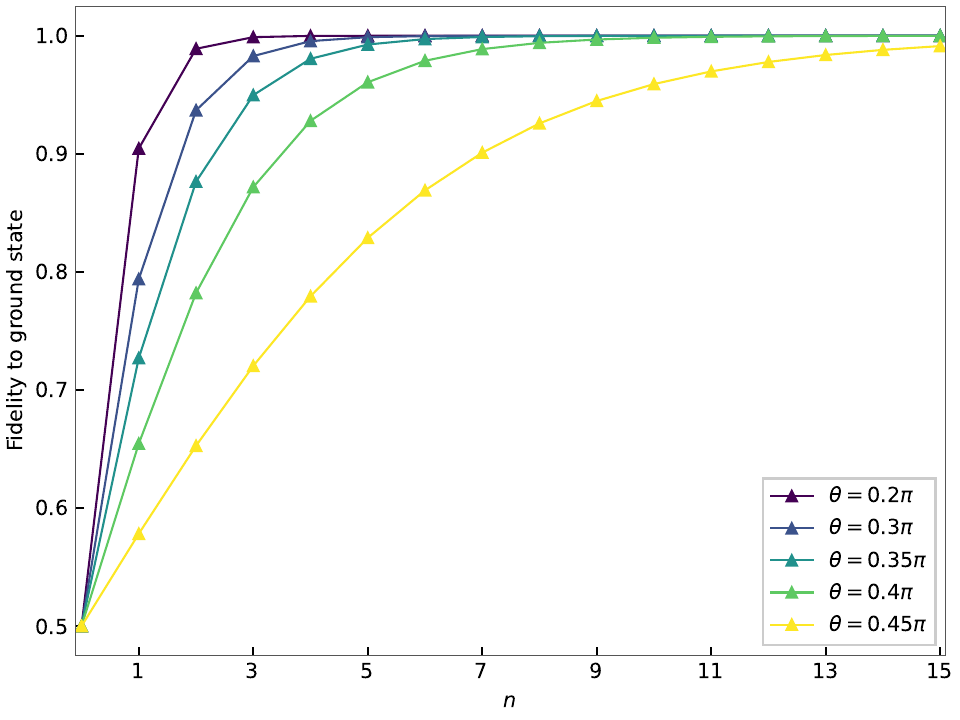}
    \caption{Simulation of imaginary time evolution $e^{-\tau H_z}$ in Graphix with repeated applications of pattern $\mathcal{P}_\text{NU}$. Using the \pythoninline{BranchSelector} class in Graphix, we post-select each iteration of the pattern on the outcome set $\{0,0\}$ to simulate the successful application of the time evolution gate,  given by Eqs.~\eqref{eq:nunit_gate}--\eqref{eq:nunit_subgates}. Starting with the initial state $\ket{+}$, the evolution tilts it towards ground state $\ket{0}$ of the Hamiltonian $-Z$. The figure depicts the fidelity to the ground state, expressed  $\abs{\braket{0}{\text{out}}}^2$, as a function of the number of repetitions $n$ for various measurement angles $\theta$ on the $\planeXZ$ plane.}
    \label{fig:ite_var_theta}
\end{figure}

The authors in \cite{ARS24:nunit} show that by concatenating the gate $M_{0,0}$ $n$ times, we obtain a gate implementing  the imaginary time evolution $e^{-\tau H_z}$, with $\tau=\frac{n}{2}\log a$ and the Hamiltonian $H_z=-Z$. In Fig.~\ref{fig:ite_var_theta}, we simulate this in Graphix by applying $\mathcal{P}_\text{NU}$ with outcomes $\{0,0\}$ repeatedly on an input qubit prepared in the $\ket{+}$ state and achieve a gradual projection on the ground state $\ket{0}$ of $H_z$. The measurement angle $\theta$ acts as a knob to control the convergence speed --- the closer the measurement axis is to the $Z$ axis, the faster the input state evolves towards the ground state of the Hamiltonian. 

While in Fig.~\ref{fig:ite_var_theta} we show simulation results from repeated applications of $M_{s_1, s_2}$ while selecting the $\{0,0\}$ outcome, when using a QPU in the real world we need to deal with the fact that the gate is probabilistic. Furthermore, unlike in unitary MBQC, undesired random outcomes of non-unitary gates cannot be corrected with unitary Pauli operators. Therefore the authors in \cite{ARS24:nunit} also introduce a probability amplification protocol to enhance the total likelihood of implementing the gate $M_{0,0}$, whereby an undesired outcome is compensated by updating the measurement angle of the subsequent pattern:
\begin{equation} \label{eq:nunit_theta}
    \theta' = \tan^{-1}\left(\frac{-1}{a^2}\right).
\end{equation}
Note that the definition of the gate $M_{s_1, s_2}$ in Eq.~\eqref{eq:nunit_gate} implies that an $s_2=1$ outcome can always be corrected by applying an X gate --- as a result, we follow \cite{ARS24:nunit} and ignore this case. The angle correction described in Eq.~\eqref{eq:nunit_theta} is thus applied only on an $s_1=1$ outcome. Given an input state $\ket{\psi} = \cos{\frac{\beta}{2}}\ket{0} + e^{i\varphi}\sin{\frac{\beta}{2}}\ket{1}$, repeating this process eventually yields a correct implementation of $M_{0,0}$ with a probability approaching
\begin{equation} \label{eq:nunit_pmax}
    p_{\text{max}}(\ket{\psi},a) = \cos^2\left(\frac{\beta}{2}\right) + a^{-2}\sin^2\left(\frac{\beta}{2}\right).
\end{equation}

\begin{figure}
    \centering
    \includegraphics[width=1.0\linewidth]{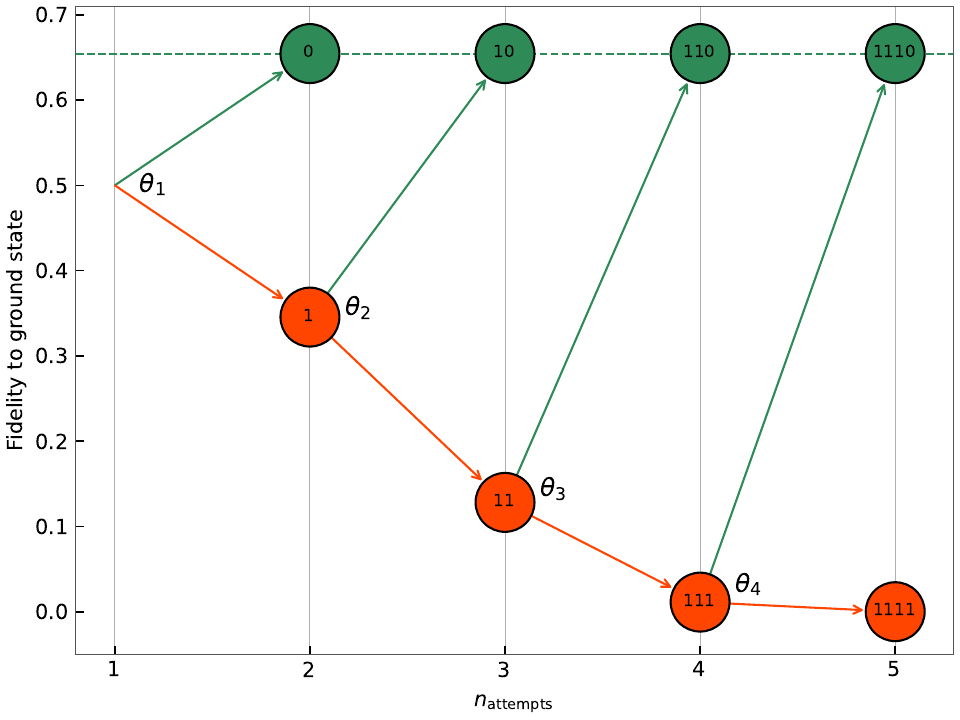}
    \caption{Fidelity to the ground state $\ket{0}$ after each attempt to apply gate $M_{0,0}$ in a probability amplification round with $\theta=0.4\pi$ and input state $\ket{+}$. The unsuccessful outcomes are marked in orange and the successful ones in green. The angles $\theta_n$ denote the corrected angles computed for each attempt using Eq.~\eqref{eq:nunit_theta}, with $\theta_1=\theta$. The green dashed line depicts the new fidelity to the ground state, $\abs{\braket{0}{\text{out}}}^2$, after successfully applying $M_{0,0}$.}
    \label{fig:ite_p_amp}
\end{figure}

In Fig.~\ref{fig:ite_p_amp}, we simulate one probability amplification round in Graphix with 4 attempts to apply the gate $M_{0,0}$. Instead of using an ad-hoc \pythoninline{FixedBranchSelector} to fix the measurement outcomes, this time we perform a realistic simulation with random branch selection and re-apply $\mathcal{P}_\text{NU}$ with the updated angle until we get $s_1=0$. The round ends if we obtain $s_1=0$ after any attempt, as this means we have applied the desired gate. Fig.~\ref{fig:ite_p_amp} verifies this since the state evolves toward the ground state by exactly the same increase in fidelity in each of these branches. However, we observe that in case of successive $s_1=1$ outcomes, the state evolves quickly towards the $\ket{1}$ state. In the above example, the fidelity to the ground state drops to the order of $10^{-5}$ after only four unsuccessful attempts. 

After consecutive application of Eq.~\eqref{eq:nunit_theta}, the value of $\theta$ during a probability amplification round rapidly approaches $0$. In fact for the example in Fig.~\ref{fig:ite_p_amp}, the corrected measurement angle is at the order of $10^{-5}$ after five attempts. This means that the planar measurement of node 0 approaches the Z axis. At the same time, since there is a chance that the state evolves rapidly towards $\ket{1}$, in this case the probability of obtaining a $0$ outcome from this measurement and thus implementing $M_{0,0}$ vanishes within a few attempts.

Successful probability amplification correctly implements the non-unitary gate $M_{0,0}$. If we therefore implement it consecutively and succeed each time, we obtain the same imaginary time evolution curves as in Fig.~\ref{fig:ite_var_theta}. However, Eq.~\eqref{eq:nunit_pmax} reflects the fact that each implementation fails with a finite probability that depends on the initial state $\ket{\psi}$ and the angle parametrizing the gate. An imaginary time evolution implementation with $n$ applications of $M_{0,0}$ using probability amplification will therefore succeed with the overall probability
\begin{equation} \label{eq:nunit_psuccess}
    p_\text{success} = \prod_{i=1}^n p_\text{max}(\ket{\psi_i},a),
\end{equation}
where $\ket{\psi_i}$ is the state preceding the $i$-th application of $M_{0,0}$.

\begin{figure}
    \centering
    \includegraphics[width=1.0\linewidth]{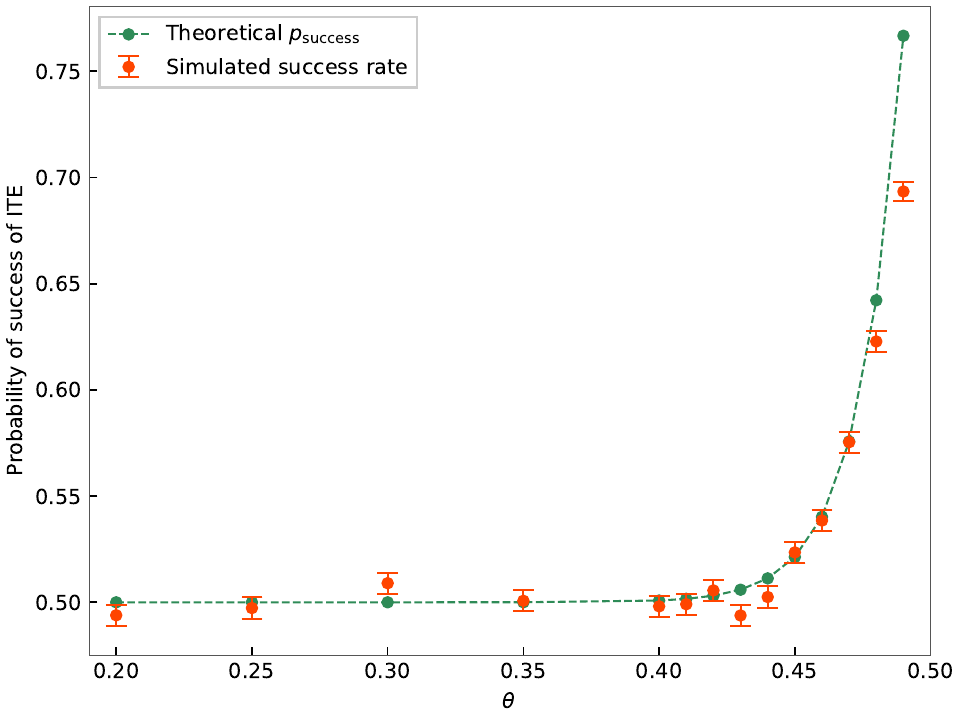}
    \caption{The overall probability of successfully implementing imaginary time evolution $e^{-\tau H_z}$ with probability amplification as a function of $\theta$ in units of $\pi$, using input state $\ket{+}$ and $n=10$. The theoretical curve is computed using Eqs.~\eqref{eq:nunit_pmax}--\eqref{eq:nunit_psuccess}, and the empirical curve shows the mean success rate with standard error computed from 10000 simulations of time evolution in Graphix.}
    \label{fig:ite_psuccess}
\end{figure}

Fig.~\ref{fig:ite_psuccess} depicts this probability as a function of $\theta$, comparing the analytical values from Eq.~\eqref{eq:nunit_psuccess} with the observed mean success rates after simulating time evolution with probability amplification in Graphix. We find that $p_\text{success}$ is exactly $1/2$ for $\theta < 0.4\pi$, and then rises very fast towards unity as $\theta \rightarrow 0.5\pi$. In conjunction with Fig.~\ref{fig:ite_var_theta}, this result illustrates the trade-off between evolution speed and probability of success in the selection of the measurement angle $\theta$. While the evolution towards the ground state is much slower for high $\theta$, the probability-amplification protocol required in a realistic MBQC experiment is much more likely to succeed. On the other hand, while the evolution gets faster with decreasing $\theta$, the procedure only succeeds half of the time. This constant success probability for $\theta < 0.4\pi$ means that reducing $\theta$ below this mark yields faster evolution without any loss in success probability. However, from a realistic perspective, the resolution of the angle values must be factored in and will most likely be the critical limitation.


\section{Conclusion and future work}
\label{sec:conclusion}
Built upon rigorous conceptual foundations and a well-structured software architecture, the current version of Graphix provides a comprehensive toolbox to reason within the MBQC model and express MBQC-native quantum algorithms. It implements a broad range of state-of-the-art algorithms to manipulate MBQC objects, and compile, optimize and simulate measurement patterns. The accompanying array of plugins to interface with, for instance, \stim, \perceval or the \qasm grammar yields a modular framework that is easily extensible and well-integrated with the quantum software ecosystem. The list of existing plugins is presented in Appendix~\ref{app:plugins}.

New software projects have built on Graphix, demonstrating the modularity and extensibility of the framework, as well as its potential to unlock new applications, the prime example being the verification of secure quantum computation in the MBQC model with the Veriphix project \cite{LMKO21,veriphix2024,veriphix_paper2026}. On this basis, further extensions could be envisioned
such as fusion-based quantum computation (FBQC) \cite{B+23:FBQC} to target quantum-optical graph state generation which has already been partially addressed by Optyx, a ZX-calculus-based Python library providing an interface with Graphix \cite{dF+24:fusionflow,K+25:optyx}. Other natural extensions could cover different qubit measurement-based models \cite{Kissinger2019:parityphaseMBQC,RYA23:MBQCString} as well as the qudit \cite{BKMM+23:quditMBQC,RD26} and continuous-variable MBQC \cite{Menicucci06:CVMBQC,GWMRvL_09,Menicucci14:ftMBQC,BD2023:CVflow} models. Furthermore, Graphix provides a robust software platform for developing and implementing new approaches to quantum error correction and fault tolerance, building on the Raussendorf--Harrington--Goyal proposal \cite{RHG06, RHG07} and more general foliated codes \cite{NB18:mb_fault_tolerance,PATV25:fault_tolerance}, while leveraging the hybrid quantum--classical nature of MBQC.

An important direction for future work is the integration of Graphix with technology-specific software frameworks and quantum hardware as reliable mid-circuit measurements and feed-forward become available, to better understand the associated software stack requirements.  Another key objective is the formal verification of Graphix \cite{RPZ18, HRHW21, ZBSLY23, LCR22}. A certified reference implementation of MBQC would provide a foundation for the formal verification of measurement-based algorithms, cryptographic protocols, and their implementations, while contributing to the broader development of a formally verified quantum software stack.

We anticipate incorporating new algorithms and concepts as the field evolves to provide increasingly comprehensive support for the MBQC framework. In the short term, development will focus on improving pattern optimization, compilation and parallelization strategies, as well as developing efficient backends for high-throughput simulations in variational workflows.

We believe that the library is now well positioned to support research on the fundamental principles and practical applications of MBQC, as well as the development of compilation workflows for quantum hardware. In addition to serving as a research platform, the library provides a pedagogical resource for students and practitioners learning about MBQC. As an active open-source project, Graphix welcomes contributions from the quantum information community to improve its usability and performance, and extend its capabilities.


\section{Author contributions}
All authors contributed to writing and testing Graphix code and participated in regular, collaborative code reviews. All authors participated in writing and reviewing the manuscript. LLMs were used to generate TikZ code for some of the figures as well as for improving the structure, grammar, and clarity of presentation in some sections. Some contributions to the Graphix library may contain code partially generated by LLMs but, as all modifications to the codebase, they were integrated only after systematic reviews and approvals by at least 2 members of the maintenance team.

\section{Acknowledgments}
We thank all past and present contributors to Graphix. Specifically Luka Music for the collaboration on the \perceval backend, William Cashman for the \pyzx interface, as well as Masato Fukushima, Daichi Sasaki, Sora Shiratani and Yuki Watanabe for starting the project and their initial work. We thank all participants in the 2024 and 2025 Graphix workshops held in Paris, as well as the Unitary Foundation and the DIM QuanTiP programme of the Région Île-de-France for funding these events. S.S. thanks Fixstars Amplify and the Unitary Foundation for their support.

We thank Ulysse Chabaud for suggesting the reference about efficient measurement-based pseudo-randomness \cite{MGDM18:pseudo}, useful discussions throughout the project as well as for critical reading of early versions of this manuscript. We also thank Harold Ollivier for guidance all along the project and for critical reading of early versions of this manuscript. M.U. warmly thanks Rajarsi Pal for the multiple discussions and his initial implementation of the circuit extraction algorithm.

This work has been co-funded by the Hybrid Quantum Initiative (HQI) supported by France 2030 under ANR grant ANR-22-PNCQ-0002 and the European Union's Horizon Europe research and innovation programme under grant agreement No. 101102140 – QIA Phase 1.

\onecolumn
\bibliographystyle{linksen}
\bibliography{graphixbib}

\newpage
\appendix

\section{Detailed contributions and history of the project}
\label{app:history}
This paper highlights the authors' contributions to the Graphix library. Throughout this Appendix, the term \emph{the authors} systematically refers to the authors of the present paper. The development of the library was initiated by Shinichi Sunami \cite{SF22:graphix} and its overall structure is greatly owed to Sunami and collaborators. Detailed contributions can be verified on the Graphix repository \cite{graphix_repo}. Here we comment on the improvements that were done on \emph{existing} functionality in the code base on June 2023. Completely new additions are discussed in the main text.

When  the authors started contributing, Graphix contained a preliminary version of the transpiler, using hard-coded command sequences for each gate, and some of these sequences did not admit flow. The authors updated all of these sequences to ensure that they admit causal flow and completely rewrote the transpiler to align its design more closely with the literature, using the J+CZ gate set as an intermediate representation.

The implementations of generalized-flow and Pauli-flow detection originally followed the algorithms published in the literature, with the time complexities of \(\order{|V|^4}\) and \(\order{|V|^5}\), respectively. The authors replaced them with the best known algorithm, which runs in \(\order{|V|^3}\) \cite{MB24:algebraic}. In addition, the implementations of the flow-finding algorithms (causal flow, generalized flow, and Pauli flow) previously relied on a recursive tail-call formulation. Since Python does not optimize tail recursion, this approach caused stack overflows on large graphs. The authors replaced it with an iterative implementation. They also completely redesigned the flow API so that flows and corrections became first-class objects that can be converted to and from open graphs and patterns.

Before the implementation of Pauli node removal optimization, the library contained a graph-state simulator relying on the equivalence between stabilizer states and local-Clifford-decorated graph states \cite{vdNDdM04}. More precisely, the implementation followed the graphical framework of \cite{Elliot08:graphicalclifford,Elliot09:graphicalmeas}. Initially, the simulator only supported measurements in the XY plane; this limitation was removed by the authors. The simulator also lacked Pauli pushing and therefore stopped as soon as it encountered the first non-Pauli measurement. The authors implemented the Pauli-pushing phase. Finally, the simulator only worked when all input nodes were prepared in the state \(\ket{+}\). This limitation ultimately motivated its replacement by the Pauli node removal optimization.

A preliminary version of the standardization algorithm existed prior to the authors' work. The initial implementation repeatedly applied incomplete rewriting rules to commute commands one by one, resulting in an algorithm running in \(\order{|V|^3}\). Moreover, this implemetation was limited to MBQC, without supporting local Clifford. The standardization algorithm presented in this paper was designed by the authors and runs in linear time.

A preliminary version of the space minimization algorithm existed prior to the authors' work. It used the causal flow of the underlying open graph instead of extracting the causal flow of the pattern itself, which could break the runnability of the pattern whenever the two differed. Furthermore, the extraction of a space-optimal pattern was tightly coupled to the space-minimization heuristic and could not be applied to a user-specified measurement order. The latter is now possible. When the heuristic failed, it may return a pattern with higher space instead of the original pattern. This behaviour was corrected by the authors.

The initial library contained a functional statevector simulator, a partially implemented density matrix backend as well as a prototype of a tensor network backend. The authors polished the statevector backend, corrected and completed the density matrix backend, as well as designed and implemented the noise model interface. The tensor network backend has been left untouched except for mild non-regression modifications and is left for future work.

The visualization algorithm was introduced in the original version of the library. The authors introduced functionality to enable finer control on the visualized objects.

From a software engineering perspective, an initial effort had been made to introduce static type checking with \mypy and \pyright. However, this effort was limited to secondary modules, which represented roughly one quarter of the code base, including some modules written by the authors. The type annotation and static type checking of the core modules (including the pattern module, the transpiler, the simulator and the backends, and the visualization module) were carried out by the authors. The authors introduced the type parametrization of classes representing MBQC concepts discussed in Subsection \ref{subsec:architecture}.
Additionally, the authors took care of modularizing the library and in particular the backends.

\section{List of plugins}
\label{app:plugins}
\begin{table}[ht]
\centering
\begin{tblr}{
    colspec = {Q[l,wd=0.25\textwidth] Q[l,wd=0.7\textwidth]},
    row{1} = {font=\bfseries},
}
\toprule
Plugin & Description \\
\midrule
{\ttfamily graphix-stim-backend}
& An efficient simulator of Clifford patterns using the \stim library. \\

{\ttfamily graphix-stim-compiler}
& A Clifford compilation pass using \stim functionalities that allows the round-trip conversion
\pythoninline{Circuit} $\rightarrow$ \pythoninline{Pattern} $\rightarrow$ \pythoninline{Circuit}. \\

{\ttfamily graphix-qasm-parser}
& A parser from \qasm circuit specifications into Graphix \pythoninline{Circuit} objects, which can then be transpiled into patterns. There is ongoing work to provide support for the full \qasm specification. \\

{\ttfamily graphix-pyzx}
& A plugin for importing and exporting ZX-diagrams using the \pyzx library. Graphix supports circuit extraction and compilation natively, but this plugin allows targeting Clifford-specialized compilation routines. \\

{\ttfamily graphix-brickworks-\\transpiler}
& A circuit-to-pattern transpiler that yields open graphs with a brickwork structure. \\

{\ttfamily graphix-mqtbench}
& An interface between Graphix and the MQT Bench
quantum circuit benchmarking suite in the Munich Quantum Toolkit. \\

{\ttfamily graphix-symbolic}
& A symbolic pattern simulator for patterns with non-numerical measurement angles using the \sympy library. \\

{\ttfamily graphix-perceval-\\backend}
& An interface to run a Graphix \pythoninline{Pattern} on Quandela's \perceval simulator backends. \\

{\ttfamily graphix-ibmq}
& An interface with IBM's \qiskit framework. \\
\bottomrule
\end{tblr}
\caption{Graphix plugins and their functionalities. All plugins are available at the TeamGraphix GitHub repository \cite{graphix_repo}}
\label{tab:graphix-plugins}
\end{table}

\section{Standardization}
\label{app:standardization}
By applying commutation rules, patterns can be rewritten such that commands appear in the following order while preserving their semantics:
\begin{itemize}
\item First, all the preparation commands \(\N_i\);
\item Then, all the entanglement commands \(\E_{ij}\),
and we may assume the same edge \((i,j)\) is never entangled twice;
\item Then, all the measurement commands \(\Mdom{_i^{\lambda, \alpha}}{s}{t}\), with their correction domains;
\item And finally, all the by-products. In Graphix, the standardization procedure puts them in this order: first all the \(\Z_i^s\) commands, then all the \(\X_i^s\) commands,
and finally all the \(\C_i^c\) commands.
We may assume that for every node \(i\), there is at most one \(\Z_i^s\), one \(\X_i^s\), and one \(\C_i^c\) command, that the domain of the signal \(s\) is not empty, and that the Clifford \(c\) is not \(I\).
\end{itemize}

A standardized pattern first constructs the entire graph state before measuring the nodes.
The order of the preparation commands and the entanglement commands is arbitrary since they commute.
All the \(\X\) and \(\Z\) corrections and all the Clifford commands applied to measured nodes are absorbed by the measurement commands.

The commutation rules are complete for the MBQC fragment (i.e., for patterns without Clifford commands).
The only non-trivial commutation rule for this fragment is: \(\E_{ij} \X_i^s \equiv \X_i^s \Z_j^s \E_{ij}\),
meaning that applying an \(X\) correction to a node applies a \(Z\) correction to its subsequent neighbors.

In the more general context of MBQC+LC (i.e., patterns with Clifford commands), there exist patterns without standard form since some single-qubit Clifford gates do not commute with \(\CZ\).
Indeed, let \(c\) be a single-qubit Clifford gate.
The pattern \(\E_{01}\C_0^c\) has a standard form if and only if \(\CZ \cdot (c \otimes \Id) = (c' \otimes d') \cdot \CZ^\epsilon\) for some single-qubit Clifford gates \(c'\) and \(d'\), and \(\epsilon \in \{0, 1\}\).
The case \(\epsilon = 0\) is impossible since multiplying both sides on the right by \(c^{-1} \otimes \Id\)
gives \(\CZ = (c' \cdot c^{-1}) \otimes d'\), expressing \(\CZ\) as a tensor product of single-qubit unitaries.
Therefore, \(\epsilon = 1\), and multiplying both sides on the right by \(\CZ\) yields the equivalent condition
\(\CZ \cdot (c \otimes \Id) \cdot \CZ = c' \otimes d'\), that is, the conjugation \(W = \CZ \cdot (c \otimes \Id) \cdot \CZ\) must be local.
We now show that this holds if and only if \(c Z c^\dag = \pm Z\).

Since \(Z\) commutes with \(\CZ\), we have \(W \cdot (Z \otimes \Id) \cdot W^\dag = \CZ \cdot (c Z c^\dag \otimes \Id) \cdot \CZ\).
Since \(c\) is Clifford, \(c Z c^\dag = \pm P\) for some Pauli \(P\). By case analysis on \(P\),
\[\CZ \cdot (P \otimes \Id) \cdot \CZ =
\begin{cases}
X \otimes Z & \text{if \(P = X\)}\\
Y \otimes Z & \text{if \(P = Y\)}\\
Z \otimes I & \text{if \(P = Z\)}
\end{cases}
\]
If \(W = c' \otimes d'\), then \(W \cdot (Z \otimes \Id) \cdot W^\dag = (c' Z {c'}^\dag) \otimes \Id\). Therefore, we must have \(P = Z\) and hence \(c Z c^\dag = \pm Z\).

Reciprocally, if \(c Z c^\dag = \pm Z\), then we consider the two possible signs.
If \(c Z c^\dag = Z\), then \(c \in \{\Id, Z, \Sa, \Sa^\dag\} = \mathcal D\).
Since both \(c \in \mathcal D\) and \(\CZ\) are diagonal in the computational basis, they commute: 
\(\CZ \cdot (c \otimes \Id) \cdot \CZ = c \otimes \Id\), therefore \(\E_{01}\C_0^c \equiv \C_0^c \E_{01}\).
If instead \(c Z c^\dag = - Z\), then \(c \in \{X, Y, X \Sa, X \Sa^\dag\} = X \mathcal D\).
Since \(c\) flips the computational basis, the conditional phase from each \(\CZ\) application acts on complementary basis states, producing a \(\Z\) correction on qubit \(1\), \ie,
\(\CZ \cdot (c \otimes \Id) \cdot \CZ = c \otimes Z\), therefore \(\E_{01}\C_0^c \equiv \C_0^c \C_1^{Z} \E_{01}\).

We have shown that \(\E_{01}\C_0^c\) has a standard form if and only if \(c\) is one of the 8 single-qubit Clifford gates
in the set \(\mathcal D \cup X \mathcal D\). For the other 16 Clifford gates such that \(c Z c^\dag \neq \pm Z\),
there is no standard form for \(\E_{01}\C_0^c\).

Clifford commands on measured nodes can be absorbed into the measurements. Indeed, we have
\begin{itemize}
\item \(\Mdom{_i^{\lambda,\alpha}}{s}{t} \C_i^{\Ha} \equiv \Mdom{_i^{\lambda',\alpha'}}{t}{s}\),
with
\[
(\lambda', \alpha') = \begin{cases}
  (\planeYZ, -\alpha) & \text{if \(\lambda = \planeXY\)}\\
  (\planeXY, -\alpha) & \text{if \(\lambda = \planeYZ\)}\\
  (\planeXZ, \frac \pi 2 - \alpha) & \text{if \(\lambda = \planeXZ\)}
\end{cases}
\]
\item \(\Mdom{_i^{\lambda,\alpha}}{s}{t} \C_i^{\Sa} \equiv \Mdom{_i^{\lambda',\alpha'}}{s}{s \bigtriangleup t}\),
with
\[
(\lambda', \alpha') = \begin{cases}
  (\planeXY, \alpha + \frac {3\pi} 2) & \text{if \(\lambda = \planeXY\)}\\
  (\planeXZ, \alpha) & \text{if \(\lambda = \planeYZ\)}\\
  (\planeYZ, - \alpha) & \text{if \(\lambda = \planeXZ\)}
\end{cases}
\]
and where $s \bigtriangleup t := \bigoplus_{j\in D_X \bigtriangleup D_Z} t_j$, given $s := \bigoplus_{j\in D_X} s_j$ and $t := \bigoplus_{j\in D_Z} t_j$, with $D_X$ and $D_Z$ the signal domains.
\item \(\Mdom{_i^{\lambda,\alpha}}{s}{t} \C_i^{Z} \equiv \Mdom{_i^{\lambda,\alpha'}}{s}{t}\),
with
\[
\alpha' = \begin{cases}
  \alpha + \frac \pi 2 & \text{if \(\lambda = \planeXY\)}\\
  - \alpha & \text{if \(\lambda \in \{\planeYZ, \planeXZ\}\)}
\end{cases}
\]
\end{itemize}
Absorption rules for other Clifford gates follow from the HSZ decomposition (see Appendix~\ref{app:clifford_gates} for a list of such decompositions).

It remains to show how Clifford commands commute with by-products. Still relying on the HSZ decomposition, we have
\begin{itemize}
\item \(\X_i^s \Z_i^t \C_i^{\Ha} \equiv \C_i^{\Ha} \X_i^t \Z_i^s\),
\item \(\X_i^s \Z_i^t \C_i^{\Sa} \equiv \C_i^{\Sa} \X_i^s \Z_i^{s \bigtriangleup t}\),
\item \(\X_i^s \Z_i^t \C_i^{Z} \equiv \C_i^{Z} \X_i^s \Z_i^t\),
\end{itemize}

In Graphix, the \pythoninline{StandardizedPattern} class represents a pattern in standardized form.
This class provides direct access to partitioned lists of commands categorized by type.
An instance of \pythoninline{StandardizedPattern} can be created from a \pythoninline{Pattern} via the \pythoninline{StandardizedPattern.from_pattern} class method.
Because other optimization passes and flow-analysis algorithms operate directly on \pythoninline{StandardizedPattern} instances, it is often more efficient to convert patterns into this form in advance.

This method implements the following algorithm through a single pass over the pattern.
The algorithm initializes the following data structures:
\begin{itemize}
\item A list \(\mathcal N\) of commands \(\N\), initially empty;
\item A set \(\mathcal E\) of entanglement commands, initially empty;
\item A list \(\mathcal M\) of commands \(\M\), initially empty;
\item Maps \(\mathcal X\) and \(\mathcal Z\) from nodes to their correction domains, initially empty;
\item A mapping \(\mathcal C\) from every node to a Clifford operation, initially \(I\).
\end{itemize}
Then, for every command in the pattern, taken in execution order (right to left):
\begin{itemize}
\item Preparation: \(\mathcal N \leftarrow \N_i \cdot \mathcal N\), \ie, \(\N_i\) is added to the list \(\mathcal N\). The last command is written on the left to be consistent with the usual pattern notation. However, in practice, the command \(\N_i\) is simply added at the end of the list.
\item Entanglement \(\E_{ij}\):
\begin{itemize}
\item \(\E_{ij}\) is incorporated into the entanglement set by symmetric difference, removing command if it is already there:
\(\mathcal E \leftarrow \mathcal E \bigtriangleup \{\E_{ij}\} \).
\item The \(\X\) corrections \(\mathcal X(i)\) accumulated for \(i\), resp. \(j\), are propagated to the \(Z\) corrections \(\mathcal Z(j)\) applied to \(j\) (resp. \(i\)):
\(\mathcal Z(j) \leftarrow \mathcal Z(j) \bigtriangleup \mathcal X(i)\) and
\(\mathcal Z(i) \leftarrow \mathcal Z(i) \bigtriangleup \mathcal X(j)\).
\item If \(\mathcal C(i) \in \X \mathcal D\), then \(\mathcal C(j) \leftarrow Z \cdot \mathcal C(j)\).
\item If \(\mathcal C(j) \in \X \mathcal D\), then \(\mathcal C(i) \leftarrow Z \cdot \mathcal C(i)\).
\item If \(\mathcal C(i), \mathcal C(j) \notin \mathcal D \cup \X \mathcal D\), then reject the pattern: there is no standardized form without ancilla.
\end{itemize}
\item Measurement \(\Mdom{_i^{\lambda,\alpha}}{s}{t}\):
\begin{itemize}
\item Using the absorption rules described above for the Clifford commands, compute \(s',t',\lambda',\alpha'\) such that
\[\Mdom{_i^{\lambda,\alpha}}{s}{t} \cdot \C_i^{\mathcal C(i)} \equiv \Mdom{_i^{\lambda',\alpha'}}{s'}{t'}\]
\item \(\mathcal M \leftarrow \Mdom{_i^{\lambda',\alpha'}}{s' \bigtriangleup \mathcal X(i)}{t' \bigtriangleup \mathcal Z(i)} \cdot \mathcal M\).
\end{itemize}
\item By-products \(\X_i^s\) or \(\Z_i^t\):
\begin{itemize}
\item Applying the commutation rules given above between Clifford commands and by-products, compute \(s'\) and \(t'\) such that
\[\X_i^s\Z_i^t\C_i^{\mathcal C(i)} \equiv \C_i^{\mathcal C(i)}\X_i^{s'}\Z_i^{t'}\]
(with \(t = \emptyset\) for an \(\X\) command and \(s = \emptyset\) for a \(\Z\) command),
\item \(\mathcal X(i) \leftarrow \mathcal X(i) \bigtriangleup s'\) and \(\mathcal Z(i) \leftarrow \mathcal Z(i) \bigtriangleup t'\).
\end{itemize}
\item Clifford \(\C_i^c\): \(\mathcal C(i) \leftarrow c \cdot \mathcal C(i)\).
\end{itemize}
After processing all commands, the resulting pattern \(
  \left(\prod_{i \in \mathrm{dom}(\mathcal C)} \C_i^{\mathcal C(i)}\right)
  \cdot
  \left(\prod_{i \in \mathrm{dom}(\mathcal X)} \X_i^{\mathcal X(i)}\right)
  \cdot
  \left(\prod_{i \in \mathrm{dom}(\mathcal Z)} \Z_i^{\mathcal Z(i)}\right)
  \cdot
  \mathcal M
  \cdot
  \mathcal E
  \cdot
  \mathcal N
  \)
is in standardized form and remains equivalent to the original pattern, provided the original pattern was runnable.
This computation takes linear time relative to the length of the input pattern.
The runnability of the original pattern must be verified beforehand: the standardization process does not detect non-runnable patterns and may inadvertently transform them into runnable ones.

The \pythoninline{to_pattern} method transforms an instance of \pythoninline{StandardizedPattern} into an instance of \pythoninline{Pattern} with standardized command ordering.
For convenience, \pythoninline{Pattern.standardize} performs the round-trip conversion from \pythoninline{Pattern} to \pythoninline{StandardizedPattern} and back.

\section{Flow definitions}
\label{app:flow_definitions}
In this Appendix we give the formal definitions of the various kind of flows for the sake of completeness.

\begin{definition}[Causal flow \cite{DK06:determinism}]
An open graph $(G, I, O, \lambda: O^c \rightarrow  {\planeXY})$ has causal flow if there exists a map $c: O^c \rightarrow I^c$ and a
strict partial order $\prec$ over $V$ such that for all $i \in O^c$:

\begin{itemize}
    \item (C1) Nodes $i$ and $c(i)$ are neighbors,
    \item (C2) $i \prec c(i)$,
    \item (C3) $i \prec j$ for all neighbors $j \neq i$ of node $c(i)$.
\end{itemize}
\label{def:causal_flow}
\end{definition}

Causal flow is a sufficient but not necessary condition for robust determinism. In other words,
there exist open graphs without causal flow which support patterns realizing unitary transformations.

\begin{definition}[Generalized flow \cite{BKMP07:gflow,BMBdF+21}]\label{def:gflow}
An open graph \((G, I, O, \lambda : O^c \rightarrow \{\planeXY, \planeXZ, \planeYZ\})\) has generalized flow (\textit{gflow}) if there exists a map $g: O^c \rightarrow \mathcal P(I^c)$ from measured qubits to a subset of prepared qubits, and a strict partial order $\prec$ over $V$, such that for all $i \in O^c$:

\begin{itemize}
    \item (G1) If $j \in g(i)$ and $i \neq j$, then $i \prec j$.
    \item (G2) If $j \in \odd(g(i))$ and $i \neq j$, then $i \prec j$.
    \item (G3) If $\lambda(i) = \planeXY$, then $i \notin g(i)$ and $i \in \odd(g(i))$.
    \item (G4) If $\lambda(i) = \planeXZ$, then $i \in g(i)$ and $i \in \odd(g(i))$.
    \item (G5) If $\lambda(i) = \planeYZ$, then $i \in g(i)$ and $i \notin \odd(g(i))$.
\end{itemize}
Here $\odd(S)$ is the \emph{odd neighborhood} of $S$ in the graph $G$, i.e., the set of vertices having an odd number of neighbors with the elements in $S$.
\end{definition}

Gflow is a sufficient and necessary condition for robust determinism on open graphs \emph{without Pauli measurements}. Every causal flow is a gflow, with $g(i) = \{f(i)\}$ and the same partial order.

\begin{remark}
    Since (G4) and (G5) impose that $i\in g(i)$ and $g: O^c \rightarrow \mathcal{P}(I^c)$, inputs of open graphs with gflow must be measured in the $\planeXY$ plane (or not be measured at all, in which case they are outputs as well).
    \label{rmk:input_nodes_flow}
\end{remark}

\begin{definition}[Pauli flow \cite{BKMP07:gflow}]\label{def:pauli_flow}
An open graph \((G, I, O, \lambda : O^c \rightarrow \{\planeXY, \planeXZ, \planeYZ, \axisX, \axisY, \axisZ\})\) has Pauli flow if there exists a map
$p: O^c \rightarrow \mathcal{P}(I^c)$ and a strict partial order $\prec$ over $V$ such that for all $i \in O^c$:
\begin{itemize}
    \item (P1) If $j \in p(i)$, and $i \neq j$, and $\lambda(j) \notin \{ \axisX, \axisY\}$, then $i \prec j$.
    \item (P2) If $j \in \odd(p(i))$, and $i \neq j$, and $\lambda(j) \notin \{\axisY, \axisZ\}$, then $i \prec j$.
    \item (P3) If $\neg (i \prec j)$, and $i \neq j$, and $\lambda(j) = \axisY$, then $j \in p(i) \; \mathrm{iff} \; j \in
    \odd(p(i))$.
    \item (P4) If $\lambda(i) = \planeXY$, then $i \notin p(i)$ and $i \in \odd(p(i))$.
    \item (P5) If $\lambda(i) = \planeXZ$, then $i \in p(i)$ and $i \in \odd(p(i))$.
    \item (P6) If $\lambda(i) = \planeYZ$, then $i \in p(i)$ and $i \notin \odd(p(i))$.
    \item (P7) If $\lambda(i) = {\axisX}$, then $i \in \odd(p(i))$.
    \item (P8) If $\lambda(i) = {\axisZ}$, then $i \in p(i)$.
    \item (P9) If $\lambda(i) = {\axisY}$, then, either:
        \begin{itemize}
            \item $i \notin p(i)$ and $i \in \odd(p(i))$, or
            \item $i \in p(i)$ and $i \notin \odd(p(i))$.
        \end{itemize}
\end{itemize}
\end{definition}

Pauli flow is a sufficient and necessary condition for robust determinism on open graphs. Furthermore, note that the definition of Pauli flow just adds (P3) and (P7-P9) to the definition of gflow, therefore, a gflow is a Pauli flow.

\section{Flow extraction from an $\mathbf{xz}$-correction strategy}
\label{app:flow_from_xz}
In this Appendix we detail how Graphix implements the flow extraction routines from an $\mathbf{xz}$-correction strategy, without recurring to the flow-finding algorithms on the underlying open graph. Our starting point is an \pythoninline{XZCorrections} object bundling:
\begin{itemize}
    \item The \pythoninline{OpenGraph} object on which the $\mathbf{xz}$-corrections are defined.
    \item Two dictionaries $\mathbf{x}$, $\mathbf{z}$ mapping every measured node $i$ to the set of nodes on which an $\X$ correction (respectively, $\Z$) must be applied when the $s_i = 1$ outcome occurs.
    \item A partition of the open graph’s nodes in sequence of sets (denoted \emph{layers}) representing a partial measurement order. By convention, if the open graph has any output nodes, they are in layer 0. For all other layers, if $i < j$, then nodes in layer $j$ are measured before those in layer $i$.
\end{itemize}

We assume that the \pythoninline{XZCorrections} instance is well formed. That is, it defines a runnable computation and its fields are consistent with each other.

Our goal is to reconstruct a \pythoninline{CausalFlow}, \pythoninline{GFlow} and \pythoninline{PauliFlow} instances. We recall that all of them comprise an \pythoninline{OpenGraph} object, a dictionary representing the correction function, and a partition of the open graph’s nodes in sequence of sets representing the partial order layering. The open graph and the partial order layering from the \pythoninline{XZCorrections} instance are directly assigned to the corresponding fields of the new flow object. In the following subsections, we explain how to reconstruct the correction function in each case.

\subsection{Causal flow}

We recall Eq.~\eqref{eq:flow_to_corrections_causal}:
\begin{align}
    \mathbf x_{\text c}(i) &= c(i),\\
    \mathbf z_{\text c}(i) &= N_G(c(i)) \setminus \{i\}.
\end{align}
We can trivially assign the map $\mathbf x_{\text c}$ to the flow's correction function. Then, we iterate over the open graph's nodes following the partial order and we assert that conditions (C1-C3) in Definition \ref{def:causal_flow} are fulfilled. We also assert that nodes are measured on the $\planeXY$ plane. If any assertion fails, we conclude that the given $\mathbf{xz}$-correction strategy is not compatible with a causal flow.

\subsection{Generalised flow}
We recall Eq.~\eqref{eq:flow_to_corrections_g}:
\begin{align}
    \mathbf x_{\text g}(i) &= g(i)\setminus \{i\},\\
    \mathbf z_{\text g}(i) &= \operatorname{Odd}(g(i)) \setminus \{i\}
\end{align}
In this case, a measured node $i$ may belong to $g(i)$, yet not appear in the $\mathbf x_g$ correction map. Following propositions (G4-G5) in Definition \ref{def:gflow} we add $i$ to $\mathbf x_{\text g}(i)$ if $\lambda(i) = \planeXZ | \planeYZ$, and assign the resulting map to the flow's correction function. Similar to the causal flow routine, we iterate over the open graph's nodes following the partial order and we assert that conditions (G1-G5) in Definition \ref{def:gflow} are fulfilled. If any assertion fails, we conclude that the given $\mathbf{xz}$-correction strategy is not compatible with a gflow. Note that in this case it is not necessary to check that all measurements are \emph{planar} measurements (in opposition to Pauli measurements) since this is guaranteed by the class hierarchy of the measurement objects (see discussion in Subsection \ref{subsec:architecture}). 

\subsection{Pauli flow}
We recall Eq.\eqref{eq:flow_to_corrections_pauli}:
\begin{align}
    \mathbf x_{\text p}(i) &= p(i)\cap \{j | i \prec j\},\\
    \mathbf z_{\text p}(i) &= \operatorname{Odd}(p(i)) \cap \{j | i \prec j\},
\end{align}
Here, the difficulty stems from the observation that anachronistic corrections (i.e., nodes $j \in p(i)$ that are not in the future of $i$, $j \preceq i$) do not appear in the $\mathbf x_p$ correction map. As per (P1) in Definition \ref{def:pauli_flow}, this may concern nodes measured along the $\axisX$ and $\axisY$ axes. We reconstruct the correction function $p$ as follows.

For each measured node $i$:
\begin{itemize}
    \item We set $p(i) := \mathbf x_{\text p} (i)$.
    \item If $\lambda(i) = \planeXZ|\planeYZ|\axisZ$, we update $p(i) = p(i) \cap \{i\}$ as per (P5-P6, P8). Note that if $i$ is also an input node we contradict Remark \ref{rmk:input_nodes_flow}, therefore we abort and conclude that the $\mathbf{xz}$-corrections strategy is not compatible with a Pauli flow.
    \item We define and solve a linear system of equations over $\mathbb F_2$ encoding the constraints in (P1-P9) concerning the odd neighbourhood of $p(i)$, $A \mathbf x = \mathbf v$. Each entry in the unknowns vector $\mathbf x$ is associated with a candidate node $j$ such that $j \in I^c, \lambda(j) = \axisX|\axisY, j \preceq i$. Each row of $A$ is associated to a node $g$ in the open graph, except for nodes measured in $\axisZ$, and it is the indicator vector of which candidate nodes are in the neighbourhood of $g$ (the adjacency vector of $g$ restricted to candidates nodes). The product $A_{g, :} \cdot \mathbf x$ signals whether $g$ belongs to the odd neighbourhood of the candidates set.
    \begin{itemize}
        \item For every future node $g \succ i$ and the output nodes, the corresponding entry in the vector $\mathbf v$ is 1 if $N_{G}(g) \cap p(i)$ has an odd number of elements and $g \notin \mathbf z_{\text{p}}(i)$, or vice versa, and 0 otherwise. If the node is measured along $\axisY$ or $\axisZ$ we do not add an equation. 
        \item For every node $g \preceq i$,  the corresponding entry in the vector $\mathbf v$ is 1 if $N_{G}(g) \cap p(i)$ has an odd number of elements, and 0 otherwise (note that these nodes never appear on $\mathbf z_{\text p}(i)$). If the node $g$ is measured along $\axisY$ and it is a candidate node, we flip the entry $A_{g, g}$ to account for (P3). If the node is measured along $\axisZ$ we do not add an equation.
        \item For the node $i$ itself, if $\lambda(i) = \planeXY|\planeXZ|\axisX|\axisY$ the corresponding entry in the vector $\mathbf v$ is 0 if $N_{G}(g) \cap p(i)$ has an odd number of elements, 1 otherwise, and vice versa if $\lambda(i) = \planeYZ$. If the node $i$ is measured along $\axisY$ and it is a candidate node, we flip the entry $A_{i, i}$ to account for (P9). If the node is measured along $\axisZ$ we do not add an equation.
    \end{itemize}

    \item If the system of equations does not have a solution, conclude that the $\mathbf{xz}$-corrections strategy is not compatible with a Pauli flow. Else, we update $p(i)$ with the nodes $j$ such that $\mathbf x_j = 1$ in the solution and assign it to the flow's correction function.
\end{itemize}


\section{Proof of max-space optimality for patterns with causal flow}
\label{app:max_space_cf}

In this appendix we show that the optimal max-space of a pattern is achieved by any measurement total order that extends
the causal flow partial order, provided that we compute the
corresponding pattern by preparing each qubit immediately before one of its neighbours is measured.

Indeed, consider such a measurement order. Suppose a measurement decreases the number of prepared qubits. Before the next measurement that also decreases the number of prepared qubits, at least one preparation must occur, namely the preparation of \(c(u)\), where \(u\) is the second measured node. Indeed, if the second measurement is performed on \(u\), then its flow successor \(c(u)\) cannot yet have been measured, because \(u<c(u)\) in the flow order. Since \(c(u)\) is adjacent to \(u\), it must already have been prepared before \(u\) is measured. Therefore, the preparation of \(c(u)\) occurs between the two measurements. Since the space never decreases twice in succession, its maximum value is at most one greater than its final value, which is the number of output nodes. Conversely, this bound is optimal. If there are no isolated output nodes, every output node must be prepared before one of its neighbours is measured. In particular, before the last measurement takes place, every output node has already been prepared. Therefore, immediately before the last measurement, the prepared qubits consist of all output qubits together with the qubit being measured. Hence the number of prepared qubits is at least the number of output nodes plus one.

\section{List of Clifford gates}
\label{app:clifford_gates}
The following table enumerates the 24 single-qubit Clifford gates.

The first column describes each gate in the \(\{X,Y,Z,S,S^\dag,H\}\) gate set, which forms a compact basis and is useful, for instance, when exporting to \qasm 3.

The second column describes each gate in the \(\{H,S,Z\}\) gate set, which is used, for instance, in the standardization procedure (see Appendix~\ref{app:standardization}).

The third column gives a pattern implementation, where node 0 is the input and the last node (either 0, 1, 2, or 3) is the output.

All these decompositions are proven to be optimal: there is no shorter decomposition within the corresponding gate set, and no pattern implementation using fewer qubits.
\begin{tabular}{|l|l|l|}
\hline
Clifford gate &
HSZ decomposition &
Pattern implementation
\\\hline
\(I\) &
\(I\) &
\emph{No operation}\\
\(X\) &
\(H Z H\) &
\(X_{2}^{1}\,M_{1}^{-X}\,Z_{2}^{0}\,M_{0}\,E_{1,2}\,E_{0,1}\,N_{2}\,N_{1}\)\\
\(Y\) &
\(H Z H Z\) &
\(X_{2}^{1}\,M_{1}^{-X}\,Z_{2}^{0}\,M_{0}^{-X}\,E_{1,2}\,E_{0,1}\,N_{2}\,N_{1}\)\\
\(Z\) &
\(Z\) &
\(X_{2}^{1}\,M_{1}\,Z_{2}^{0}\,M_{0}^{-X}\,E_{1,2}\,E_{0,1}\,N_{2}\,N_{1}\)\\
\(S\) &
\(S\) &
\(X_{2}^{1}\,M_{1}\,Z_{2}^{0}\,M_{0}^{-Y}\,E_{1,2}\,E_{0,1}\,N_{2}\,N_{1}\)\\
\(S^\dag\) &
\(S Z\) &
\(X_{2}^{1}\,M_{1}\,Z_{2}^{0}\,M_{0}^{+Y}\,E_{1,2}\,E_{0,1}\,N_{2}\,N_{1}\)\\
\(H\) &
\(H\) &
\(X_{1}^{0}\,M_{0}\,E_{0,1}\,N_{1}\)\\
\(S^\dag H S^\dag\) &
\(H S H\) &
\(X_{2}^{1}\,M_{1}^{-Y}\,X_{2}^{0}\,Z_{2}^{0}\,M_{0}\,E_{1,2}\,E_{0,1}\,N_{2}\,N_{1}\)\\
\(H Z\) &
\(H Z\) &
\(X_{1}^{0}\,M_{0}^{-X}\,E_{0,1}\,N_{1}\)\\
\(S^\dag X\) &
\(H Z H S\) &
\(X_{2}^{1}\,M_{1}^{-X}\,Z_{2}^{0}\,M_{0}^{-Y}\,E_{1,2}\,E_{0,1}\,N_{2}\,N_{1}\)\\
\(S X\) &
\(S H Z H\) &
\(X_{2}^{1}\,M_{1}^{-X}\,Z_{2}^{0}\,M_{0}^{+Y}\,E_{1,2}\,E_{0,1}\,N_{2}\,N_{1}\)\\
\(H Y\) &
\(Z H Z\) &
\(X_{3}^{2}\,M_{2}^{-X}\,Z_{3}^{1}\,M_{1}^{-X}\,X_{3}^{0}\,M_{0}\,E_{2,3}\,E_{1,2}\,E_{0,1}\,N_{3}\,N_{2}\,N_{1}\)\\
\(H X\) &
\(Z H\) &
\(X_{3}^{2}\,M_{2}\,Z_{3}^{1}\,M_{1}^{-X}\,X_{3}^{0}\,M_{0}\,E_{2,3}\,E_{1,2}\,E_{0,1}\,N_{3}\,N_{2}\,N_{1}\)\\
\(S^\dag H S\) &
\(H S H Z\) &
\(X_{2}^{1}\,M_{1}^{-Y}\,X_{2}^{0}\,Z_{2}^{0}\,M_{0}^{-X}\,E_{1,2}\,E_{0,1}\,N_{2}\,N_{1}\)\\
\(S H S^\dag\) &
\(S H Z S\) &
\(X_{2}^{1}\,M_{1}^{+Y}\,X_{2}^{0}\,Z_{2}^{0}\,M_{0}^{-X}\,E_{1,2}\,E_{0,1}\,N_{2}\,N_{1}\)\\
\(S H S\) &
\(S H S\) &
\(X_{2}^{1}\,M_{1}^{+Y}\,X_{2}^{0}\,Z_{2}^{0}\,M_{0}\,E_{1,2}\,E_{0,1}\,N_{2}\,N_{1}\)\\
\(H S^\dag\) &
\(H S Z\) &
\(X_{1}^{0}\,M_{0}^{+Y}\,E_{0,1}\,N_{1}\)\\
\(H S^\dag X\) &
\(Z H S\) &
\(X_{3}^{2}\,M_{2}^{+Y}\,X_{3}^{1}\,Z_{3}^{1}\,M_{1}^{-X}\,X_{3}^{0}\,M_{0}\,E_{2,3}\,E_{1,2}\,E_{0,1}\,N_{3}\,N_{2}\,N_{1}\)\\
\(H S X\) &
\(Z H Z S\) &
\(X_{3}^{2}\,M_{2}^{-Y}\,X_{3}^{1}\,Z_{3}^{1}\,M_{1}^{-X}\,X_{3}^{0}\,M_{0}\,E_{2,3}\,E_{1,2}\,E_{0,1}\,N_{3}\,N_{2}\,N_{1}\)\\
\(H S\) &
\(H S\) &
\(X_{1}^{0}\,M_{0}^{-Y}\,E_{0,1}\,N_{1}\)\\
\(S H\) &
\(S H\) &
\(X_{2}^{1}\,M_{1}^{+Y}\,X_{2}^{0}\,Z_{2}^{0}\,M_{0}^{+Y}\,E_{1,2}\,E_{0,1}\,N_{2}\,N_{1}\)\\
\(S^\dag H\) &
\(S Z H\) &
\(X_{2}^{1}\,M_{1}^{-Y}\,X_{2}^{0}\,Z_{2}^{0}\,M_{0}^{-Y}\,E_{1,2}\,E_{0,1}\,N_{2}\,N_{1}\)\\
\(S H Y\) &
\(S Z H Z\) &
\(X_{2}^{1}\,M_{1}^{-Y}\,X_{2}^{0}\,Z_{2}^{0}\,M_{0}^{+Y}\,E_{1,2}\,E_{0,1}\,N_{2}\,N_{1}\)\\
\(S^\dag H Y\) &
\(S H Z\) &
\(X_{2}^{1}\,M_{1}^{+Y}\,X_{2}^{0}\,Z_{2}^{0}\,M_{0}^{-Y}\,E_{1,2}\,E_{0,1}\,N_{2}\,N_{1}\)\\
\hline
\end{tabular}





\end{document}